\documentclass[a4paper,10pt]{article}
\usepackage{amssymb}
\usepackage{amsmath}
\usepackage[utf8]{inputenc}
\usepackage{fullpage}
\usepackage{boxedminipage}
\usepackage{listings}
\usepackage{minitoc}
\usepackage{graphicx}

\makeatletter \@addtoreset{equation}{section} \makeatother
\renewcommand{\theequation}{\thesection.\arabic{equation}}
\newif\ifpdf \ifx\pdfoutput\undefined \pdffalse
\else \pdfoutput=1 \pdftrue \fi \ifpdf \else \fi
\input epsf
\def\be{\begin{equation}}
\def\ee{\end{equation}}
\def\bea{\begin{eqnarray}}
\def\eea{\end{eqnarray}}

\begin{document}

\begin{titlepage}

    \thispagestyle{empty}
    \begin{flushright}
        \hfill{CERN-PH-TH/182} \\\hfill{UCLA/07/TEP/22}\\\hfill{SU-ITP-2007-21}
    \end{flushright}

    \vspace{5pt}
    \begin{center}
        { \Huge{\textbf{Extremal Black Hole\\\vspace{10pt}and Flux Vacua Attractors}}}\vspace{25pt}
        \vspace{55pt}

         {\large{\bf Stefano Bellucci$^\clubsuit$, Sergio Ferrara$^{\diamondsuit\clubsuit\flat}$, Renata Kallosh$^{\sharp}$ and\ Alessio Marrani$^{\heartsuit\clubsuit}$}}

        \vspace{15pt}

        {$\clubsuit$ \it INFN - Laboratori Nazionali di Frascati, \\
        Via Enrico Fermi 40,00044 Frascati, Italy\\
        \texttt{bellucci,marrani@lnf.infn.it}}

        \vspace{10pt}

        {$\diamondsuit$ \it Physics Department,Theory Unit, CERN, \\
        CH 1211, Geneva 23, Switzerland\\
        \texttt{sergio.ferrara@cern.ch}}

        \vspace{10pt}
         {$\flat$ \it Department of Physics and Astronomy,\\
        University of California, Los Angeles, CA USA\\
        \texttt{ferrara@physics.ucla.edu}}

        \vspace{10pt}
         {$\sharp$ \it Physics Department,\\
        Stanford University, Stanford, CA 94305-4060, USA\\
        \texttt{kallosh@stanford.edu}}
         \vspace{10pt}

        {$\heartsuit$ \it Museo Storico della Fisica e\\
        Centro Studi e Ricerche ``Enrico Fermi"\\
        Via Panisperna 89A, 00184 Roma, Italy}

        \vspace{50pt}
        \noindent \textit{Contribution to the Proceedings of the Winter School on Attractor Mechanism 2006 (SAM2006),\\20--24 March 2006, INFN--LNF, Frascati, Italy}
\end{center}


\begin{abstract}
These lectures provide a pedagogical, introductory review of the
so-called Attractor Mechanism (AM) at work in two different
$4$-dimensional frameworks: extremal black holes in $\mathcal{N}=2$ supergravity and $%
\mathcal{N}=1$ flux compactifications. In the first case, AM
determines the stabilization of scalars at the black hole event
horizon purely in terms of the electric and magnetic charges,
whereas in the second context the AM is responsible for the
stabilization of the universal axion-dilaton and of the (complex
structure) moduli purely in terms of the RR and NSNS fluxes. Two
equivalent approaches to AM, namely the so-called ``criticality
conditions'' and ``New Attractor'' ones, are analyzed in detail in
both frameworks, whose analogies and differences are discussed. Also
a stringy analysis of both frameworks (relying on
Hodge-decomposition techniques) is performed, respectively
considering Type IIB compactified on $CY_{3}$ and its
orientifolded version, associated with $\frac{CY_{3}\times T^{2}}{\mathbb{Z}%
_{2}}$. Finally, recent results on the $U$-duality orbits and moduli
spaces of non-BPS extremal black hole attractors in $3\leqslant
\mathcal{N}\leqslant 8$, $d=4$ supergravities are reported.
\end{abstract}

\end{titlepage}
\baselineskip6 mm \newpage\tableofcontents\newpage

\section{Introduction\label{Intro}}

After the original papers \cite{FKS}-\nocite{Strom,FK1,FK2}\cite{FGK} from
mid 90's dealing mostly with the Bogomol'ny-Prasad-Sommerfeld\newline
(BPS) black holes (BHs), \textit{extremal BH attractors} have been recently
widely investigated \cite{Sen-old1}--\nocite
{GIJT,Sen-old2,K1,TT,G,GJMT,Ebra1,K2,Ira1,Tom,
BFM,AoB-book,FKlast,Ebra2,BFGM1,rotating-attr,K3,Misra1,Lust2,BFMY,CdWMa,
DFT07-1,BFM-SIGRAV06,Cer-Dal,ADFT-2,Saraikin-Vafa-1,Ferrara-Marrani-1,TT2,
ADOT-1,fm07,CCDOP,Misra2,Astefanesei,Anber,Myung1,CFM1,BMOS-1,Hotta,Gao,
PASCOS07,Sen-review,Belhaj1,AFMT1,Gaiotto1,BFMS1,GLS1}\cite{ANYY1} (see also
\cite{OSV}--\nocite{Sen2,Verlinde1,OVV,DGOV,ANV,Pestun1,GSV}\cite{Fiol1}).
Such a \textit{renaissance} is mainly due to the (re)discovery of new
classes of solutions to the attractor equations corresponding to non-BPS
horizon geometries: certain configurations of moduli stabilized near the
horizon of extremal BHs exist which break supersymmetry. In addition, the
stabilization of moduli in the context of string theory has become a central
issue of string cosmology. The attractor equations used in the past for
stabilizing moduli near the horizon of an extremal BH have turned out to be
useful in the context of flux vacua.

In this introduction we will first briefly remind the basic structure of the
BPS BH attractors in $\mathcal{N}=2$, $d=4$ supergravity. After that, we
will outline the main features of the recent developments in non-BPS
extremal BH attractors and flux vacua, a detailed description of which will
be given in the subsequent sections.

An horizon extremal BH attractor geometry is in general supported by
particular configurations of the $1\times \left( 2n_{V}+2\right) $
symplectic vector of the BH field-strength fluxes, \textit{i.e.} of the BH
magnetic and electric charges:
\begin{equation}
Q\equiv \left( p^{\Lambda },q_{\Lambda }\right) ,~~~~~p^{\Lambda }\equiv
\frac{1}{4\pi }\int_{S_{\infty }^{2}}\mathcal{F}^{\Lambda },~~~q_{\Lambda
}\equiv \frac{1}{4\pi }\int_{S_{\infty }^{2}}\mathcal{G}_{\Lambda
},~~~\Lambda =0,1,...,n_{V},  \label{Gamma-tilde}
\end{equation}
where, in the case of $\mathcal{N}=2$, $d=4$ supergravity, $n_{V}$ denotes
the number of Abelian vector supermultiplets coupled to the supergravity one
(containing the Maxwell vector $A^{0}$, usually named \textit{graviphoton}).
Here $\mathcal{F}^{\Lambda }=dA^{\Lambda }$ and $\mathcal{G}_{\Lambda }$ is
the ``dual'' field-strength two-form \cite{3,4}.

\textit{BPS BH attractor equations} fix the values of all moduli near BH
horizon in terms of the electric and magnetic charges. The most compact form
of these equations was given in \cite{Behrndt:1996jn}, where the K\"{a}hler
invariant period $\left( Y^{\Lambda }\,,F_{\Lambda }\left( Y\right) \right) $
was introduced by multiplying the covariantly holomorphic period $V(z,\bar{z}%
)$ (see Eq. (\ref{PI}) below) on the (complex conjugate of the) $\mathcal{N}%
=2$, $d=4$ \textit{central charge function} $\bar{Z}$, so that
\begin{equation}
\bar{Z}V\equiv \left( Y^{\Lambda }\,,F_{\Lambda }\left( Y\right) \right)
\end{equation}
where $Y^{\Lambda }=Y^{\Lambda }\left( z,\bar{z}\right) $ and $V=\left(
L^{\Lambda }\,,M_{\Lambda }\right) $. In terms of such variables, the BPS
attractor equations are very simple and state that at the BH horizon the
moduli $\left( z,\bar{z}\right) $ depend on electric and magnetic charges so
that equations
\begin{equation}
Y^{\Lambda }-\bar{Y}^{\Lambda }=ip^{\Lambda }\ ,\qquad F_{\Lambda }\left(
Y\right) -\bar{F}_{\Lambda }\left( \bar{Y}\right) =iq_{\Lambda }
\label{BPSattr}
\end{equation}
are satisfied, and their solution defines moduli as functions of charges
\begin{equation}
z_{cr}=z_{cr}(p,q)\,,\qquad \bar{z}_{cr}=\bar{z}_{cr}(p,q).
\end{equation}
BPS attractors equations (\ref{BPSattr}) are equivalent to the condition of
unbroken supersymmetry: $DZ=0$.

A simple way to derive the BH attractor equations, which also gives a clear
link to their use in the context of flux vacua, is by using the language of
string theory compactified on a Calabi-Yau threefold ($CY_{3}$) \cite{Moore}%
. One starts with the Hodge decomposition of the 3-form flux (see Eq. (\ref
{decomp3}) below)
\begin{equation}
\mathcal{H}_{3}=-2Im[\bar{Z}\hat{\Omega}_{3}-\bar{D}^{i}\bar{Z}D_{i}\hat{%
\Omega}_{3}]=\int_{S_{\infty }^{2}}\widehat{\mathcal{F}}^{+},  \label{Hodge}
\end{equation}
where $\hat{\Omega}_{3}$ is the covariantly holomorphic 3-form of the $%
CY_{3} $, $\widehat{\mathcal{F}}^{+}$ is the self-dual 5-form of type IIB
string theory and $S_{\infty }^{2}$ is the 2-sphere at infinity, as in the
definition (\ref{Gamma-tilde}) (see \textit{e.g.} \cite{4}). By integration
over a symplectic basis of 3-cycles of $CY_{3}$ the decomposition (\ref
{Hodge}) can be brought to the form (see Eq. (\ref{SKG-final1}) below)
\begin{equation}
Q^{T}=-2Im[\bar{Z}V-\bar{D}^{i}\bar{Z}D_{i}V].  \label{HodgeInt}
\end{equation}
By inserting the condition of unbroken supersymmetry $D_{i}Z=0$ into the
identities (\ref{Hodge}) and (\ref{HodgeInt}), one obtains the BPS extremal
BH attractor Eqs. (\ref{BPSattr}) in a stringy framework:
\begin{equation}
\mathcal{H}_{3}=-2Im[\bar{Z}\hat{\Omega}_{3}]_{DZ=0},  \label{BPS1}
\end{equation}
or equivalently:
\begin{equation}
{Q}^{T}=-2Im\left[ \overline{Z}V\right] _{DZ=0}.  \label{BPSattr-2}
\end{equation}
This attractor Eq. presents a particular case of the criticality condition
for the so-called effective BH potential, $\partial _{i}V_{BH}=0$, where
(see definition (\ref{VBH1}) below)
\begin{equation}
V_{BH}(z,\bar{z})\equiv |Z|^{2}+g^{i\overline{j}}\left( D_{i}Z\right)
\overline{D}_{\overline{j}}\overline{Z}.
\end{equation}

Another important feature of the BPS attractors is the relation between the
second derivative of $V_{BH}$ at the critical points $\partial V_{BH}=0$ and
the metric $g_{i\overline{j}}$ of the scalar manifold (usually called
\textit{moduli space} in string theory), namely
\begin{equation}
\left( \partial _{i}\overline{\partial }_{\overline{j}}V_{BH}\right)
_{_{\partial V_{BH}=0}}=2\left( g_{i\overline{j}}V_{BH}\right) _{\partial
V_{BH}=0}.  \label{stability}
\end{equation}
Since $V_{BH}$ at the supersymmetric critical point $DZ=0$ (with
non-vanishing entropy) is strictly positive ($\left. V_{BH}\right|
_{DZ=0}=|Z|_{DZ=0}^{2}>0$), Eq. (\ref{stability}) implies that all BPS
attractors are stable, at least as long as the metric of the moduli space is
strictly positive definite. Note that in the BPS case the condition of
non-vanishing entropy requires $\left. Z\right| _{DZ=0}\neq 0$.

The recent developments with \textit{non-BPS BH attractors} can be described
shortly as follows. For extremal non-BPS BH solutions of $\mathcal{N}=2$, $%
d=4$ supergravity one finds the mechanism of stabilization of moduli near BH
horizon with some properties of the same nature as in BPS case, and some
properties somewhat different.

Many nice features of the BH attractors in the past were associated with the
unbroken supersymmetry of BPS BHs. During the last few years the basic
reason for the attractor behaviour of extremal BHs has been discovered to be
geometrical\footnote{%
We are grateful to A. Linde for this insight, see also \cite{K2}.}: extremal
BHs (regardless their supersymmetry-preserving features) all have moduli
which acquire fixed values at the BH horizon independent of their values at
infinity! Their values at the horizon depend only on the electric and
magnetic BH charges. The existence of an infinite throat in the space-time
geometry of extremal BHs leads to an evolution towards the horizon such that
the moduli forget their initial conditions at (spatial) infinity \cite{K2}.
Since a Schwarzschild-type BH geometry with non-vanishing horizon area is
never extremal, this phenomenum never takes place when solving the equations
of motion for scalar fields in such a background: their values at the
horizon depend on the initial conditions of the radial dynamical evolution,
because there are no coordinate systems with infinite distance from the
event horizon.

A simple qualifier of both BPS and non-BPS attractors remains valid in the
form of a critical point of the BH potential:
\begin{equation}
\partial V_{BH}=0\ :\qquad \mathit{for\;BPS:}\quad DZ=0,\qquad \mathit{%
for\;non-BPS:}\quad DZ\neq 0.
\end{equation}

The non-BPS attractor Eqs. in the form generalizing Eqs. (\ref{BPSattr}) can
be given separately for the cases $Z\neq 0$ and $Z=0$ \footnote{%
The non-BPS BH attractor Eqs. $\partial V_{BH}=0$ with the condition $Z=0,$ $%
DZ\neq 0$ will be discussed later in the lectures.}.

In the case $Z\neq 0$ one finds (see Eq. (\ref{alg-non-BPS-Z<>0-2}) below)
\begin{equation}
Q^{T}=-2Im\left\{ \left[ \overline{Z}V-\frac{i}{2}\frac{Z}{\left| Z\right|
^{2}}C^{\overline{i}\overline{j}\overline{k}}\left( \overline{D}_{\overline{j%
}}\overline{Z}\right) \left( \overline{D}_{\overline{k}}\overline{Z}\right)
\overline{D}_{\overline{i}}\overline{V}\right] \right\} _{non-BPS,Z\neq 0}.
\label{BPS-non-BPS}
\end{equation}
Here one starts with the identity (\ref{HodgeInt}) and replaces the second
term using the expression for it derived from the non-BPS $Z\neq 0$
criticality condition $\partial V_{BH}=0$. The attractor Eq. (\ref
{BPS-non-BPS}) is a clear generalization of the BPS attractor Eq. (\ref
{BPSattr})-(\ref{BPSattr-2}) with $Z\neq 0$: at $DZ=0$ the second term in
the right-hand side (r.h.s.) of Eq. (\ref{BPS-non-BPS}) vanishes and it
reduces exactly to Eq. (\ref{BPSattr-2}) or its detailed form given by Eq. (%
\ref{BPSattr}).

For both classes ($Z\neq 0$ and $Z=0$) of non-BPS attractors the critical
value of $V_{BH}$ remains positive, since by definition $V_{BH}$ is a real,
positive function in the scalar manifold. However, the universal BPS
stability condition (\ref{stability}) is not valid anymore and one has to
study this issue separately\footnote{%
In case that the critical Hessian matrix has some ``massless modes'' (%
\textit{i.e.} vanishing eigenvalues), one has to look at higher-order
covariant derivatives of $V_{BH}$ evaluated at the considered point, and
study their sign. Depending on the configurations of the BH charges, one can
obtain stable or unstable critical points.
\par
Examples in \ literature of investigations beyond the Hessian level can be
found in \cite{TT,K3,Misra1}. A detailed analysis of the stability of
critical points of $V_{BH}$ in (the large volume limit of) compactifications
of Type IIA superstrings on $CY_{3}$s has been recently given in \cite{TT2}.
\par
The issue of stability of non-BPS critical points of $V_{BH}$ in homogeneous
(not necessarily symmetric) $\mathcal{N}=2$, $d=4$ special K\"{a}hler
geometries has been treated exhaustively in \cite{fm07}. It was derived that
all non-BPS critical points of $V_{BH}$ in such geometries are stable, up to
a certain number of ``flat'' directions (present at all order in the
covariant differentiation of $V_{BH}$), which span a certain moduli space,
pertaining to the considered class of solutions of the attractor Eqs.. The
results of \cite{fm07} hold in general for any theory (not necessarily
involving supersymmetry) in which gravity is coupled to Abelian gauge
vectors and with a scalar sigma model endowed with homogeneous geometry (see
further below in the present lectures).}.

In the present review we will consider only critical points of $V_{BH}$ ($%
\frac{1}{2}$-BPS as well as non-BPS) which are \textit{non-degenerate},
\textit{i.e.} with a finite, non-vanishing horizon area, corresponding to
the so-called \textit{``large''} BHs\footnote{%
For further elucidations, we refer the reader \textit{e.g.} to the recent
lectures of Sen \cite{Sen-review}, where important aspects of BH attractors
are presented: the microscopic string theory counting of states explaining
the macroscopic BH entropy and the treatment of higher-derivative terms in
the actions.}.

Due to the so-called Attractor Mechanism (AM) \cite{FKS}-\nocite
{Strom,FK1,FK2}\cite{FGK}, the Bekenstein-Hawking entropy \cite{BH1}\textbf{%
\ } of \textit{``large''} extremal BHs can be obtained by extremizing $%
V_{BH}\left( \phi ,{Q}\right) $, where ``$\phi $'' now denotes the set of
real scalars relevant for the AM, and $Q$ is defined by Eq. (\ref
{Gamma-tilde}). In $\mathcal{N}=2$, $d=4$ supergravity, non-degenerate
attractor horizon geometries correspond to BH solitonic states belonging to $%
\frac{1}{2}$-BPS ``short massive multiplets'' or to non-BPS ``long massive
multiplets'' , respectively. The BPS bound \cite{BPS} requires that
\begin{equation}
M_{ADM}\geq |Z|,
\end{equation}
where $M_{ADM}$ denotes the Arnowitt-Deser-Misner (ADM) mass \cite{ADM}. At
the event horizon, extremal BPS BHs do saturate such a bound, whereas the
non-BPS ones satisfy\footnote{%
Here and in what follows, the subscript ``$H$'' will denote values at the BH
event horizon.}
\begin{equation}
\begin{array}{l}
\frac{1}{2}\text{\textit{-BPS}:~~}0<\left| Z\right| _{H}=M_{ADM,H}; \\
~ \\
\text{\textit{non-BPS~}}\left\{
\begin{array}{l}
Z\neq 0\text{:~~}0<\left| Z\right| _{H}<M_{ADM,H}; \\
\\
Z=0\text{:~~}0=\left| Z\right| _{H}<M_{ADM,H},
\end{array}
\right. \text{ }
\end{array}
\end{equation}
where $M_{ADM,H}$ is obtained by extremizing $V_{BH}\left( \phi ,{Q}\right) $
with respect to its dependence on the scalars:
\begin{equation}
M_{ADM,H}\left( {Q}\right) =\sqrt{\left. V_{BH}\left( \phi ,{Q}\right)
\right| _{\partial _{\phi }V_{BH}=0}}.
\end{equation}

The (purely) charge-dependent BH entropy $S_{BH}$ is given by the
Bekenstein-Hawking entropy-area formula \cite{BH1,FGK}
\begin{equation}
S_{BH}\left( {Q}\right) =\frac{A_{H}\left( {Q}\right) }{4}=\pi \left.
V_{BH}\left( \phi ,{Q}\right) \right| _{\partial _{\phi }V_{BH}=0}=\pi
V_{BH}\left( \phi _{H}\left( {Q}\right) ,{Q}\right) ,  \label{BHEA}
\end{equation}
where $A_{H}$ is the area of the BH event horizon.

Non-degenerate, non-supersymmetric (non-BPS) extremal BH (and black string)
attractors arise also in $\mathcal{N}=2$, $d=5$, $6$ supergravity and in $%
\mathcal{N}>2$, $d=4$, $5$, $6$ extended supergravities (see \textit{e.g.}
\cite
{FG,FG2,Ferrara-Maldacena,Larsen-review,FKlast,ADFT,CFM1,CCDOP,PASCOS07,AFMT1}%
, and Refs. therein). In the present lectures we will focus on extremal BH
attractors in $\mathcal{N}=2$, $d=4$ ungauged supergravity coupled to
Abelian vector multiplets, where the scalar manifold parameterized by the
scalars is endowed with the so-called special K\"{a}hler (SK) geometry (see
Sect. \ref{SKG-gen}).\bigskip \smallskip

\textit{Flux vacua} (FV) became recently one of the new playgrounds for
string theory, in general and in particular in the context of moduli
stabilization (for an introduction to flux compactifications, see \textit{%
e.g.} \cite
{GMPT,Grana-review,Louis-review,Douglas-Kachru-review,Blumenhagen-review}
and Refs. therein).

The advances of observational cosmology and the emergence of the so-called
\textit{``standard cosmological model''} enforce on string
theory/supergravity a responsibility to address the current and future
observations. This requires a solution of the problem of moduli
stabilization. In the early Universe during inflation, all string theory
moduli but the inflaton have to be stabilized, in order to produce an
effective four-dimensional General Relativity and also in order for
inflation to explain the cosmic microwave background observations. At the
present time, all moduli have to be stabilized in a four-dimensional de
Sitter space to explain dark energy and acceleration of the Universe which
took place during the last few billion years.

The procedure of moduli stabilization in string theory consists of few steps.

One of the steps is the stabilization of moduli by fluxes in type IIB string
theory, determining $d=4$ FV, with effective $\mathcal{N}=1$ local
supersymmetry and complex structure moduli stabilized; an important feature
of such a procedure is the non-stabilization of the K\"{a}hler moduli.
However, the largest contribution to the counting of the Calabi-Yau vacua in
the so-called \textit{String Landscape} comes from the diversity of FV.

Thus, it is still interesting to study the mechanism of stabilization of the
axion-dilaton and complex structure moduli in FV, ignoring the K\"{a}hler
moduli. We will deal with such a scenario, in the particular case in which
the geometry of the complex structure moduli is SK, and not simply
K\"{a}hler. In such a framework, it turns out that the Eqs. determining the
FV configurations are closely related to the abovementioned extremal BH
attractor Eqs. of $\mathcal{N}=2$, $d=4$ supergravity.

In the studies of FV one can start with an F-theory flux compactification on
an elliptically fibered Calabi-Yau fourfold $CY_{4}$ in the orientifold
limit in which $CY_{4}=\frac{CY_{3}\times T^{2}}{\mathbb{Z}_{2}}$, where $%
T^{2}$ is the two-torus. In type IIB string theory, this is equivalent to
compactifying on the orientifold limit of $CY_{3}$. The resulting low
energy, $d=4$ effective theory is ${\mathcal{N}}=1$ supergravity, where the
information on string theory choice of compactification is encoded into a
flux superpotential $W$ and a K\"{a}hler potential $K$. As explained above,
we assume that both flux superpotential $W$ and K\"{a}hler potential $K$
depend only on the complex structure (CS) moduli of $CY_{4}$, spanning the
CS moduli space $M$. Because of the orientifold limit of $CY_{4}$, $M$ has
the product structure ${\mathcal{M}}={\mathcal{M}}_{CS}(CY_{3})\times {%
\mathcal{M}}_{\tau }$ (see Eq. (\ref{MM}) further below), where ${\mathcal{M}%
}_{CS}(CY_{3})$ (simply named ${\mathcal{M}}_{CS}$ further below) is the CS
moduli space of $CY_{3}$ and ${\mathcal{M}}_{\tau }$ is the moduli space of
the elliptic curve $T^{2}$ spanned by the axion-dilaton $\tau $ (named $%
t^{0} $ in the treatment of the present lectures). Let us here just mention
that in order to stabilize also the K\"{a}hler moduli of $CY_{4}$, one
should incorporate the non-perturbative string effects (see \textit{e.g.}
\cite{Kachru:2003aw}), which however we will not discuss here.

The potential in the effective $\mathcal{N}=1$, $d=4$ supergravity theory,
in the Planckian units set equal to one, is given by \cite
{Bagger-Witten,Cremmer-et-al}
\begin{equation}
V_{\mathcal{N}=1}=e^{K}\left(
\sum_{A=0}^{h_{2,1}(CY_{3})}|D_{A}W|^{2}-3|W|^{2}\right)
=\sum_{A=0}^{h_{2,1}(CY_{3})}|D_{A}Z|^{2}-3|Z|^{2}\ ,  \label{14-Nov-1}
\end{equation}
where $A=0$ refers to the axion-dilaton $\tau \equiv t^{0}$ and $A=i\in
\{1,\cdots ,h_{2,1}(CY_{3})\}$ to the CS moduli $t^{i}$ of $CY_{3}$ ($%
h_{2,1}\equiv dim\left( H^{2,1}\left( CY_{3}\right) \right) $; see Subsect.
\ref{CY3-Orientifolds}). We defined $Z\equiv e^{\frac{K}{2}}W$, as for the
extremal BH attractors in $\mathcal{N}=2$, $d=4$ supergravity, even if the
analogy is only formal, because in the present $d=4$ framework with $%
\mathcal{N}=1$ local supersymmetry there is no central charge at all.
\newline

The real K\"{a}hler potential of the effective $\mathcal{N}=1$, $d=4$
supergravity theory reads (see Eq. (\ref{FV-KK}) below)
\begin{equation}
K=-ln\langle \Omega _{4},\bar{\Omega}_{4}\rangle =-ln(\langle \Omega _{1},%
\bar{\Omega}_{1}\rangle )-ln(\langle \Omega _{3},\bar{\Omega}_{3}\rangle )\ ,
\end{equation}
where $\Omega _{4}$ is a nowhere vanishing holomorphic 4-form defined on $%
CY_{4}$. In the orientifold limit, $\Omega _{4}$ is a product of an
appropriate holomorphic 3-form $\Omega _{3}$ of $CY_{3}$ and the holomorphic
1-form $\Omega _{1}$ of the torus $T^{2}$ (see Subsubsect. \ref{1-3-4-Forms}%
).

The flux holomorphic superpotential $W$ is defined as a section of the line
bundle ${\mathcal{L}}$ by \cite{Gukov:1999ya,Denef-Douglas-1} (see Eq. (\ref
{FV-WW}) below)
\begin{equation}
W\equiv Ze^{-{\frac{K}{2}}}=\langle \mathcal{F}_{4},\Omega _{4}\rangle
\equiv \int_{CY_{4}}\mathcal{F}_{4}\wedge \Omega _{4}\ ,
\end{equation}
where $\mathcal{F}_{4}\in H^{4}(CY_{4})$ is the 4-form flux.

In generic local \textit{``}flat'' coordinates of $M$ (with $0$ and $A$%
-indices respectively referring to the axion-dilaton and CS moduli of $%
CY_{3} $), $\mathcal{F}_{4}$ enjoys the following Hodge decomposition, which
we present here in terms of $Z$ for the sake of comparison with its \textit{%
``}BH-counterpart'' (\textit{i.e.} the Hodge decomposition (\ref{Hodge}) of
the 3-form flux $\mathcal{H}_{3}$) ($\hat{\Omega}_{4}\equiv e^{\frac{1}{2}%
K}\Omega _{4}$; see definition (\ref{Omega4-hat}) and Eq. (\ref{FV-decomp2})
below):
\begin{equation}
\frak{F}_{4}=2Re\left[ \overline{Z}\hat{\Omega}_{4}-\delta ^{A\overline{B}%
}\left( \overline{D}_{\overline{B}}\overline{Z}\right) D_{A}\hat{\Omega}%
_{4}+\delta ^{A\overline{B}}\left( \overline{D}_{\underline{\overline{0}}}%
\overline{D}_{\overline{B}}\overline{Z}\right) D_{\underline{0}}D_{A}\hat{%
\Omega}_{4}\right] .  \label{G-dec}
\end{equation}
By imposing the supersymmetry-preserving condition $DZ=0$ (formally
identical to the one appearing in the abovementioned theory of extremal BH
attractors in $\mathcal{N}=2$, $d=4$ supergravity), the identity (\ref{G-dec}%
) becomes a supersymmetric FV Attractor Eq.. Indeed, the left-hand side
(l.h.s.) depends on fluxes and the r.h.s. depends on axion-dilaton and on CS
moduli of $CY_{3}$; thus, the solution stabilizes the axion-dilaton and the
CS moduli of $CY_{3}$ purely in terms of fluxes: beside Eq. (\ref{BPS1}),
one gets
\begin{equation}
\frak{F}_{4}=2Re\left[ \overline{Z}\hat{\Omega}_{4}+\delta ^{A\overline{B}%
}\left( \overline{D}_{\underline{\overline{0}}}\overline{D}_{\overline{B}}%
\overline{Z}\right) D_{\underline{0}}D_{A}\hat{\Omega}_{4}\right] _{DZ=0}.
\label{FV-SUSY-Attractor-Eqs}
\end{equation}

Let us now compare the supersymmetric FV Attractor Eqs. (\ref
{FV-SUSY-Attractor-Eqs}) with their \textit{``}BH-counterpart'', \textit{i.e.%
} with the BPS extremal BH Attractor Eqs. (\ref{BPS1}).

In the case of FV, instead of the imaginary part we have a real part of a
somewhat analogous expression: this is due to the fact that for FV one has a
4-form flux $\frak{F}_{4}$ on a $CY_{3}$ orientifold rather than of a 3-form
flux $\mathcal{H}_{3}$ on $CY_{3}$.

The other significant difference is in the second term in the r.h.s. of Eq. (%
\ref{FV-SUSY-Attractor-Eqs}). This term, absent in the BH case, is
proportional to the second-order covariant derivative of $Z$ along the $\tau
$ direction and one of the directions pertaining to the CS moduli of $CY_{3}$%
. In the BH case there is a relation $D_{i}D_{j}Z=iC_{ijk}\overline{D}^{k}%
\overline{Z}$ (see the second of Eqs. (\ref{SKG-rels1}) below), and
therefore the second covariant derivative of $Z$ is not an independent term
for BH, differently from $D_{\underline{0}}D_{I}Z$ in the FV case, which is
an independent term in the decomposition of forms.

The absence of such a term in the BPS extremal BH Attractor Eqs. (\ref{BPS1}%
) does not allow for non-degenerate BPS extremal BH attractors with
vanishing central charge: indeed, on the BH side $Z=0$ and $DZ=0$ yield $%
V_{BH}=0$. This limit case corresponds to a classical \textit{``small}''
extremal BH, exhibiting a naked singularity because the area of the BH event
horizon vanishes. The Attractor Mechanism in such a case simply ceases to
hold, because for $Z=0$ and $DZ=0$ the BPS extremal BH Attractor Eqs. (\ref
{BPSattr}) admit as unique solution $Q=0$.

The same does not happen on the FV side. Indeed, by substituting $Z=0$ and $%
DZ=0$ into the Hodge decomposition (\ref{G-dec}) does not generate any
unconsistency: the second term in the r.h.s. of Eq. (\ref
{FV-SUSY-Attractor-Eqs}) provides a consistent solution for superymmetric
Minkowski vacua (with $DZ=0$ and $V_{\mathcal{N}=1}=0$). Of course, more
general supersymmetric solutions with $Z\neq 0$ are allowed, and they
correspond to supersymmetric AdS FV (see \textit{e.g.} \cite
{G,Soroush:2007ed}).\bigskip \smallskip

The aim of the present paper is to show the Attractor Mechanism at work in
two completely different $d=4$ frameworks: extremal BH in $\mathcal{N}=2$
supergravity and $\mathcal{N}=1$ flux compactifications.

The plan of the paper is as follows.

In Sect. \ref{SKG-gen} we recall the fundamentals of the special K\"{a}hler
geometry, underlying the vector multiplets' scalar manifold of $\mathcal{N}%
=2 $, $d=4$ ungauged supergravity, as well as the complex structure moduli
space of certain $\mathcal{N}=1$, $d=4$ supergravities obtained by
consistently orientifolding of $\mathcal{N}=2$ theories, such as Type IIB
compactified on $\frac{CY_{3}\times T^{2}}{\mathbb{Z}_{2}}$.

Sect. \ref{extremal-BH} gives an introduction to the issue of the Attractor
Mechanism in the framework where it was originally discovered by Ferrara,
Kallosh and Strominger \cite{FKS}-\nocite{Strom,FK1,FK2}\cite{FGK}, namely
in the stabilization of the vector multiplet' scalars near the event horizon
of an extremal, static, spherically symmetric and asymptotically flat BH in $%
\mathcal{N}=2$, $d=4$ ungauged supergravity.

Subsect. \ref{Sect2} presents the so-called \textit{``criticality
conditions''} approach to the Attractor Mechanism, in which the purely
charge-dependent stabilized configurations of the scalars at the BH horizon
can be computed as the critical points of a certain real positive BH
effective potential $V_{BH}$, whose classification is given in Subsubsect.
\ref{classification-crit-BH}. The stability of the critical points of $%
V_{BH} $ is then analyzed in Subsect. \ref{Sect4}, both in the general case
of $n_{V}$ moduli (Subsubsect. \ref{Sect4-n}) and in the 1-modulus case
(Subsubsect. \ref{Sect4-1}).

Subsect. \ref{General-Formulation} presents another, equivalent approach to
the Attractor Mechanism, recently named \textit{``New Attractor''} approach.
In Subsubsect. \ref{BH-New-Attractor} it is exploited in a general $\mathcal{%
N}=2$, $d=4$ supergravity framework, by substituting the (various classes
of) criticality conditions of $V_{BH}$ into some geometrical identities of
special K\"{a}hler geometry, expressing nothing but a change of basis
between \textit{``dressed''} and \textit{``undressed''} charges, and derived
in Subsubsect. \ref{Derivation-SKG-identities}.

Subsect. \ref{Micro-BH} implements the \textit{``New Attractor''} approach
in a stringy framework, namely in Type IIB compactified on $CY_{3}$. In
Subsubsect. \ref{Stringy-BH-New-Attractor} the (various classes of)
criticality conditions of $V_{BH}$ are inserted into some general identities
(equivalent to the identities derived in Subsubsect. \ref
{Derivation-SKG-identities}), expressing the decomposition of the real,
K\"{a}hler gauge-invariant 3-form flux $\mathcal{H}_{3}$ along the third
Dalbeault cohomogy of $CY_{3}$, and derived in Subsubsect. \ref
{Hodge-decomposition}.

Sect. \ref{Flux-Vacua-Attractors} deals with the Attractor Mechanism in a
completely different framework, namely in $\mathcal{N}=1$, $d=4$ ungauged
supergravity obtained by consistently orientifolding the $\mathcal{N}=2$
theory, and thus mantaining a special K\"{a}hler geometry of the manifold of
the scalars surviving the orientifolding. In the considered example of Type
IIB associated with $\frac{CY_{3}\times T^{2}}{\mathbb{Z}_{2}}$, the
Attractor Mechanism determines the stabilization of the universal
axion-dilaton and of the complex structure moduli in terms of the
Ramond-Ramond (RR) and Neveu-Schwarz-Neveu-Schwarz (NSNS) fluxes.

Subsect. \ref{CY3-Orientifolds} introduces the fundamentals of the geometry
of (the moduli space of) $CY_{3}$ orientifolds: the vielbein and the metric
tensor (Subsubsect. \ref{Bein and Metric}), the relevant 1-, 3- and 4-forms
(Subsubsect. \ref{1-3-4-Forms}), and the Hodge decomposition of the real,
K\"{a}hler gauge-invariant 4-form flux $\frak{F}_{4}$, unifying the RR
3-form flux $\frak{F}_{3}$ with the NSNS 3-form flux $\frak{H}_{3}$
(Subsubsect. \ref{FV-Hodge-Decomposition}).

Subsect. \ref{FV-Criticality-Conditions-approach} presents the so-called
\textit{``criticality conditions''} approach to the Attractor Mechanism in
flux vacua (FV) compactifications of the kind considered above, in which the
(complex structure) moduli space is endowed with special K\"{a}hler
geometry. The purely flux-dependent stabilized vacuum configurations of the
axion-dilaton and complex structure moduli can be computed as the critical
points of a certain real (not necessarily positive) FV effective potential $%
V_{\mathcal{N}=1}$. Since (differently from its BH $\mathcal{N}=2$
counterpart $V_{BH}$) $V_{\mathcal{N}=1}$ has no definite sign, the FV
attractor configurations can correspond to a de Sitter (dS, $V_{\mathcal{N}%
=1}>0$), Minkowski ($V_{\mathcal{N}=1}=0$) or anti-de Sitter (AdS, $V_{%
\mathcal{N}=1}<0$) vacuum.

Finally, Subsect. \ref{SUSY-Flux-Vacua-Attractors} implements the \textit{%
``New Attractor''} approach to the Attractor mechanism in the considered
class of FV compactifications, in the case of supersymmetric vacuum
configurations. The supersymmetric criticality conditions of $V_{\mathcal{N}%
=1}$ are inserted into the Hodge decomposition of the 4-form flux $\frak{F}%
_{4}$, and the resulting supersymmetric FV Attractor Eqs. lead to the
classification of the supersymmetric FV into three general families.

Two Appendices, containing technical details, conclude the lectures.%
\setcounter{equation}0

\section{\label{SKG-gen}Special K\"{a}hler Geometry}

In the present Section we briefly recall the fundamentals of the SK geometry
underlying the scalar manifold $\mathcal{M}_{n_{V}}$ of $\mathcal{N}=2$, $%
d=4 $ supergravity, $n_{V}$ being the number of Abelian vector
supermultiplets coupled to the supergravity multiplet ($dim_{\mathbb{C}}%
\mathcal{M}_{n_{V}}=n_{V}$) (see \textit{e.g.} \cite
{dWVP,CKV,CVP,dWLVP,Strominger-SKG,Castellani1,DFF,dWVVP,3,4}).

It is convenient to switch from the Riemannian $2n_{V}$-dim.
parameterization of $\mathcal{M}_{n_{V}}$ given by the local real
coordinates $\left\{ \phi ^{a}\right\} _{a=1,...,2n_{V}}$ to the K\"{a}hler $%
n_{V}$-dim. holomorphic/antiholomorphic parameterization given by the local
complex coordinates $\left\{ z^{i},\overline{z}^{\overline{i}}\right\} _{i,%
\overline{i}=1,...,n_{V}}$. This corresponds to the following \textit{%
unitary Cayley transformation}:
\begin{equation}
z^{k}\equiv \frac{\varphi ^{2k-1}+i\varphi ^{2k}}{\sqrt{2}},~~k=1,...,n_{V}.
\label{unit-transf}
\end{equation}

The metric structure of $\mathcal{M}_{n_{V}}$ is given by the covariant
(special) K\"{a}hler metric tensor $g_{i\overline{j}}\left( z,\overline{z}%
\right) =\partial _{i}\overline{\partial }_{\overline{j}}K\left( z,\overline{%
z}\right) $, $K\left( z,\overline{z}\right) $ being the real K\"{a}hler
potential.

Usually, the $n_{V}\times n_{V}$ Hermitian matrix $g_{i\overline{j}}$ is
assumed to be non-degenerate (\textit{i.e.} invertible, with non-vanishing
determinant and rank $n_{V}$) and with strict positive Euclidean signature (%
\textit{i.e.} with all strictly positive eigenvalues)\textit{\ globally} in $%
\mathcal{M}_{n_{V}}$. We will so assume, even though we will be concerned
mainly with the properties of $g_{i\overline{j}}$ at those peculiar points
of $\mathcal{M}_{n_{V}}$ which are critical points of $V_{BH}$.

It is worth here remarking that various possibilities arise when going
beyond the assumption of \textit{global strict regular }$g_{i\overline{j}}$,
namely:

- (\textit{locally}) \textit{not strictly regular} $g_{i\overline{j}}$,
\textit{i.e. }a (\textit{locally}) non-invertible metric tensor, with some
strictly positive and some vanishing eigenvalues (rank $<n_{V}$);

- (\textit{locally}) \textit{non-regular non-degenerate} $g_{i\overline{j}}$%
, \textit{i.e. }a (\textit{locally}) invertible metric tensor with \textit{%
pseudo-Euclidean signature}, namely with some strictly positive and some
strictly negative eigenvalues (rank $=n_{V}$);

- (\textit{locally}) \textit{non-regular degenerate} $g_{i\overline{j}}$,
\textit{i.e. }a (\textit{locally}) non-invertible metric tensor with some
strictly positive, some strictly negative, and some vanishing eigenvalues
(rank $<n_{V}$).

The \textit{local} violation of strict regularity of $g_{i\overline{j}}$
would produce some kind of ``phase transition'' in the SKG endowing $%
\mathcal{M}_{n_{V}}$, corresponding to a breakdown of the $1$-dim. effective
Lagrangian picture (see \cite{FGK}, \cite{2}, and also \cite{AoB-book} and
\cite{ADFT}) of $d=4$ (extremal) BHs obtained by integrating all massive
states of the theory out, unless new massless states appear \cite{FGK}.

The previously mentioned $\mathcal{N}=2$, $d=4$ covariantly holomorphic
\textit{central charge function} is defined as
\begin{equation}
\begin{array}{l}
Z\left( z,\overline{z};q,p\right) \equiv Q\epsilon V\left( z,\overline{z}%
\right) =q_{\Lambda }L^{\Lambda }\left( z,\overline{z}\right) -p^{\Lambda
}M_{\Lambda }\left( z,\overline{z}\right) =e^{\frac{1}{2}K\left( z,\overline{%
z}\right) }Q\epsilon \Pi \left( z\right) = \\
\\
=e^{\frac{1}{2}K\left( z,\overline{z}\right) }\left[ q_{\Lambda }X^{\Lambda
}\left( z\right) -p^{\Lambda }F_{\Lambda }\left( z\right) \right] \equiv e^{%
\frac{1}{2}K\left( z,\overline{z}\right) }W\left( z;q,p\right) ,
\end{array}
\label{Z}
\end{equation}
where $\epsilon $ is the $\left( 2n_{V}+2\right) $-dim. square symplectic
metric (subscripts denote dimensions of square sub-blocks)
\begin{equation}
\epsilon \equiv \left(
\begin{array}{ccc}
0_{n_{V}+1} &  & -\mathbb{I}_{n_{V}+1} \\
&  &  \\
\mathbb{I}_{n_{V}+1} &  & 0_{n_{V}+1}
\end{array}
\right) ,  \label{Omega}
\end{equation}
and $V\left( z,\overline{z}\right) $ and $\Pi \left( z\right) $ respectively
stand for the $\left( 2n_{V}+2\right) \times 1$ covariantly holomorphic
(K\"{a}hler weights $\left( 1,-1\right) $) and holomorphic (K\"{a}hler
weights $\left( 2,0\right) $) period vectors in symplectic basis:
\begin{equation}
\begin{array}{c}
\overline{D}\overline{_{i}}V\left( z,\overline{z}\right) =\left( \overline{%
\partial }_{\overline{i}}-\frac{1}{2}\overline{\partial }_{\overline{i}%
}K\right) V\left( z,\overline{z}\right) =0,~~~D_{i}V\left( z,\overline{z}%
\right) =\left( \partial _{i}+\frac{1}{2}\partial _{i}K\right) V\left( z,%
\overline{z}\right) \\
\\
\Updownarrow \\
\\
V\left( z,\overline{z}\right) =e^{\frac{1}{2}K\left( z,\overline{z}\right)
}\Pi \left( z\right) ,~~\overline{D}\overline{_{i}}\Pi \left( z\right) =%
\overline{\partial }_{\overline{i}}\Pi \left( z\right) =0,~~~D_{i}\Pi \left(
z\right) =\left( \partial _{i}+\partial _{i}K\right) \Pi \left( z\right) ,
\\
\\
\Pi \left( z\right) \equiv \left(
\begin{array}{c}
X^{\Lambda }\left( z\right) \\
\\
F_{\Lambda }\left( X\left( z\right) \right)
\end{array}
\right) =exp\left( -\frac{1}{2}K\left( z,\overline{z}\right) \right) \left(
\begin{array}{c}
L^{\Lambda }\left( z,\overline{z}\right) \\
\\
M_{\Lambda }\left( z,\overline{z}\right)
\end{array}
\right) ,
\end{array}
\label{PI}
\end{equation}
with $X^{\Lambda }\left( z\right) $ and $F_{\Lambda }\left( X\left( z\right)
\right) $ being the holomorphic sections of the $U(1)$ line (Hodge) bundle
over $\mathcal{M}_{n_{V}}$. $W\left( z;q,p\right) $ is the so-called \textit{%
holomorphic }$\mathcal{N}=2$, $d=4$\textit{\ central charge function}, also
named $\mathcal{N}=2$ \textit{superpotential}.

Up to some particular choices of local symplectic coordinates in $\mathcal{M}%
_{n_{V}}$, the covariant symplectic holomorphic sections $F_{\Lambda }\left(
X\left( z\right) \right) $ may be seen as derivatives of an \textit{%
holomorphic prepotential} function $F$ (with K\"{a}hler weights $\left(
4,0\right) $):
\begin{equation}
F_{\Lambda }\left( X\left( z\right) \right) =\frac{\partial F\left( X\left(
z\right) \right) }{\partial X^{\Lambda }}.  \label{prepotential}
\end{equation}
In $\mathcal{N}=2$, $d=4$ supergravity the holomorphic function $F$ is
constrained to be homogeneous of degree $2$ in the contravariant symplectic
holomorphic sections $X^{\Lambda }\left( z\right) $, \textit{i.e.} (see \cite
{4} and Refs. therein)
\begin{equation}
2F\left( X\left( z\right) \right) =X^{\Lambda }\left( z\right) F_{\Lambda
}\left( X\left( z\right) \right) .  \label{hom-prop-F}
\end{equation}

The normalization of the holomorphic period vector $\Pi \left( z\right) $ is
such that
\begin{equation}
K\left( z,\overline{z}\right) =-ln\left[ i\left\langle \Pi \left( z\right) ,%
\overline{\Pi }\left( \overline{z}\right) \right\rangle \right] \equiv -ln%
\left[ i\Pi ^{T}\left( z\right) \epsilon \overline{\Pi }\left( \overline{z}%
\right) \right] =-ln\left\{ i\left[ \overline{X}^{\Lambda }\left( \overline{z%
}\right) F_{\Lambda }\left( z\right) -X^{\Lambda }\left( z\right) \overline{F%
}_{\Lambda }\left( \overline{z}\right) \right] \right\} ,  \label{norm-PI}
\end{equation}
where $\left\langle \cdot ,\cdot \right\rangle $ stands for the symplectic
scalar product defined by $\epsilon $.

Note that under a K\"{a}hler transformation
\begin{equation}
K\left( z,\overline{z}\right) \longrightarrow K\left( z,\overline{z}\right)
+f\left( z\right) +\overline{f}\left( \overline{z}\right)
\end{equation}
($f\left( z\right) $ being a generic holomorphic function), the holomorphic
period vector transforms as
\begin{equation}
\Pi \left( z\right) \longrightarrow \Pi \left( z\right) e^{-f\left( z\right)
}\Leftrightarrow X^{\Lambda }\left( z\right) \longrightarrow X^{\Lambda
}\left( z\right) e^{-f\left( z\right) }.
\end{equation}
This means that, at least locally, the contravariant holomorphic symplectic
sections $X^{\Lambda }\left( z\right) $ can be regarded as a set of
homogeneous coordinates on $\mathcal{M}_{n_{V}}$, provided that the Jacobian
complex $n_{V}\times n_{V}$ holomorphic matrix
\begin{equation}
e_{i}^{a}\left( z\right) \equiv \frac{\partial }{\partial z^{i}}\left( \frac{%
X^{a}\left( z\right) }{X^{0}\left( z\right) }\right) ,\text{ ~}a=1,...,n_{V}
\end{equation}
is invertible. If this is the case, then one can introduce the local
projective symplectic coordinates
\begin{equation}
t^{a}\left( z\right) \equiv \frac{X^{a}\left( z\right) }{X^{0}\left(
z\right) },
\end{equation}
and the SKG of $\mathcal{M}_{n_{V}}$ turns out to be based on the
holomorphic prepotential $\mathcal{F}\left( t\right) \equiv \left(
X^{0}\right) ^{-2}F\left( X\right) $. By using the $t$-coordinates, Eq. (\ref
{norm-PI}) can be rewritten as follows ($\mathcal{F}_{a}\left( t\right)
=\partial _{a}\mathcal{F}\left( t\right) $, $\overline{t}^{a}=\overline{t^{a}%
}$, $\overline{\mathcal{F}}_{a}\left( \overline{t}\right) =\overline{%
\mathcal{F}_{a}\left( t\right) }$):
\begin{equation}
K\left( t,\overline{t}\right) =-ln\left\{ i\left| X^{0}\left( z\left(
t\right) \right) \right| ^{2}\left[ 2\left( \mathcal{F}\left( t\right) -%
\overline{\mathcal{F}}\left( \overline{t}\right) \right) -\left( t^{a}-%
\overline{t}^{a}\right) \left( \mathcal{F}_{a}\left( t\right) +\overline{%
\mathcal{F}}_{a}\left( \overline{t}\right) \right) \right] \right\} .
\end{equation}
By performing a K\"{a}hler gauge-fixing with $f\left( z\right) =ln\left(
X^{0}\left( z\right) \right) $, yielding that $X^{0}\left( z\right)
\longrightarrow 1$, one thus gets
\begin{equation}
\left. K\left( t,\overline{t}\right) \right| _{X^{0}\left( z\right)
\longrightarrow 1}=-ln\left\{ i\left[ 2\left( \mathcal{F}\left( t\right) -%
\overline{\mathcal{F}}\left( \overline{t}\right) \right) -\left( t^{a}-%
\overline{t}^{a}\right) \left( \mathcal{F}_{a}\left( t\right) +\overline{%
\mathcal{F}}_{a}\left( \overline{t}\right) \right) \right] \right\} .
\label{K-t-X0=1}
\end{equation}
In particular, one can choose the so-called \textit{special coordinates},
i.e. the system of local projective $t$-coordinates such that
\begin{equation}
e_{i}^{a}\left( z\right) =\delta _{i}^{a}\Leftrightarrow t^{a}\left(
z\right) =z^{i}\left( +c^{i},\text{~}c^{i}\in \mathbb{C}\right) .
\end{equation}
Thus, Eq. (\ref{K-t-X0=1}) acquires the form
\begin{equation}
\left. K\left( t,\overline{t}\right) \right| _{X^{0}\left( z\right)
\longrightarrow 1,e_{i}^{a}\left( z\right) =\delta _{i}^{a}}=-ln\left\{ i%
\left[ 2\left( \mathcal{F}\left( z\right) -\overline{\mathcal{F}}\left(
\overline{z}\right) \right) -\left( z^{j}-\overline{z}^{\overline{j}}\right)
\left( \mathcal{F}_{j}\left( z\right) +\overline{\mathcal{F}}_{\overline{j}%
}\left( \overline{z}\right) \right) \right] \right\} .
\end{equation}

Moreover, it should be recalled that $Z$ has K\"{a}hler weights $\left( p,%
\overline{p}\right) =\left( 1,-1\right) $, and therefore its
K\"{a}hler-covariant derivatives read
\begin{equation}
D_{i}Z=\left( \partial _{i}+\frac{1}{2}\partial _{i}K\right) Z,~~\overline{D}%
_{\overline{i}}Z=\left( \overline{\partial }_{\overline{i}}-\frac{1}{2}%
\overline{\partial }_{\overline{i}}K\right) Z.  \label{DiZ}
\end{equation}

The fundamental differential relations of SK geometry are\footnote{%
Actually, there are different (equivalent) defining approaches to SK
geometry. For subtleties and further elucidation concerning such an issue,
see \textit{e.g.} \cite{Craps1} and \cite{Craps2}.} (see \textit{e.g.} \cite
{4}):
\begin{equation}
\left\{
\begin{array}{l}
D_{i}Z=Z_{i}; \\
\\
D_{i}Z_{j}=iC_{ijk}g^{k\overline{k}}\overline{D}_{\overline{k}}\overline{Z}%
=iC_{ijk}g^{k\overline{k}}\overline{Z}_{\overline{k}}; \\
\\
D_{i}\overline{D}_{\overline{j}}\overline{Z}=D_{i}\overline{Z}_{\overline{j}%
}=g_{i\overline{j}}\overline{Z}; \\
\\
D_{i}\overline{Z}=0,
\end{array}
\right.  \label{SKG-rels1}
\end{equation}
where the first relation is nothing but the definition of the so-called
\textit{matter charges} $Z_{i}$, and the fourth relation expresses the
K\"{a}hler-covariant holomorphicity of $Z$. $C_{ijk}$ is the rank-3,
completely symmetric, covariantly holomorphic tensor of SK geometry (with
K\"{a}hler weights $\left( 2,-2\right) $) (see \textit{e.g.}\footnote{%
Notice that the third of Eqs. (\ref{C}) correctly defines the Riemann tensor
$R_{i\overline{j}k\overline{l}}$, and it is actual the opposite of the one
which may be found in a large part of existing literature. Such a
formulation of the so-called \textit{SKG constraints} is well defined,
because, as we will mention at the end of Sect. \ref{Sect4}, it yields
negative values of the constant scalar curvature of ($n_{V}=1$-dim.)
homogeneous symmetric compact SK manifolds.} \cite{4,Castellani1,DFF}):
\begin{equation}
\begin{array}{l}
\left\{
\begin{array}{l}
C_{ijk}=\left\langle D_{i}D_{j}V,D_{k}V\right\rangle =e^{K}\left( \partial
_{i}\mathcal{N}_{\Lambda \Sigma }\right) D_{j}X^{\Lambda }D_{k}X^{\Sigma }=
\\
\\
=e^{K}\left( \partial _{i}X^{\Lambda }\right) \left( \partial _{j}X^{\Sigma
}\right) \left( \partial _{k}X^{\Xi }\right) \partial _{\Xi }\partial
_{\Sigma }F_{\Lambda }\left( X\right) \equiv e^{K}W_{ijk},\text{~~}\overline{%
\partial }_{\overline{l}}W_{ijk}=0; \\
\\
C_{ijk}=D_{i}D_{j}D_{k}\mathcal{S},~~\mathcal{S}\equiv -iL^{\Lambda
}L^{\Sigma }Im\left( F_{\Lambda \Sigma }\right) ,~~F_{\Lambda \Sigma }\equiv
\frac{\partial F_{\Lambda }}{\partial X^{\Sigma }},F_{\Lambda \Sigma }\equiv
F_{\left( \Lambda \Sigma \right) }~; \\
\\
C_{ijk}=-ig_{i\overline{l}}\overline{f}_{\Lambda }^{\overline{l}%
}D_{j}D_{k}L^{\Lambda },~~~\overline{f}_{\Lambda }^{\overline{l}}\left(
\overline{D}\overline{L}_{\overline{s}}^{\Lambda }\right) \equiv \delta _{%
\overline{s}}^{\overline{l}};
\end{array}
\right. \\
\\
\overline{D}_{\overline{i}}C_{jkl}=0\text{ (\textit{covariant holomorphicity}%
)}; \\
\\
R_{i\overline{j}k\overline{l}}=-g_{i\overline{j}}g_{k\overline{l}}-g_{i%
\overline{l}}g_{k\overline{j}}+C_{ikp}\overline{C}_{\overline{j}\overline{l}%
\overline{p}}g^{p\overline{p}}\text{ (usually named \textit{SKG constraints})%
}; \\
\\
D_{[i}C_{j]kl}=0,
\end{array}
\label{C}
\end{equation}
where the last property is a consequence, through the SKG constraints and
the covariant holomorphicity of $C_{ijk}$, of the Bianchi identities for the
Riemann tensor $R_{i\overline{j}k\overline{l}}$ (see \textit{e.g.} \cite
{Castellani1}), and square brackets denote antisymmetrization with respect
to enclosed indices. For later convenience, it is here worth writing the
expression for the holomorphic covariant derivative of $C_{ijk}$:
\begin{equation}
D_{i}C_{jkl}=D_{(i}C_{j)kl}=\partial _{i}C_{jkl}+\left( \partial
_{i}K\right) C_{jkl}+\Gamma _{ij}^{~~m}C_{mkl}+\Gamma
_{ik}^{~~m}C_{mjl}+\Gamma _{il}^{~~m}C_{mjk}.  \label{DC}
\end{equation}

It is worth recalling that in a generic K\"{a}hler geometry $R_{i\overline{j}%
k\overline{l}}$ reads
\begin{equation}
\begin{array}{l}
R_{i\overline{j}k\overline{l}}=g^{m\overline{n}}\left( \overline{\partial }_{%
\overline{l}}\overline{\partial }_{\overline{j}}\partial _{m}K\right)
\partial _{i}\overline{\partial }_{\overline{n}}\partial _{k}K-\overline{%
\partial }_{\overline{l}}\partial _{i}\overline{\partial }_{\overline{j}%
}\partial _{k}K=g_{k\overline{n}}\partial _{i}\overline{\Gamma }_{\overline{l%
}\overline{j}}^{~~\overline{n}}=g_{n\overline{l}}\overline{\partial }_{%
\overline{j}}\Gamma _{ki}^{~~n}, \\
\\
\overline{R_{i\overline{j}k\overline{l}}}=R_{j\overline{i}l\overline{k}}%
\text{ \ \ \ (\textit{reality})}, \\
\\
\Gamma _{ij}^{~~l}=-g^{l\overline{l}}\partial _{i}g_{j\overline{l}}=-g^{l%
\overline{l}}\partial _{i}\overline{\partial }_{\overline{l}}\partial
_{j}K=\Gamma _{\left( ij\right) }^{~~l},
\end{array}
\text{\ }  \label{Riemann}
\end{equation}
where $\Gamma _{ij}^{~~l}$ stand for the Christoffel symbols of the second
kind of the K\"{a}hler metric $g_{i\overline{j}}$.

In the first of Eqs. (\ref{C}), a fundamental entity, the so-called kinetic
matrix $\mathcal{N}_{\Lambda \Sigma }\left( z,\overline{z}\right) $ of $%
\mathcal{N}=2$, $d=4$ supergravity, has been introduced (see also Eq. (\ref
{L}) further below). It is an $\left( n_{V}+1\right) \times \left(
n_{V}+1\right) $ complex symmetric, moduli-dependent, K\"{a}hler
gauge-invariant matrix defined by the following fundamental \textit{%
Ans\"{a}tze} of SKG, solving the \textit{SKG constraints} (given by the
third of Eqs. (\ref{C})):
\begin{equation}
M_{\Lambda }=\mathcal{N}_{\Lambda \Sigma }L^{\Sigma },~~D_{i}M_{\Lambda }=%
\overline{\mathcal{N}}_{\Lambda \Sigma }D_{i}L^{\Sigma }.  \label{Ans1}
\end{equation}
By introducing the $\left( n_{V}+1\right) \times \left( n_{V}+1\right) $
complex matrices ($I=1,...,n_{V}+1$)
\begin{equation}
f_{I}^{\Lambda }\left( z,\overline{z}\right) \equiv \left( \overline{D}_{%
\overline{i}}\overline{L}^{\Lambda }\left( z,\overline{z}\right) ,L^{\Lambda
}\left( z,\overline{z}\right) \right) ,\text{ \ }h_{I\Lambda }\left( z,%
\overline{z}\right) \equiv \left( \overline{D}_{\overline{i}}\overline{M}%
_{\Lambda }\left( z,\overline{z}\right) ,M_{\Lambda }\left( z,\overline{z}%
\right) \right) ,
\end{equation}
the \textit{Ans\"{a}tze} (\ref{Ans1}) univoquely determine $\mathcal{N}%
_{\Lambda \Sigma }\left( z,\overline{z}\right) $ as
\begin{equation}
\mathcal{N}_{\Lambda \Sigma }\left( z,\overline{z}\right) =h_{I\Lambda
}\left( z,\overline{z}\right) \circ \left( f^{-1}\right) _{\Sigma
}^{I}\left( z,\overline{z}\right) ,  \label{N}
\end{equation}
where $\circ $ denotes the usual matrix product, and $\left( f^{-1}\right)
_{\Sigma }^{I}f_{I}^{\Lambda }=\delta _{\Sigma }^{\Lambda }$, $\left(
f^{-1}\right) _{\Lambda }^{I}f_{J}^{\Lambda }=\delta _{J}^{I}$.

The covariantly holomorphic $\left( 2n_{V}+2\right) \times 1$ period vector $%
V\left( z,\overline{z}\right) $ is \textit{symplectically orthogonal} to all
its K\"{a}hler-covariant derivatives:
\begin{equation}
\left\{
\begin{array}{l}
\left\langle V\left( z,\overline{z}\right) ,D_{i}V\left( z,\overline{z}%
\right) \right\rangle =0; \\
\\
\left\langle V\left( z,\overline{z}\right) ,\overline{D}_{\overline{i}%
}V\left( z,\overline{z}\right) \right\rangle =0; \\
\\
\left\langle V\left( z,\overline{z}\right) ,D_{i}\overline{V}\left( z,%
\overline{z}\right) \right\rangle =0; \\
\\
\left\langle V\left( z,\overline{z}\right) ,\overline{D}_{\overline{i}}%
\overline{V}\left( z,\overline{z}\right) \right\rangle =0.
\end{array}
\right.  \label{ortho-rels}
\end{equation}
Morover, it holds that
\begin{eqnarray}
&&
\begin{array}{l}
g_{i\overline{j}}\left( z,\overline{z}\right) =-i\left\langle D_{i}V\left( z,%
\overline{z}\right) ,\overline{D}_{\overline{j}}\overline{V}\left( z,%
\overline{z}\right) \right\rangle = \\
\\
=-2Im\left( \mathcal{N}_{\Lambda \Sigma }\left( z,\overline{z}\right)
\right) D_{i}L^{\Lambda }\left( z,\overline{z}\right) \overline{D}_{%
\overline{i}}\overline{L}^{\Sigma }\left( z,\overline{z}\right) =2Im\left(
F_{\Lambda \Sigma }\left( z\right) \right) D_{i}L^{\Lambda }\left( z,%
\overline{z}\right) \overline{D}_{\overline{i}}\overline{L}^{\Sigma }\left(
z,\overline{z}\right) ;
\end{array}
\label{ortho1} \\
&&  \notag \\
&&
\begin{array}{l}
\left\langle V\left( z,\overline{z}\right) ,D_{i}\overline{D}_{\overline{j}%
}V\left( z,\overline{z}\right) \right\rangle =iC_{ijk}g^{k\overline{k}%
}\left\langle V\left( z,\overline{z}\right) ,\overline{D}_{\overline{k}}%
\overline{V}\left( z,\overline{z}\right) \right\rangle =0.
\end{array}
\label{ortho2}
\end{eqnarray}
\setcounter{equation}0

\section{\label{extremal-BH}Extremal Black Hole Attractor Equations\newline
in $\mathcal{N}=2$, $d=4$ (ungauged) Supergravity}

\setcounter{equation}0
\def\theequation{3.\arabic{subsection}.\arabic{equation}}
\subsection{\label{Sect2}Black Hole Effective Potential and ``\textit{%
Criticality Conditions'' Approach}}

In $\mathcal{N}=2$, $d=4$ supergravity the ``effective BH potential'' reads
\cite{FK1,FK2,4}
\begin{equation}
V_{BH}\left( z,\overline{z};q,p\right) =\left| Z\right| ^{2}\left( z,%
\overline{z};q,p\right) +g^{i\overline{j}}\left( z,\overline{z}\right)
D_{i}Z\left( z,\overline{z};q,p\right) \overline{D}_{\overline{j}}\overline{Z%
}\left( z,\overline{z};q,p\right) =I_{1}\left( z,\overline{z};q,p\right)
\geqslant 0,  \label{VBH1}
\end{equation}
where $I_{1}$ is the first, positive-definite real invariant $I_{1}$ of SK
geometry (see \textit{e.g.} \cite{K3,4}). It should be noticed that $V_{BH}$
can also be obtained by left-multiplying the SKG vector identity (\ref
{SKG-identities1}) by the $1\times \left( 2n_{V}+2\right) $ complex
moduli-dependent vector $-\frac{1}{2}Q\mathcal{M}\left( \mathcal{N}\right) $%
; indeed, since the matrix $\mathcal{M}\left( \mathcal{N}\right) $ is
symplectic, one finally gets \cite{FK1,FK2,4}
\begin{equation}
V_{BH}\left( z,\overline{z};q,p\right) =-\frac{1}{2}Q\mathcal{M}\left(
\mathcal{N}\right) Q^{T}.  \label{VBH2}
\end{equation}
It is interesting to remark that the result (\ref{VBH2}) can be elegantly
obtained from the SK geometry identities (\ref{SKG-identities1}) by making
use of the following relation (see \cite{FKlast}, where a generalization for
$\mathcal{N}>2$-extended supergravities is also given):\textbf{\ }
\begin{equation}
\frac{1}{2}\left( \mathcal{M}\left( \mathcal{N}\right) +i\Omega \right)
\mathcal{V}=i\Omega \mathcal{V}\Leftrightarrow \mathcal{M}\left( \mathcal{N}%
\right) \mathcal{V}=i\Omega \mathcal{V},
\end{equation}
where $\mathcal{V}$ is a $\left( 2n_{V}+2\right) \times \left(
n_{V}+1\right) $ matrix defined as follows:
\begin{equation}
\mathcal{V}\equiv \left( V,\overline{D}_{\overline{1}}\overline{V},...,%
\overline{D}_{\overline{n_{V}}}\overline{V}\right) .
\end{equation}

By differentiating Eq. (\ref{VBH1}) with respect to the moduli, the
criticality conditions of $V_{BH}$ can be easily shown to acquire the form
\cite{FGK}
\begin{equation}
D_{i}V_{BH}=\partial _{i}V_{BH}=0\Leftrightarrow 2\overline{Z}D_{i}Z+g^{j%
\overline{j}}\left( D_{i}D_{j}Z\right) \overline{D}_{\overline{j}}\overline{Z%
}=0.  \label{AEs1}
\end{equation}
These are the what one should rigorously refer to as the $\mathcal{N}=2$, $%
d=4$ supergravity Attractor Eqs. (AEs).

In the present work, we will call \textit{AEs} also some \textit{geometrical
identities} evaluated along the criticality conditions of the relevant
``effective potential''. Indeed, both for extremal BHs attractors in $%
\mathcal{N}=2$, $d=4$ supergravity and for FV attractors in
$\mathcal{N}=1$, $d=4$ supergravity (\textit{at least} for the one
coming from some peculiar compactifications of superstrings: see
Sect. \ref{Flux-Vacua-Attractors}), there exist two different
approaches to determining the attractors:

\textit{i}) the so-called \textit{``criticality conditions'' approach},
based on the direct solution of the conditions giving the stationary points
of the relevant ``effective potential'';

\textit{ii}) the so-called \textit{``new attractor'' approach}, based on the
solution of some fundamental geometrical identities evaluated along the
criticality conditions of the relevant ``effective potential''.

Such two approaches are completely equivalent. Dependingly on the considered
frameworks, it can be convenient to exploit one approach rather than the
other one (see \textit{e.g.} \cite{BFMY} for an explicit case).

By using the relations (\ref{SKG-rels1}), the $\mathcal{N}=2$ AEs (\ref{AEs1}%
) can be recast as follows \cite{FGK}:
\begin{equation}
D_{i}V_{BH}=\partial _{i}V_{BH}=2\overline{Z}D_{i}Z+iC_{ijk}g^{j\overline{l}%
}g^{k\overline{m}}\left( \overline{D}_{\overline{l}}\overline{Z}\right)
\overline{D}_{\overline{m}}\overline{Z}.  \label{AEs2}
\end{equation}
Eqs. (\ref{AEs2}) are nothing but the relations between the $\mathcal{N}=2$
central charge function $Z$ (\textit{graviphoton charge}) and the $n_{V}$
\textit{matter charges} $Z_{i}$ (coming from the $n_{V}$ Abelian vector
supermultiplets), holding at the critical points of $V_{BH}$ in the SK
scalar manifold $\mathcal{M}_{n_{V}}$. As it is seen, the tensor $C_{ijk}$
plays a key role.

It is known that static, spherically symmetric, asymptotically flat extremal
BHs in $d=4$ are described by an effective $d=1$ Lagrangian (\cite{FGK},
\cite{2}, and also \cite{AoB-book} and \cite{ADFT}), with an effective
scalar potential and effective fermionic ``mass terms'' terms controlled by
the field-strength fluxes vector $Q$ defined by Eq. (\ref{Gamma-tilde}). The
\textit{``apparent'' gravitino mass} is given by $Z$, whereas the $%
n_{V}\times n_{V}$ \textit{gaugino mass matrix} $\Lambda _{ij}$ reads (see
the second Ref. of \cite{DFF})
\begin{equation}
\Lambda _{ij}=-iD_{i}Z_{j}=C_{ijk}g^{k\overline{k}}\overline{Z}_{\overline{k}%
}=\Lambda _{\left( ij\right) }.
\end{equation}
Note that $\Lambda _{ij}$ is part of the holomorphic/anti-holomorphic form
of the $2n_{V}\times 2n_{V}$ covariant Hessian of $Z$, which is nothing but
the holomorphic/anti-holomorphic form of the scalar mass matrix. The \textit{%
supersymmetry order parameters}, related to the mixed gravitino-gaugino
couplings, are given by the \textit{matter charge( function)s} $D_{i}Z=Z_{i}$
(see the first of Eqs. (\ref{SKG-rels1})).

By assuming that the K\"{a}hler potential is regular, i.e. that $\left|
K\right| <\infty $ globally in $\mathcal{M}_{N_{V}}$ (or \textit{at least}
at the critical points of $V_{BH}$), one gets that
\begin{equation}
\partial _{i}V_{BH}=0\Leftrightarrow 2\overline{W}D_{i}W+ie^{K}W_{ijk}g^{j%
\overline{l}}g^{k\overline{m}}\left( \overline{D}_{\overline{l}}\overline{W}%
\right) \overline{D}_{\overline{m}}\overline{W}=0.  \label{AEs3}
\end{equation}
\setcounter{equation}0
\def\theequation{3.1.\arabic{subsubsection}.\arabic{equation}}
\subsubsection{\label{classification-crit-BH}Classification of Critical
Points of $V_{BH}$}

Starting from the general structure of the criticality conditions (\ref{AEs3}%
) and assuming also the \textit{non-degeneracy} (\textit{i.e.} $\left.
V_{BH}\right| _{\partial V_{BH}=0}>0$) \textit{condition}, the critical
points of $V_{BH}$ can be classified in three general classes, analyzed in
the next three Subsubsubsects.. \setcounter{equation}0
\def\theequation{3.1.1.1.\arabic{equation}}
\paragraph{Supersymmetric ($\frac{1}{2}$-BPS)}

~

The \textit{supersymmetric (}$\frac{1}{2}$\textit{-BPS)} critical points of $%
V_{BH}$ are determined by the constraints (sufficient but not necessary
conditions for Eqs. (\ref{AEs3}))
\begin{equation}
Z\neq 0,D_{i}Z=0,\forall i=1,...,n_{V}.  \label{BPS-conds}
\end{equation}
The horizon ADM squared mass at $\frac{1}{2}$-BPS critical points of $V_{BH}$
saturates the BPS bound:\textbf{\ }
\begin{equation}
M_{ADM,H,\frac{1}{2}-BPS}^{2}=V_{BH,\frac{1}{2}-BPS}=\left[ \left| Z\right|
^{2}+g^{i\overline{j}}\left( D_{i}Z\right) \overline{D}_{\overline{j}}%
\overline{Z}\right] _{\frac{1}{2}-BPS}=\left| Z\right| _{\frac{1}{2}%
-BPS}^{2}>0.  \label{V-BPS}
\end{equation}

Considering the $\mathcal{N}=2$, $d=4$ supergravity Lagrangian in a static,
spherically symmetric, asymptotically flat extremal BH background, and
denoting by $\psi $ and $\lambda ^{i}$ respectively the gravitino and
gaugino fields, it is easy to see that such a Lagrangian contains terms of
the form (see the second and third Refs. of \cite{DFF})
\begin{equation}
\begin{array}{l}
Z\psi \psi ; \\
\\
C_{ijk}g^{k\overline{k}}\left( \overline{D}_{\overline{k}}\overline{Z}%
\right) \lambda ^{i}\lambda ^{j}; \\
\\
\left( D_{i}Z\right) \lambda ^{i}\psi .
\end{array}
\label{22apr2}
\end{equation}
Thus, the conditions (\ref{BPS-conds}) imply the gaugino mass term and the $%
\lambda \psi $ term to vanish at the $\frac{1}{2}$-BPS critical points of $%
V_{BH}$ in $\mathcal{M}_{n_{V}}$. It is interesting to remark that the
gravitino ``apparent mass'' term $Z\psi \psi $ is in general non-vanishing,
also when evaluated at the considered $\frac{1}{2}$-BPS attractors; this is
ultimately a consequence of the fact that the extremal BH horizon geometry
at the $\frac{1}{2}$-BPS (as well as at the non-BPS) attractors is
Bertotti-Robinson $AdS_{2}\times S^{2}$ \cite{BR1,BR2,BR3}.
\setcounter{equation}0
\def\theequation{3.1.1.2.\arabic{equation}}
\paragraph{Non-supersymmetric (non-BPS) with $Z\neq 0$}

~

The \textit{non-supersymmetric (}non\textit{-BPS) }critical points of $%
V_{BH} $ with non-vanishing central charge are determined by the constraints
\begin{equation}
Z\neq 0,D_{i}Z\neq 0,\text{\textit{at least} for some }i\in \left\{
1,...,n_{V}\right\} ,  \label{non-BPS-Z<>0-conds}
\end{equation}
which, substituted in Eqs. (\ref{AEs3}), yield:
\begin{equation}
\begin{array}{c}
D_{i}Z=-\frac{i}{2\overline{Z}}C_{ijk}g^{j\overline{l}}g^{k\overline{m}%
}\left( \overline{D}_{\overline{l}}\overline{Z}\right) \overline{D}_{%
\overline{m}}\overline{Z},~~\forall i=1,...,n_{V}; \\
\Updownarrow \\
\overline{D}_{\overline{i}}\overline{Z}=\frac{i}{2Z}\overline{C}_{\overline{i%
}\overline{j}\overline{k}}g^{l\overline{j}}g^{m\overline{k}}\left(
D_{l}Z\right) D_{m}Z,~~\forall \overline{i}=\overline{1},...,\overline{n_{V}}%
,
\end{array}
\label{non-BPS-Z<>0-fund}
\end{equation}
in turn implying that
\begin{equation}
\begin{array}{l}
g^{i\overline{i}}\left( D_{i}Z\right) \overline{D}_{\overline{i}}\overline{Z}%
=-\frac{i}{2\overline{Z}}C_{ijk}g^{i\overline{i}}g^{j\overline{l}}g^{k%
\overline{m}}\left( \overline{D}_{\overline{i}}\overline{Z}\right) \left(
\overline{D}_{\overline{l}}\overline{Z}\right) \overline{D}_{\overline{m}}%
\overline{Z}= \\
\\
=\frac{i}{2Z}\overline{C}_{\overline{i}\overline{j}\overline{k}}g^{i%
\overline{i}}g^{l\overline{j}}g^{m\overline{k}}\left( D_{i}Z\right) \left(
D_{l}Z\right) D_{m}Z.
\end{array}
\label{non-BPS-Z<>0-fund2}
\end{equation}
Such critical points are \textit{non-supersymmetric} ones (\textit{i.e.}
they do \textit{not} preserve any of the 8 supersymmetry degrees of freedom
of the asymptotical Minkowski background), and they correspond to an
extremal, non-BPS BH background. They are commonly named \textit{non-BPS} $%
Z\neq 0$ \textit{critical points of }$V_{BH}$.

AEs (\ref{AEs3}) and conditions (\ref{non-BPS-Z<>0-conds}) imply
\begin{equation}
\left( C_{ijk}\right) _{non-BPS,Z\neq 0}\neq 0,~~~\text{for some }\left(
i,j,k\right) \in \left\{ 1,...,n_{V}\right\} ^{3}.
\end{equation}

By using Eq. (\ref{non-BPS-Z<>0-fund}) and the so-called SK geometry
constraints (see the third of Eqs. (\ref{C})), the horizon ADM squared mass
corresponding to non-BPS $Z\neq 0$ critical points of $V_{BH}$ can be
elaborated as follows:
\begin{eqnarray}
&&
\begin{array}{l}
M_{ADM,H,non-BPS,Z\neq 0}^{2}=V_{BH,non-BPS,Z\neq 0}=\left[ \left| Z\right|
^{2}+g^{i\overline{j}}\left( D_{i}Z\right) \overline{D}_{\overline{j}}%
\overline{Z}\right] _{non-BPS,Z\neq 0}= \\
\\
=\left\{ \left| Z\right| ^{2}\left[
\begin{array}{l}
1+\frac{1}{4\left| Z\right| ^{4}}R_{k\overline{r}n\overline{s}}g^{k\overline{%
m}}g^{t\overline{r}}g^{n\overline{l}}g^{u\overline{s}}\left( D_{t}Z\right)
\left( D_{u}Z\right) \left( \overline{D}_{\overline{l}}\overline{Z}\right)
\overline{D}_{\overline{m}}\overline{Z}+ \\
\\
+\frac{1}{2\left| Z\right| ^{4}}\left[ g^{i\overline{j}}\left( D_{i}Z\right)
\overline{D}_{\overline{j}}\overline{Z}\right] ^{2}
\end{array}
\right] \right\} _{non-BPS,Z\neq 0}.
\end{array}
\notag \\
&&  \label{V-non-BPS-Z<>0-1}
\end{eqnarray}
As far as $g_{i\overline{j}}$ is strictly positive-definite globally (or
\textit{at least} at the non-BPS $Z\neq 0$ critical points of $V_{BH}$), $%
M_{ADM,H,non-BPS,Z\neq 0}^{2}$ does \textit{not} saturate the BPS bound (
\cite{K1}, \cite{K2}, \cite{Tom}):
\begin{equation}
\begin{array}{l}
M_{ADM,H,non-BPS,Z\neq 0}^{2}=V_{BH,non-BPS,Z\neq 0}= \\
\\
=\left[ \left| Z\right| ^{2}+g^{i\overline{j}}\left( D_{i}Z\right) \overline{%
D}_{\overline{j}}\overline{Z}\right] _{non-BPS,Z\neq 0}>\left| Z\right|
_{non-BPS,Z\neq 0}^{2}.
\end{array}
\label{V-non-BPS-Z<>0-2}
\end{equation}
Starting from Eq. (\ref{V-non-BPS-Z<>0-1}), one can introduce and further
elaborate the so-called \textit{non-BPS }$Z\neq 0$\textit{\ supersymmetry
breaking order parameter} as follows:
\begin{eqnarray}
&&
\begin{array}{l}
\left( 0<\right) \mathcal{O}_{non-BPS,Z\neq 0}\equiv \left[ \frac{g^{i%
\overline{j}}\left( D_{i}Z\right) \overline{D}_{\overline{j}}\overline{Z}}{%
\left| Z\right| ^{2}}\right] _{non-BPS,Z\neq 0}= \\
\\
=-\left[ \frac{i}{2\overline{Z}\left| Z\right| ^{2}}C_{ijk}g^{i\overline{i}%
}g^{j\overline{l}}g^{k\overline{m}}\left( \overline{D}_{\overline{i}}%
\overline{Z}\right) \left( \overline{D}_{\overline{l}}\overline{Z}\right)
\overline{D}_{\overline{m}}\overline{Z}\right] _{non-BPS,Z\neq 0}= \\
\\
=\left[ \frac{i}{2Z\left| Z\right| ^{2}}\overline{C}_{\overline{i}\overline{j%
}\overline{k}}g^{i\overline{i}}g^{l\overline{j}}g^{m\overline{k}}\left(
D_{i}Z\right) \left( D_{l}Z\right) D_{m}Z\right] _{non-BPS,Z\neq 0},
\end{array}
\notag \\
&&  \label{UCLApre}
\end{eqnarray}
where Eqs. (\ref{non-BPS-Z<>0-fund2}) were used. Since it holds that
\begin{eqnarray}
&&
\begin{array}{l}
\left[ \frac{g^{i\overline{j}}\left( D_{i}Z\right) \overline{D}_{\overline{j}%
}\overline{Z}}{\left| Z\right| ^{2}}\right] _{non-BPS,Z\neq 0}= \\
\\
=\left\{ \frac{1}{4\left| Z\right| ^{4}}R_{k\overline{r}n\overline{s}}g^{k%
\overline{m}}g^{t\overline{r}}g^{n\overline{l}}g^{u\overline{s}}\left(
D_{t}Z\right) \left( D_{u}Z\right) \left( \overline{D}_{\overline{l}}%
\overline{Z}\right) \overline{D}_{\overline{m}}\overline{Z}+\frac{1}{2}\left[
\frac{g^{i\overline{j}}\left( D_{i}Z\right) \overline{D}_{\overline{j}}%
\overline{Z}}{\left| Z\right| ^{2}}\right] ^{2}\right\} _{non-BPS,Z\neq 0},
\end{array}
\notag \\
&&
\end{eqnarray}
$\mathcal{O}_{non-BPS,Z\neq 0}$ defined by Eq. (\ref{UCLApre}) can
equivalently be rewritten as follows:
\begin{equation}
\mathcal{O}_{non-BPS,Z\neq 0}=\left[ \frac{1}{4\left| Z\right| ^{4}}g^{i%
\overline{j}}C_{ikn}\overline{C}_{\overline{j}\overline{r}\overline{s}}g^{n%
\overline{l}}g^{k\overline{m}}g^{t\overline{r}}g^{u\overline{s}}\left(
D_{t}Z\right) \left( D_{u}Z\right) \left( \overline{D}_{\overline{l}}%
\overline{Z}\right) \overline{D}_{\overline{m}}\overline{Z}\right]
_{non-BPS,Z\neq 0}.  \label{23apr1}
\end{equation}
Eqs. (\ref{23apr1}) imply that
\begin{gather}
\begin{array}{l}
M_{ADM,H,non-BPS,Z\neq 0}^{2}=V_{BH,non-BPS,Z\neq 0}=\left| Z\right|
_{non-BPS,Z\neq 0}^{2}\left[ 1+\mathcal{O}_{non-BPS,Z\neq 0}\right] = \\
\\
=\left\{ \left| Z\right| ^{2}\left[ 3-2\frac{\mathcal{R}\left( Z\right) }{%
g^{i\overline{j}}C_{ikn}\overline{C}_{\overline{j}\overline{r}\overline{s}%
}g^{n\overline{l}}g^{k\overline{m}}g^{t\overline{r}}g^{u\overline{s}}\left(
D_{t}Z\right) \left( D_{u}Z\right) \left( \overline{D}_{\overline{l}}%
\overline{Z}\right) \overline{D}_{\overline{m}}\overline{Z}}\right] \right\}
_{non-BPS,Z\neq 0},
\end{array}
\notag  \label{23apr3-2} \\
\end{gather}
where the \textit{sectional curvature} (see \textit{e.g.} \cite
{Riemann-Finsler} and \cite{Differential-Geometry})
\begin{equation}
\mathcal{R}\left( Z\right) \equiv R_{i\overline{j}k\overline{l}}g^{i%
\overline{i}}g^{j\overline{j}}g^{k\overline{k}}g^{l\overline{l}}\left(
D_{j}Z\right) \left( D_{l}Z\right) \left( \overline{D}_{\overline{i}}%
\overline{Z}\right) \overline{D}_{\overline{k}}\overline{Z}
\label{sectional-curvature}
\end{equation}
was introduced.\medskip

Now, by using the relations of SK geometry it is possible to show that
\begin{equation}
\begin{array}{c}
\overline{D}_{\overline{m}}D_{i}C_{jkl}=\left[ \overline{D}_{\overline{m}%
},D_{i}\right] C_{jkl}=\overline{D}_{\overline{m}}D_{(i}C_{j)kl}=\overline{D}%
_{\overline{m}}D_{(i}C_{jkl)}=3C_{p(kl}C_{ij)n}g^{n\overline{n}}g^{p%
\overline{p}}\overline{C}_{\overline{n}\overline{p}\overline{m}}-4g_{\left(
l\right| \overline{m}}C_{\left| ijk\right) } \\
\\
\Updownarrow \\
\\
C_{p(kl}C_{ij)n}g^{n\overline{n}}g^{p\overline{p}}\overline{C}_{\overline{n}%
\overline{p}\overline{m}}=\frac{4}{3}g_{\left( l\right| \overline{m}%
}C_{\left| ijk\right) }+\overline{E}_{\overline{m}\left( ijkl\right) },
\end{array}
\label{UCLA1}
\end{equation}
where we introduced the rank-5 tensor
\begin{gather}
\begin{array}{l}
\overline{E}_{\overline{m}ijkl}=\overline{E}_{\overline{m}\left( ijkl\right)
}\equiv \frac{1}{3}\overline{D}_{\overline{m}}D_{i}C_{jkl}=\frac{1}{3}%
\overline{D}_{\overline{m}}D_{(i}C_{jkl)}=C_{p(kl}C_{ij)n}g^{n\overline{n}%
}g^{p\overline{p}}\overline{C}_{\overline{n}\overline{p}\overline{m}}-\frac{4%
}{3}g_{\left( l\right| \overline{m}}C_{\left| ijk\right) }= \\
\\
=g^{n\overline{n}}R_{\left( i\right| \overline{m}\left| j\right| \overline{n}%
}C_{n\left| kl\right) }+\frac{2}{3}g_{\left( i\right| \overline{m}}C_{\left|
jkl\right) },
\end{array}
\notag \\
\end{gather}
where the SK geometry constraints were used, as well. Now, by recalling the
criticality conditions (\ref{AEs3}) of $V_{BH}$, and by using Eq. (\ref
{non-BPS-Z<>0-fund}), one gets that at non-BPS, $Z\neq 0$ critical points of
$V_{BH}$ it holds that

\begin{eqnarray}
&&
\begin{array}{l}
2\overline{Z}D_{i}Z= \\
\\
=\frac{i}{4Z^{2}}E_{i\left( \overline{s}\overline{n}\overline{t}\overline{u}%
\right) }g^{p\overline{n}}g^{q\overline{s}}g^{r\overline{t}}g^{v\overline{u}%
}\left( D_{p}Z\right) \left( D_{q}Z\right) \left( D_{r}Z\right) D_{v}Z+ \\
\\
+\frac{i}{3Z^{2}}\left( D_{i}Z\right) \overline{C}_{\overline{n}\overline{t}%
\overline{u}}g^{p\overline{n}}g^{r\overline{t}}g^{v\overline{u}}\left(
D_{p}Z\right) \left( D_{r}Z\right) D_{v}Z.
\end{array}
\notag \\
&&  \label{UCLA4}
\end{eqnarray}
By using Eqs. (\ref{UCLApre}), (\ref{UCLA4}) and (\ref{UCLApre}), with a
little effort it is thus possible to compute that
\begin{eqnarray}
&&
\begin{array}{l}
M_{ADM,H,non-BPS,Z\neq 0}^{2}=V_{BH,non-BPS,Z\neq 0}=\left| Z\right|
_{non-BPS,Z\neq 0}^{2}\left[ 1+\mathcal{O}_{non-BPS,Z\neq 0}\right] = \\
\\
=\left| Z\right| _{non-BPS,Z\neq 0}^{2}\left\{ 4-\frac{3}{4}\left[ \frac{1}{%
\left| Z\right| ^{2}}\frac{E_{i\left( \overline{k}\overline{l}\overline{m}%
\overline{n}\right) }g^{i\overline{j}}g^{k\overline{k}}g^{l\overline{l}}g^{m%
\overline{m}}g^{n\overline{n}}\left( \overline{D}_{\overline{j}}\overline{Z}%
\right) \left( D_{k}Z\right) \left( D_{l}Z\right) \left( D_{m}Z\right) D_{n}Z%
}{N_{3}\left( Z\right) }\right] _{non-BPS,Z\neq 0}\right\} ,
\end{array}
\notag \\
&&  \label{UCLA10}
\end{eqnarray}
where we defined the complex cubic form
\begin{equation}
N_{3}\left( Z\right) \equiv \overline{C}_{\overline{i}\overline{j}\overline{k%
}}g^{i\overline{i}}g^{j\overline{j}}g^{k\overline{k}}\left( D_{i}Z\right)
\left( D_{j}Z\right) D_{k}Z.  \label{UCLA9}
\end{equation}

Now, by comparing Eq. (\ref{UCLA10}) with Eq. (\ref{23apr3-2}) and by
recalling the definition (\ref{sectional-curvature}), one obtains the
following relations to hold at the non-BPS, $Z\neq 0$ critical points of $%
V_{BH}$:
\begin{eqnarray}
&&
\begin{array}{l}
\frac{3}{4}\left[ \frac{1}{\left| Z\right| ^{2}}\frac{E_{i\left( \overline{k}%
\overline{l}\overline{m}\overline{n}\right) }g^{i\overline{j}}g^{k\overline{k%
}}g^{l\overline{l}}g^{m\overline{m}}g^{n\overline{n}}\left( \overline{D}_{%
\overline{j}}\overline{Z}\right) \left( D_{k}Z\right) \left( D_{l}Z\right)
\left( D_{m}Z\right) D_{n}Z}{N_{3}\left( Z\right) }\right] _{non-BPS,Z\neq
0}-1= \\
\\
=\left[ \frac{\mathcal{R}\left( Z\right) }{2\left| Z\right| ^{2}g^{t%
\overline{u}}\left( D_{t}Z\right) \overline{D}_{\overline{u}}\overline{Z}}%
\right] _{non-BPS,Z\neq 0}= \\
\\
=2\left[ \frac{\mathcal{R}\left( Z\right) }{g^{i\overline{j}}C_{ikn}%
\overline{C}_{\overline{j}\overline{r}\overline{s}}g^{n\overline{l}}g^{k%
\overline{m}}g^{t\overline{r}}g^{u\overline{s}}\left( D_{t}Z\right) \left(
D_{u}Z\right) \left( \overline{D}_{\overline{l}}\overline{Z}\right)
\overline{D}_{\overline{m}}\overline{Z}}\right] _{non-BPS,Z\neq 0}.
\end{array}
\notag \\
&&  \label{UCLA11}
\end{eqnarray}
\medskip

Let us now consider the case of homogeneous symmetric SK manifolds, in which
the K\"{a}hler-invariant Riemann-Christoffel tensor $R_{i\overline{j}k%
\overline{l}}$ is covariantly constant\footnote{%
Indeed, due to the reality of $R_{i\overline{j}k\overline{l}}$ in any K\"{a}%
hler manifold, it holds that
\begin{equation*}
D_{m}R_{i\overline{j}k\overline{l}}=0\Leftrightarrow \overline{D}_{\overline{%
m}}R_{i\overline{j}k\overline{l}}=0.
\end{equation*}
}. From this it follows that \cite{CVP}: \textbf{\ }
\begin{equation}
D_{m}R_{i\overline{j}k\overline{l}}=0\Leftrightarrow
D_{i}C_{jkl}=D_{(i}C_{j)kl}=0\Rightarrow \overline{D}_{\overline{m}%
}D_{i}C_{jkl}=0\Leftrightarrow D_{m}\overline{D}_{\overline{i}}\overline{C}_{%
\overline{j}\overline{k}\overline{l}}=0.  \label{UCLA12}
\end{equation}
This implies the \textit{global} vanishing of the tensor $\overline{E}_{%
\overline{i}jklm}$, yielding \cite{CVP}
\begin{equation}
C_{p(kl}C_{ij)n}g^{n\overline{n}}g^{p\overline{p}}\overline{C}_{\overline{n}%
\overline{p}\overline{m}}=\frac{4}{3}g_{\left( l\right| \overline{m}%
}C_{\left| ijk\right) }\Leftrightarrow g^{n\overline{n}}R_{\left( i\right|
\overline{m}\left| j\right| \overline{n}}C_{n\left| kl\right) }=-\frac{2}{3}%
g_{\left( i\right| \overline{m}}C_{\left| jkl\right) }.
\label{UCLA13}
\end{equation}
By recalling Eqs. (\ref{UCLA4}) and (\ref{UCLA9}), one obtains the following
noteworthy relation, holding in homogeneous symmetric SK manifolds:
\begin{equation}
\left( Z\left| Z\right| ^{2}\right) _{non-BPS,Z\neq 0}=\frac{i}{6}\left[
N_{3}\left( Z\right) \right] _{non-BPS,Z\neq 0},  \label{UCLA14}
\end{equation}
implying that $\left[ \frac{N_{3}\left( Z\right) }{Z}\right] _{non-BPS,Z\neq
0}$ has vanishing real part and strictly negative imaginary part, given by $%
-6\left| Z\right| _{non-BPS,Z\neq 0}^{2}$. By recalling Eq. (\ref{23apr1}),
Eq. (\ref{UCLA14}) implies the value of the \textit{supersymmetry breaking
order parameter} at non-BPS, $Z\neq 0$ critical points of $V_{BH}$ in
homogeneous symmetric SK manifolds to be
\begin{equation}
\mathcal{O}_{non-BPS,Z\neq 0}=3\Longrightarrow \Delta _{non-BPS,Z\neq 0}=0.
\label{UCLA15}
\end{equation}
By recalling Eq. (\ref{UCLA10}), one thus finally gets that
\begin{equation}
M_{ADM,H,non-BPS,Z\neq 0}^{2}=V_{BH,non-BPS,Z\neq 0}=4\left| Z\right|
_{non-BPS,Z\neq 0}^{2}=\frac{2}{3}i\left[ \frac{N_{3}\left( Z\right) }{Z}%
\right] _{non-BPS,Z\neq 0},  \label{UCLA16}
\end{equation}
where in the last step we used the relation (\ref{UCLA14}). The result $%
V_{BH,non-BPS,Z\neq 0}=4\left| Z\right| _{non-BPS,Z\neq 0}^{2}$ has been
firstly obtained, by exploiting group-theoretical methods, in \cite{BFGM1}.

Finally, by recalling Eq. (\ref{UCLA11}) and using Eqs. (\ref{UCLA14}) and (%
\ref{UCLA16}), one obtains the following relation, holding for homogeneous
symmetric SK manifolds:
\begin{equation}
\left. \mathcal{R}\left( Z\right) \right| _{non-BPS,Z\neq 0}=-6\left|
Z\right| _{non-BPS,Z\neq 0}^{4}.  \label{UCLA17}
\end{equation}

It is worth pointing out that, while Eq. (\ref{UCLA12}) (holding
globally) are peculiar to homogeneous symmetric SK manifolds, Eqs.
(\ref{UCLA14})-(\ref{UCLA17}) hold in general also for homogeneous
non-symmetric SK manifolds, in which the Riemann-Christoffel tensor $R_{i%
\overline{j}k\overline{l}}$ (and thus, through the SK constraints, $C_{ijk}$%
) is \textit{not} covariantly constant. Indeed, as obtained in \cite{DFT07-1}
for all the considered non-BPS, $Z\neq 0$ critical points of $V_{BH}$ in
homogeneous non-symmetric SK manifolds it holds that
\begin{equation}
\left[ E_{i\left( \overline{k}\overline{l}\overline{m}\overline{n}\right)
}g^{i\overline{j}}g^{k\overline{k}}g^{l\overline{l}}g^{m\overline{m}}g^{n%
\overline{n}}\left( \overline{D}_{\overline{j}}\overline{Z}\right) \left(
D_{k}Z\right) \left( D_{l}Z\right) \left( D_{m}Z\right) D_{n}Z\right]
_{non-BPS,Z\neq 0}=0,  \label{UCLA20}
\end{equation}
which actually seems to be the most general (necessary and sufficient)
condition in order for Eqs. (\ref{UCLA14})-(\ref{UCLA17}) to hold. Finally,
it should be stressed that in \cite{TT} the result (\ref{UCLA15}) and thus $%
V_{BH,non-BPS,Z\neq 0}=4\left| Z\right| _{non-BPS,Z\neq 0}^{2}$ was obtained
for a generic SK geometry with a cubic holomorphic prepotential
(corresponding to the large volume limit of Type IIA on Calabi-Yau
threefolds), at least for the non-BPS, $Z\neq 0$ critical points of $V_{BH}$
satisfying the Ansatz
\begin{equation}
z_{non-BPS,Z\neq 0}^{i}=p^{i}t\left( p,q\right) ,~\forall i=1,...,n_{V},
\label{UCLA21}
\end{equation}
where the $z_{non-BPS,Z\neq 0}^{i}$s are the critical moduli, and $t\left(
p,q\right) $ is a purely charge-dependent quantity.\medskip

Furthermore, it is worth noticing that the general criticality conditions (%
\ref{AEs1}) of $V_{BH}$ can be recognized to be the general Ward identities
relating the gravitino mass $Z$, the gaugino masses $D_{i}D_{j}Z$ and the
supersymmetry-breaking order parameters $D_{i}Z$ in a generic spontaneously
broken supergravity theory \cite{9}. Indeed, away from $\frac{1}{2}$-BPS
critical points (\textit{i.e.} for $D_{i}Z\neq 0$ for some $i$), the AEs (%
\ref{AEs1}) can be re-expressed as follows (see also \cite{Saraikin-Vafa-1}%
):
\begin{equation}
\left( \mathbf{M}_{ij}h^{j}\right) _{\partial V_{BH}=0}=0,  \label{EQ}
\end{equation}
with
\begin{equation}
\mathbf{M}_{ij}\equiv D_{i}D_{j}Z+2\frac{\overline{Z}}{\left[ g^{k\overline{k%
}}\left( D_{k}Z\right) \left( \overline{D}_{\overline{k}}\overline{Z}\right) %
\right] }\left( D_{i}Z\right) \left( D_{j}Z\right) ,\text{ }(\text{%
K\"{a}hler weights }\left( 1,-1\right) ),
\end{equation}
and
\begin{equation}
h^{j}\equiv g^{j\overline{j}}\overline{D}_{\overline{j}}\overline{Z},\text{ }%
(\text{K\"{a}hler weights }\left( -1,1\right) ).
\end{equation}
For a non-vanishing contravariant vector $h^{j}$ (\textit{i.e. }away from $%
\frac{1}{2}$-BPS critical points, as pointed out above), Eq. (\ref{EQ})
admits a solution iff the $n_{V}\times n_{V}$ complex symmetric matrix $%
\mathbf{M}_{ij}$ has vanishing determinant (implying that it has at most $%
n_{V}-1$ non-vanishing eigenvalues) at the considered (non-BPS) critical
points of $V_{BH}$ (however, notice that $\mathbf{M}_{ij}$\ is symmetric but
not necessarily Hermitian, thus in general its eigenvalues are not
necessarily real)\smallskip . Such a reasoning holds for all non-BPS
critical points of $V_{BH}$, \textit{i.e.} for the classes II and III of the
presented classification.

In general, non-BPS $Z\neq 0$ critical points of $V_{BH}$ in $\mathcal{M}%
_{n_{V}}$ are not necessarily stable, because the $2n_{V}\times 2n_{V}$
(covariant) Hessian matrix (in $\left( z,\overline{z}\right) $-coordinates)
of $V_{BH}$ evaluated at such points is not necessarily strictly
positive-definite. An explicit condition of stability of non-BPS $Z\neq 0$
critical points of $V_{BH}$ can be worked out in the $n_{V}=1$ case (see
\cite{BFM}, \cite{AoB-book}, \cite{BFMY}).

In general, Eqs. (\ref{22apr2}) and conditions (\ref{non-BPS-Z<>0-conds})
imply the gaugino mass term, the $\lambda \psi $ term and the gravitino
``apparent mass'' term $Z\psi \psi $ to be non-vanishing, when evaluated at
the considered non-BPS $Z\neq 0$ critical points of $V_{BH}$.
\setcounter{equation}0
\def\theequation{3.1.1.3.\arabic{equation}}
\paragraph{Non-supersymmetric (non-BPS) with $Z=0$}

~

The \textit{non-supersymmetric (non-BPS) }critical points of $V_{BH}$ with
vanishing central charge are determined by the constraints
\begin{equation}
Z=0,D_{i}Z\overset{Z=0}{=}\partial _{i}Z\neq 0,\text{\textit{at least} for
some }i\in \left\{ 1,...,n_{V}\right\} ,  \label{non-BPS-Z=0-conds}
\end{equation}
which, substituted in Eqs. (\ref{AEs3}), yield:
\begin{equation}
C_{ijk}g^{j\overline{l}}g^{k\overline{m}}\left( \overline{D}_{\overline{l}}%
\overline{Z}\right) \overline{D}_{\overline{m}}\overline{Z}\overset{Z=0}{=}%
C_{ijk}g^{j\overline{l}}g^{k\overline{m}}\left( \overline{\partial }_{%
\overline{l}}\overline{Z}\right) \overline{\partial }_{\overline{m}}%
\overline{Z}=0,~~\forall i=1,...,n_{V}.  \label{non-BPS-Z=0-fund}
\end{equation}
Such critical points are \textit{non-supersymmetric} ones, but, differently
from the class II considered above, they correspond to an extremal, non-BPS
BH background in which the horizon $\mathcal{N}=2$, $d=4$ supersymmetry
algebra is not centrally extended. They are commonly named \textit{non-BPS} $%
Z=0$ \textit{critical points of }$V_{BH}$.

The horizon ADM squared mass corresponding to non-BPS $Z=0$ critical points
of $V_{BH}$ does \textit{not} saturate the BPS bound (\cite{K1}, \cite{K2},
\cite{Tom}):
\begin{equation}
\begin{array}{l}
M_{ADM,H,non-BPS,Z=0}^{2}=V_{BH,non-BPS,Z=0}= \\
\\
=\left\{ g^{i\overline{j}}\left( \partial _{i}Z\right) \overline{\partial }_{%
\overline{j}}\overline{Z}\right\} _{non-BPS,Z=0}>\left( \left| Z\right|
^{2}\right) _{non-BPS,Z=0}=0,
\end{array}
\label{V-non-BPS}
\end{equation}
as far as $g_{i\overline{j}}$ is strictly positive-definite globally (or
\textit{at least} at the considered critical points of $V_{BH}$). Eqs. (\ref
{non-BPS-Z=0-fund}) suggest the following sub-classification of non-BPS $Z=0$
critical points of $V_{BH}$:

III.1) Critical points determined by the conditions
\begin{equation}
\left\{
\begin{array}{l}
Z=0, \\
\\
D_{i}Z\overset{Z=0}{=}\partial _{i}Z\neq 0,\text{\textit{at least} for some }%
i\in \left\{ 1,...,n_{V}\right\} , \\
\\
C_{ijk}=0,\forall i,j,k,
\end{array}
\right.  \label{non-BPS-Z=0-1}
\end{equation}
directly solving Eqs. (\ref{non-BPS-Z=0-fund}) and thus AEs (\ref{AEs3}).
This is the only possible case for $n_{V}=1$.

In particular, non-BPS $Z=0$ critical points of $V_{BH}$ do not exist at all
in the $n_{V}=1$ case of the so-called ``\textit{d-SK geometries''}, whose
stringy origin is \textit{e.g.} Type IIA on $CY_{3}$ in the large volume
limit of $CY_{3}$ (see e.g. \cite{TT}). Indeed, in such a case in special
projective coordinates (with K\"{a}hler gauge fixed such that $X^{0}\equiv 1$%
) the holomorphic prepotential $\mathcal{F}$ and $W_{ijk}$ respectively read
\begin{equation}
\begin{array}{l}
\mathcal{F}=d_{ijk}z^{i}z^{j}z^{k}; \\
\\
C_{ijk}=e^{K}d_{ijk},
\end{array}
\end{equation}
and thus, for $\left| K\right| <\infty $ at least at the considered critical
points of $V_{BH}$, the third of conditions (\ref{non-BPS-Z=0-1}) cannot be
satisfied.

III.2) Critical points determined by the conditions
\begin{equation}
\left\{
\begin{array}{l}
Z=0, \\
\\
D_{i}Z\overset{Z=0}{=}\partial _{i}Z\neq 0,~~\text{\textit{at least} for
some }i\in \left\{ 1,...,n_{V}\right\} , \\
\\
C_{ijk}\neq 0,~~\text{\textit{at least }for some }\left( i,j,k\right) \in
\left\{ 1,...,n_{V}\right\} ^{3},
\end{array}
\right.  \label{non-BPS-Z=0-2}
\end{equation}
for which Eqs. (\ref{non-BPS-Z=0-fund}), and thus AEs (\ref{AEs3}), are not
trivially solved.

In general, non-BPS $Z=0$ critical points of $V_{BH}$ in $\mathcal{M}%
_{n_{V}} $ are not necessarily stable, because the $2n_{V}\times 2n_{V}$
(covariant) Hessian matrix (in $\left( z,\overline{z}\right) $-coordinates)
of $V_{BH}$ evaluated at such points is not necessarily strictly
positive-definite. An explicit condition of stability of non-BPS $Z=0$
critical points of $V_{BH}$ can be worked out in the $n_{V}=1$ case \cite
{BFMY}.

In general, Eqs. (\ref{22apr2}) and conditions (\ref{non-BPS-Z=0-conds})
imply the the $\lambda \psi $ term to be non-vanishing and the gravitino
``apparent mass'' term $Z\psi \psi $ to vanish, when evaluated at the
considered non-BPS $Z=0$ critical points of $V_{BH}$, characterized by
vanishing ( class III.1) or non-vanishing (class III.2) gaugino mass terms.

Non-BPS $Z=0$ attractors in the so-called $st^{2}$ and $stu$ models \cite
{BKRSW,Duff-stu} have been recently studied in \cite{BMOS-1}, and their
relation with the $\frac{1}{2}$-BPS attractors has been analyzed in light of
the uplift to $\mathcal{N}=8$, $d=4$ supergravity.

\setcounter{equation}0
\def\theequation{3.\arabic{subsection}.\arabic{equation}}
\subsection{\label{Sect4}Stability of Critical Points of $V_{BH}$}

\setcounter{equation}0
\def\theequation{3.2.\arabic{subsubsection}.\arabic{equation}}
\subsubsection{\label{Sect4-n}$n_{V}$-Moduli}

In order to decide whether a critical point of $V_{BH}$ is an attractor in
strict sense, one has to consider the following condition:
\begin{equation}
H_{\mathbb{R}}^{V_{BH}}\equiv H_{ab}^{V_{BH}}\equiv D_{a}D_{b}V_{BH}>0~~~%
\text{at}~~~D_{c}V_{BH}=\frac{\partial V_{BH}}{\partial \phi ^{c}}=0\text{~~~%
}\forall c=1,...,2n_{V},  \label{stab}
\end{equation}
\textit{i.e.} the condition of (strict) positive-definiteness of the real $%
2n_{V}\times 2n_{V}$ Hessian matrix $H_{\mathbb{R}}^{V_{BH}}\equiv
H_{ab}^{V_{BH}}$ of $V_{BH}$ (which is nothing but the squared mass matrix
of the moduli) at the critical points of $V_{BH}$, expressed in the real
parameterization through the $\phi $-coordinates. Since $V_{BH}$ is
positive-definite, a stable critical point (namely, an attractor in strict
sense) is necessarily a(n at least local) minimum, and therefore it fulfills
the condition (\ref{stab}).

In general, $H_{\mathbb{R}}^{V_{BH}}$ may be block-decomposed in $%
n_{V}\times n_{V}$ real matrices:
\begin{equation}
H_{\mathbb{R}}^{V_{BH}}=\left(
\begin{array}{ccc}
\mathcal{A} &  & \mathcal{C} \\
&  &  \\
\mathcal{C}^{T} &  & \mathcal{B}
\end{array}
\right) ,  \label{Hessian-real}
\end{equation}
with $\mathcal{A}$ and $\mathcal{B}$ being $n_{V}\times n_{V}$ real
symmetric matrices:
\begin{equation}
\mathcal{A}^{T}=\mathcal{A},~\mathcal{B}^{T}=\mathcal{B}\Leftrightarrow
\left( H_{\mathbb{R}}^{V_{BH}}\right) ^{T}=H_{\mathbb{R}}^{V_{BH}}.
\end{equation}

In the local complex $\left( z,\overline{z}\right) $-parameterization, the $%
2n_{V}\times 2n_{V}$ Hessian matrix of $V_{BH}$ reads
\begin{equation}
H_{\mathbb{C}}^{V_{BH}}\equiv H_{\widehat{i}\widehat{j}}^{V_{BH}}\equiv
\left(
\begin{array}{ccc}
D_{i}D_{j}V_{BH} &  & D_{i}\overline{D}_{\overline{j}}V_{BH} \\
&  &  \\
D_{j}\overline{D}_{\overline{i}}V_{BH} &  & \overline{D}_{\overline{i}}%
\overline{D}_{\overline{j}}V_{BH}
\end{array}
\right) =\left(
\begin{array}{ccc}
\mathcal{M}_{ij} &  & \mathcal{N}_{i\overline{j}} \\
&  &  \\
\overline{\mathcal{N}_{i\overline{j}}} &  & \overline{\mathcal{M}_{ij}}
\end{array}
\right) ,  \label{Hessian-complex}
\end{equation}
where the hatted indices $\hat{\imath}$ and $\hat{\jmath}$ may be
holomorphic or antiholomorphic. $H_{\mathbb{C}}^{V_{BH}}$ is the matrix
actually computable in the SKG formalism presented in Sect. \ref{SKG-gen}
(see below, Eqs. (\ref{M-bis}) and (\ref{N-bis})).

In general, $\frac{1}{2}$-BPS critical points are (at least local) minima of
$V_{BH}$, and therefore they are stable; thus, they are \textit{attractors}
in strict sense. Indeed, the $2n_{V}\times 2n_{V}$ (covariant) Hessian
matrix $H_{\mathbb{C}}^{V_{BH}}$ evaluated at such points is strictly
positive-definite \cite{FGK} :
\begin{eqnarray}
&&
\begin{array}{l}
\left( D_{i}D_{j}V_{BH}\right) _{\frac{1}{2}-BPS}=\left( \partial
_{i}\partial _{j}V_{BH}\right) _{\frac{1}{2}-BPS}=0, \\
\\
\left( D_{i}\overline{D}_{\overline{j}}V_{BH}\right) _{\frac{1}{2}%
-BPS}=\left( \partial _{i}\overline{\partial }_{\overline{j}}V_{BH}\right) _{%
\frac{1}{2}-BPS}=2\left( g_{i\overline{j}}V_{BH}\right) _{\frac{1}{2}%
-BPS}=2\left. g_{i\overline{j}}\right| _{\frac{1}{2}-BPS}\left| Z\right| _{%
\frac{1}{2}-BPS}^{2}>0,
\end{array}
\notag \\
&&  \label{SUSY-crit}
\end{eqnarray}
where here and below the notation ``$>0$'' (``$<0$'') is understood as
strict positive-(negative-)definiteness. The Hermiticity and (strict)
positive-definiteness of the (covariant) Hessian matrix $H_{\mathbb{C}%
}^{V_{BH}}$ at the $\frac{1}{2}$-BPS critical points are due to the
Hermiticity and - assumed - (strict) positive-definiteness (actually holding
globally) of the metric $g_{i\overline{j}}$ of the SK scalar manifold being
considered.

On the other hand, non-BPS critical points of $V_{BH}$ does not
automatically fulfill the condition (\ref{stab}), and a more detailed
analysis \cite{BFGM1,AoB-book} is needed.

Using the properties of SKG, one obtains:
\begin{eqnarray}
&&\mathcal{M}_{ij}\equiv D_{i}D_{j}V_{BH}=D_{j}D_{i}V_{BH}=4i\overline{Z}%
C_{ijk}g^{k\overline{k}}\left( \overline{D}_{\overline{k}}\overline{Z}%
\right) +i\left( D_{j}C_{ikl}\right) g^{k\overline{k}}g^{l\overline{l}%
}\left( \overline{D}_{\overline{k}}\overline{Z}\right) \left( \overline{D}_{%
\overline{l}}\overline{Z}\right) ;  \notag \\
&&  \label{M-bis} \\
&&\mathcal{N}_{i\overline{j}}\equiv D_{i}\overline{D}_{\overline{j}}V_{BH}=%
\overline{D}_{\overline{j}}D_{i}V_{BH}=2\left[ g_{i\overline{j}}\left|
Z\right| ^{2}+\left( D_{i}Z\right) \left( \overline{D}_{\overline{j}}%
\overline{Z}\right) +g^{l\overline{n}}C_{ikl}\overline{C}_{\overline{j}%
\overline{m}\overline{n}}g^{k\overline{k}}g^{m\overline{m}}\left( \overline{D%
}_{\overline{k}}\overline{Z}\right) \left( D_{m}Z\right) \right] ,  \notag \\
&&  \label{N-bis}
\end{eqnarray}
with $D_{j}C_{ikl}$ given by Eq. (\ref{DC}). Clearly, evaluating Eqs. (\ref
{M-bis}) and (\ref{N-bis}) constrained by the $\frac{1}{2}$-BPS conditions $%
D_{i}Z=0,\forall i=1,...,n_{V}$, one reobtains the results (\ref{SUSY-crit}%
). Here we limit ourselves to point out that further noteworthy elaborations
of $\mathcal{M}_{ij}$ and $\mathcal{N}_{i\overline{j}}$ can be performed in
homogeneous symmetric SK manifolds, where $D_{j}C_{ikl}=0$ globally \cite
{BFGM1}, and that the K\"{a}hler-invariant $\left( 2,2\right) $-tensor $g^{l%
\overline{n}}C_{ikl}\overline{C}_{\overline{j}\overline{m}\overline{n}}$ can
be rewritten in terms of the Riemann-Christoffel tensor $R_{i\overline{j}k%
\overline{m}}$ by using the so-called ``SKG constraints'' (see the third of
Eqs. (\ref{C})) \cite{AoB-book}. Moreover, the differential Bianchi
identities for $R_{i\overline{j}k\overline{m}}$ imply $\mathcal{M}_{ij}$ to
be symmetric (see comment below Eqs. (\ref{C}) and (\ref{DC})).

Thus, one gets the following global properties:
\begin{equation}
\mathcal{M}^{T}=\mathcal{M},~~\mathcal{N}^{\dag }=\mathcal{N}\Leftrightarrow
\left( H_{\mathbb{C}}^{V_{BH}}\right) ^{T}=H_{\mathbb{C}}^{V_{BH}},
\end{equation}
implying that
\begin{equation}
\left( H_{\mathbb{C}}^{V_{BH}}\right) ^{\dag }=H_{\mathbb{C}%
}^{V_{BH}}\Leftrightarrow \mathcal{M}^{\dag }=\mathcal{M},~~\mathcal{N}^{T}=%
\mathcal{N}\Leftrightarrow \overline{\mathcal{M}}=\mathcal{M},~~\overline{%
\mathcal{N}}=\mathcal{N}.
\end{equation}
It should be stressed clearly that the symmetry but non-Hermiticity of $H_{%
\mathbb{C}}^{V_{BH}}$ actually does not matter, because what one is
ineterested in are the eigenvalues of the real form $H_{\mathbb{R}}^{V_{BH}}$%
, which is real and symmetric, and therefore admitting $2n_{V}$ \textit{real}
eigenvalues.

The relation between $H_{\mathbb{R}}^{V_{BH}}$ expressed by Eq. (\ref
{Hessian-real}) and $H_{\mathbb{C}}^{V_{BH}}$ given by Eq. (\ref
{Hessian-complex}) is expressed by the following relations between the $%
n_{V}\times n_{V}$ sub-blocks of $H_{\mathbb{R}}^{V_{BH}}$ and $H_{\mathbb{C}%
}^{V_{BH}}$ \cite{BFM,BFM-SIGRAV06}:
\begin{equation}
\left\{
\begin{array}{l}
\mathcal{M}=\frac{1}{2}\left( \mathcal{A}-\mathcal{B}\right) +\frac{i}{2}%
\left( \mathcal{C}+\mathcal{C}^{T}\right) ; \\
\\
\mathcal{N}=\frac{1}{2}\left( \mathcal{A}+\mathcal{B}\right) +\frac{i}{2}%
\left( \mathcal{C}^{T}-\mathcal{C}\right) ,
\end{array}
\right.  \label{4jan1}
\end{equation}
or its inverse
\begin{equation}
\left\{
\begin{array}{l}
\mathcal{A}=Re\mathcal{M}+Re\mathcal{N}; \\
\\
\mathcal{B}=Re\mathcal{N}-Re\mathcal{M}; \\
\\
\mathcal{C}=Im\mathcal{M}-Im\mathcal{N}.
\end{array}
\right.  \label{4jan2}
\end{equation}

The structure of the Hessian matrix gets simplified at the critical points
of $V_{BH}$, because the covariant derivatives may be substituted by the
flat ones; the critical Hessian matrices in complex
holomorphic/antiholomorphic and real local parameterizations respectively
read
\begin{eqnarray}
&&
\begin{array}{l}
\left. H_{\mathbb{C}}^{V_{BH}}\right| _{\partial V_{BH}=0}\equiv \left(
\begin{array}{ccc}
\partial _{i}\partial _{j}V_{BH} &  & \partial _{i}\overline{\partial }_{%
\overline{j}}V_{BH} \\
&  &  \\
\partial _{j}\overline{\partial }_{\overline{i}}V_{BH} &  & \overline{%
\partial }_{\overline{i}}\overline{\partial }_{\overline{j}}V_{BH}
\end{array}
\right) _{\partial V_{BH}=0}=\left(
\begin{array}{ccc}
\mathcal{M} &  & \mathcal{N} \\
&  &  \\
\overline{\mathcal{N}} &  & \overline{\mathcal{M}}
\end{array}
\right) _{\partial V_{BH}=0};
\end{array}
\label{Hessian-complex-crit} \\
&&  \notag \\
&&  \notag \\
&&
\begin{array}{l}
\left. H_{\mathbb{R}}^{V_{BH}}\right| _{\partial V_{BH}=0}=\left. \frac{%
\partial ^{2}V_{BH}}{\partial \phi ^{a}\partial \phi ^{b}}\right| _{\partial
V_{BH}=0}=\left(
\begin{array}{ccc}
\mathcal{A} &  & \mathcal{C} \\
&  &  \\
\mathcal{C}^{T} &  & \mathcal{B}
\end{array}
\right) _{\partial V_{BH}=0}.
\end{array}
\label{Hessian-real-crit}
\end{eqnarray}
Thus, the study of the condition (\ref{stab}) finally amounts to the study
of the \textit{eigenvalue problem} of the real symmetric $2n_{V}\times
2n_{V} $ critical Hessian matrix $\left. H_{\mathbb{R}}^{V_{BH}}\right|
_{\partial V_{BH}=0}$ given by Eq. (\ref{Hessian-real-crit}), which is the
Cayley-transformed of the complex (symmetric, but not necessarily Hermitian)
$2n_{V}\times 2n_{V}$ critical Hessian $\left. H_{\mathbb{C}%
}^{V_{BH}}\right| _{\partial V_{BH}=0}$ given by Eq. (\ref
{Hessian-complex-crit}). \setcounter{equation}0
\setcounter{equation}0
\def\theequation{3.2.\arabic{subsubsection}.\arabic{equation}}
\subsubsection{\label{Sect4-1}$1$-Modulus}

Once again, the situation strongly simplifies in $n_{V}=1$ SKG.

Indeed, for $n_{V}=1$ the moduli-dependent matrices $\mathcal{A}$, $\mathcal{%
B}$, $\mathcal{C}$, $\mathcal{M}$ and $\mathcal{N}$ introduced above are
simply scalar functions. In particular, $\mathcal{N}$ is real, since $%
\mathcal{C}$ trivially satisfies $\mathcal{C}=\mathcal{C}^{T}$. The
stability condition (\ref{stab}) can thus be written as
\begin{equation}
H_{\mathbb{R}}^{V_{BH}}\equiv D_{a}D_{b}V_{BH}>0\text{,~}\left(
a,b=1,2\right) ~~~\text{at}~~~D_{c}V_{BH}=\frac{\partial V_{BH}}{\partial
\phi ^{c}}=0\text{~~~}\forall c=1,2,  \label{stabi-1}
\end{equation}
and Eqs. (\ref{M-bis}) and (\ref{N-bis}) respectively simplify to
\begin{eqnarray}
&&
\begin{array}{l}
\mathcal{M}\equiv D^{2}V_{BH}=4i\overline{Z}Cg^{-1}\overline{D}\overline{Z}%
+i\left( DC\right) g^{-2}\left( \overline{D}\overline{Z}\right) ^{2};
\end{array}
\label{M-1} \\
&&  \notag \\
&&
\begin{array}{l}
\mathcal{N}\equiv D\overline{D}V_{BH}=\overline{D}DV_{BH}=2\left[ g\left|
Z\right| ^{2}+\left| DZ\right| ^{2}+\left| C\right| ^{2}g^{-3}\left|
DZ\right| ^{2}\right] ,
\end{array}
\label{N-1}
\end{eqnarray}
$DC$ being given by the case $n_{V}=1$ of Eq. (\ref{DC}):
\begin{equation}
DC=\partial C+\left[ \left( \partial K\right) +3\Gamma \right] C=\partial C+%
\left[ \left( \partial K\right) -3\partial ln\left( g\right) \right]
C=\left\{ \partial +\left[ \partial ln\left( \frac{e^{K}}{\left( \overline{%
\partial }\partial K\right) ^{3}}\right) \right] \right\} C,  \label{DC-1}
\end{equation}
where the $n_{V}=1$ Christoffel connection
\begin{equation}
\Gamma =-g^{-1}\partial g=-\partial ln\left( g\right)  \label{Christoffel-1}
\end{equation}
was used. It is easy to show that the stability condition (\ref{stabi-1})
for critical points of $V_{BH}$ in $n_{V}=1$ SKG can be equivalently
reformulated as the strict bound
\begin{equation}
\left. \mathcal{N}\right| _{\partial V_{BH}=0}>\left| \mathcal{M}\right|
_{\partial V_{BH}=0}.  \label{stab-1}
\end{equation}
Let us now see how such a bound can be further elaborated for the three
possible classes of critical points of $V_{BH}$. \setcounter{equation}0
\def\theequation{3.2.2.1.\arabic{equation}}
\paragraph{\label{stab-BPS}$\frac{1}{2}$-BPS}

~

\begin{eqnarray}
&&\mathcal{M}_{\frac{1}{2}-BPS}\equiv \left. D^{2}V_{BH}\right| _{\frac{1}{2}%
-BPS}=\left[ 3\overline{Z}D^{2}Z+g^{-1}\left( D^{3}Z\right) \overline{D}%
\overline{Z}\right] _{\frac{1}{2}-BPS}=0;  \notag \\
&&  \label{M-1-BPS} \\
&&\mathcal{N}_{\frac{1}{2}-BPS}\equiv \left. D\overline{D}V_{BH}\right| _{%
\frac{1}{2}-BPS}=\left[ 2g\left| Z\right| ^{2}+g^{-1}\left| D^{2}Z\right|
^{2}\right] _{\frac{1}{2}-BPS}=2\left( g\left| Z\right| ^{2}\right) _{\frac{1%
}{2}-BPS}.  \notag \\
&&  \label{N-1-BPS}
\end{eqnarray}
Eqs. (\ref{M-1-BPS}) and (\ref{N-1-BPS}) are nothing but the 1-modulus case
of Eq. (\ref{SUSY-crit}), and they directly satisfy the bound (\ref{stab-1}%
). Thus, consistently with what stated above, the $\frac{1}{2}$-BPS class of
critical points of $V_{BH}$ actually is a class of attractors in strict
sense (\textit{at least} local minima of $V_{BH}$). \setcounter{equation}0
\def\theequation{3.2.2.2.\arabic{equation}}
\paragraph{\label{stab-non-BPS-Z<>0}Non-BPS, $Z\neq 0$}

~

\begin{eqnarray}
&&
\begin{array}{l}
\mathcal{M}_{non-BPS,Z\neq 0}\equiv \left. D^{2}V_{BH}\right|
_{non-BPS,Z\neq 0}= \\
\\
=-2\left\{ g^{-1}\overline{Z}DZ\left[ g^{-2}\left| C\right| ^{2}Dln\left(
Z\right) +gDln\left( C\right) \right] \right\} _{non-BPS,Z\neq 0}= \\
\\
=i\left\{ Cg^{-3}\left( \overline{D}\overline{Z}\right) ^{2}\left[
g^{-2}\left| C\right| ^{2}Dln\left( Z\right) +gDln\left( C\right) \right]
\right\} _{non-BPS,Z\neq 0};
\end{array}
\notag \\
&&  \label{M-1-non-BPS-Z<>0} \\
&&  \notag \\
&&
\begin{array}{l}
\mathcal{N}_{non-BPS,Z\neq 0}\equiv \left. D\overline{D}V_{BH}\right|
_{non-BPS,Z\neq 0}=\left. \overline{D}DV_{BH}\right| _{non-BPS,Z\neq 0}= \\
\\
=2\left\{ \left| DZ\right| ^{2}\left[ 1+\frac{5}{4}g^{-3}\left| C\right| ^{2}%
\right] \right\} _{non-BPS,Z\neq 0}.
\end{array}
\notag \\
&&  \label{N-1-non-BPS-Z<>0}
\end{eqnarray}
Eq. (\ref{M-1-non-BPS-Z<>0}) yields that
\begin{eqnarray}
&&
\begin{array}{l}
\left| \mathcal{M}\right| _{non-BPS,Z\neq 0}^{2}= \\
\\
=4\left\{ \left| DZ\right| ^{4}\left[ \left| C\right| ^{4}g^{-6}+\frac{1}{4}%
g^{-4}\left| DC\right| ^{2}+2g^{-3}Re\left[ C\left( \overline{D}\overline{C}%
\right) Dln\left( Z\right) \right] \right] \right\} _{non-BPS,Z\neq 0}.
\end{array}
\notag \\
&&  \label{M-non-BPS-Z<>0}
\end{eqnarray}
By substituting Eqs. (\ref{N-1-non-BPS-Z<>0}) and (\ref{M-non-BPS-Z<>0})
into the strict inequality (\ref{stab-1}), one finally obtains the stability
condition for non-BPS, $Z\neq 0$ critical points of $V_{BH}$ in $n_{V}=1$
SKG \cite{BFM}:
\begin{gather}
\mathcal{N}_{non-BPS,Z\neq 0}>\left| \mathcal{M}\right| _{non-BPS,Z\neq 0};
\label{stab-1-1} \\
\Updownarrow  \notag \\
1+\frac{5}{4}\left( \left| C\right| ^{2}g^{-3}\right) _{non-BPS,Z\neq 0}>%
\sqrt{\left[ \left| C\right| ^{4}g^{-6}+\frac{1}{4}g^{-4}\left| DC\right|
^{2}+2g^{-3}Re\left[ C\left( \overline{D}\overline{C}\right) \left(
\overline{D}ln\overline{Z}\right) \right] \right] _{non-BPS,Z\neq 0}}.
\notag \\
\end{gather}
As it is seen from such a condition, in general $\left( DC\right)
_{non-BPS,Z\neq 0}$ is the fundamental geometrical quantity playing a key
role in determining the stability of non-BPS, $Z\neq 0$ critical points of $%
V_{BH}$ in $1$-modulus SK geometry. \setcounter{equation}0
\def\theequation{3.2.2.3.\arabic{equation}}
\paragraph{\label{stab-non-BPS-Z=0}Non-BPS, $Z=0$}

~

\begin{eqnarray}
&&
\begin{array}{l}
\mathcal{M}_{non-BPS,Z=0}\equiv \left. D^{2}V_{BH}\right| _{non-BPS,Z=0}=i%
\left[ g^{-2}\left( \partial C\right) \left( \overline{\partial }\overline{Z}%
\right) ^{2}\right] _{non-BPS,Z=0};
\end{array}
\notag \\
&&  \label{M-1-non-BPS-Z=0} \\
&&
\begin{array}{l}
\mathcal{N}_{non-BPS,Z=0}\equiv \left. D\overline{D}V_{BH}\right|
_{non-BPS,Z=0}=2\left| \partial Z\right| _{non-BPS,Z=0}^{2},
\end{array}
\notag \\
&&  \label{N-1-non-BPS-Z=0}
\end{eqnarray}
where Eqs. (\ref{non-BPS-Z=0-2}) and (\ref{non-BPS-Z=0-1}) have been used.
Eq. (\ref{M-1-non-BPS-Z=0}) yields that
\begin{equation}
\left| \mathcal{M}\right| _{non-BPS,Z=0}=\left[ g^{-2}\left| \partial
C\right| \left| \partial Z\right| ^{2}\right] _{non-BPS,Z=0}.
\label{M-non-BPS-Z=0}
\end{equation}
By substituting Eqs. (\ref{N-1-non-BPS-Z=0}) and (\ref{M-non-BPS-Z=0}) into
the strict inequality (\ref{stab-1}), one finally obtains the stability
condition for non-BPS, $Z=0$ critical points of $V_{BH}$ in $n_{V}=1$ SKG:
\begin{gather}
\mathcal{N}_{non-BPS,Z=0}>\left| \mathcal{M}\right| _{non-BPS,Z=0}; \\
\Updownarrow  \notag \\
2g_{non-BPS,Z=0}^{2}>\left| \partial C\right| _{non-BPS,Z=0}.
\label{stab-1-2}
\end{gather}
Even though the stability condition (\ref{stab-1-2}) have been obtained by
correctly using Eqs. (\ref{non-BPS-Z=0-2}) and (\ref{non-BPS-Z=0-1}),
holding at the non-BPS, $Z=0$ critical points of $V_{BH}$, in some cases it
may happen that, in the limit of approaching the non-BPS, $Z=0$ critical
point of $V_{BH}$, in $DC$ (given by Eq. (\ref{DC-1})) the ``connection
term'' $\left[ \left( \partial K\right) +3\Gamma \right] C$ is not
necessarily sub-leading with respect to the ``differential term'' $\partial
C $. Thus, the condition (\ref{stab-1-2}) can be rewritten as follows:
\begin{equation}
\begin{array}{l}
2\left( \overline{\partial }\partial K\right) _{non-BPS,Z=0}^{2}>\left|
\left\{ \partial +\left[ \left( \partial K\right) -3\partial ln\left(
g\right) \right] \right\} C\right| _{non-BPS,Z=0}= \\
\\
=\left| \left\{ \partial +\left[ \partial ln\left( \frac{e^{K}}{\left(
\overline{\partial }\partial K\right) ^{3}}\right) \right] \right\} C\right|
_{non-BPS,Z=0}.
\end{array}
\label{stab-1-3}
\end{equation}
\medskip \setcounter{equation}0
\def\theequation{3.2.2.4.\arabic{equation}}
\paragraph{Remark}

~

Let us consider the $1$-modulus stability conditions (\ref{stab-1-1}) and (%
\ref{stab-1-2})-\ref{stab-1-3}). It is immediate to realize that they are
both satisfied when the function $C$ is globally covariantly constant:
\begin{equation}
DC=\partial C+\left[ \left( \partial K\right) +3\Gamma \right] C=0,
\label{DC=0-1}
\end{equation}
\textit{i.e.} for the so-called homogeneous symmetric ($dim_{\mathbb{C}%
}=n_{V}=1$) SK geometry \cite{CKV,CVP}, univoquely associated to the coset
manifold $\frac{SU(1,1)}{U(1)}$. Such a SK manifold can be twofold
characterized as:

\textit{i}) the $n=0$ element of the irreducible rank-$1$ infinite sequence $%
\frac{SU(1,1+n)}{U(1)\otimes SU(1+n)}$ (with $n_{V}=n+rank=n+1$), or
equivalently the $n=-2$ element of the reducible rank-$3$ infinite sequence $%
\frac{SU(1,1)}{U(1)}\otimes \frac{SO(2,2+n)}{SO(2)\otimes SO(2+n)}$ (with $%
n_{V}=n+rank=n+3$). In such a case, $\frac{SU(1,1)}{U(1)}$ is endowed with a
quadratic holomorphic prepotential function reading (in a suitable
projective special coordinate, with K\"{a}hler gauge fixed such that $%
X^{0}=1 $; see \cite{BFGM1} and Refs. therein)
\begin{equation}
\mathcal{F}(z)=\frac{i}{2}\left( z^{2}-1\right) .  \label{quadr-1}
\end{equation}
By recalling the first of Eqs. (\ref{C}), such a prepotential yields $C=0$
globally (and thus Eq. (\ref{DC=0-1})), and therefore, by using the SKG
constraints (i.e. the third of Eqs. (\ref{C})) it yields also the constant
scalar curvature
\begin{equation}
\mathcal{R}\equiv g^{-2}R=-2,  \label{curv-quadr-1}
\end{equation}
where $R\equiv R_{1\overline{1}1\overline{1}}$ denotes the unique component
of the Riemann tensor. As obtained in \cite{BFGM1}, quadratic (homogeneous
symmetric) SK geometries only admit $\frac{1}{2}$-BPS and non-BPS, $Z=0$
critical points of $V_{BH}$. Thus, it can be concluded that the $1$-dim.
quadratic SK geometry determined by the prepotential (\ref{quadr-1}) admits
\textit{all} stable critical points of $V_{BH}$.

\textit{ii}) the rank-$1$ $s=t=u\equiv z$ \textit{degeneration} of the
so-called $stu$ model \cite{BKRSW,Duff-stu} ($n=0$ element of the reducible
rank-$3$ infinite sequence $\frac{SU(1,1)}{U(1)}\otimes \frac{SO(2,2+n)}{%
SO(2)\otimes SO(2+n)}$), or equivalently the rank-$1$ $s=t\equiv z$ \textit{%
degeneration} of the so-called $st^{2}$ model ($n=-1$ element of $\frac{%
SU(1,1)}{U(1)}\otimes \frac{SO(2,2+n)}{SO(2)\otimes SO(2+n)}$), or also as
an isolated case in the classification of homogeneous symmetric SK manifolds
(see \textit{e.g}. \cite{CFG}). In such a case, $\frac{SU(1,1)}{U(1)}$ is
endowed with a cubic holomorphic prepotential function reading (in a
suitable projective special coordinate, with K\"{a}hler gauge fixed such
that $X^{0}=1$; see \textit{e.g.} \cite{BFGM1} and Refs. therein)
\begin{equation}
\mathcal{F}(z)=\varrho z^{3},~\varrho \in \mathbb{C},  \label{cub-1}
\end{equation}
constrained by the condition $Im\left( z\right) <0$. It admits an uplift to
\textit{pure} $\mathcal{N}=2$ supergravity in $d=5$. By recalling the first
of Eqs. (\ref{C}), such a prepotential yields $C=6\varrho e^{K}$ (and thus
Eq. (\ref{DC=0-1})), and consequently it also yields the constant scalar
curvature
\begin{equation}
\mathcal{R}\equiv g^{-2}R=g^{-2}\left( -2g^{2}+g^{-1}\left| C\right|
^{2}\right) =-\frac{2}{3},  \label{curv-cub-1}
\end{equation}
where the SKG constraints (i.e. the third of Eqs. (\ref{C})) and the global
value\footnote{%
The global value $\left| C\right| ^{2}g^{-3}=\frac{4}{3}$ for homogeneous
symmetric cubic $n_{V}=1$ SK geometries is yielded by the $n_{V}=1$ case of
Eq. (\ref{UCLA13}).} $\left| C\right| ^{2}g^{-3}=\frac{4}{3}$ have been
used. As it can be computed (see \textit{e.g.} \cite{Saraikin-Vafa-1}), the $%
1$-dim. SK geometry determined by the prepotential (\ref{cub-1}) admits,
beside the (stable) $\frac{1}{2}$-BPS ones, stable non-BPS $Z\neq 0$
critical points of $V_{BH}$. Thus, it is another example in which \textit{all%
} critical points of $V_{BH}$ actually are attractors in a strict
sense.\smallskip

Clearly, the quadratic and cubic homogeneous symmetric $1$-modulus SK
geometries (respectively determined by holomorphic prepotentials (\ref
{quadr-1}) and (\ref{cub-1})) are not the only ones (with $n_{V}=1$)
admitting stable non-BPS critical points of $V_{BH}$. For instance, as
studied in \cite{BFMY}, the $1$-modulus SK geometries of the moduli space of
the (mirror) Fermat $CY_{3}$ \textit{quintic} $\mathcal{M}_{5}^{\prime }$
and \textit{octic} $\mathcal{M}_{8}^{\prime }$ admit, in a suitable
neighbourhood of the LG point, stable non-BPS ($Z\neq 0$) critical points of
$V_{BH}$.

It is worth remarking that recent works \cite{Ferrara-Marrani-1,
fm07,PASCOS07} gave a complete treatment of the issue of stability of
non-BPS attractors in the framework of homogeneous SKGs, finding that the
massless modes of the non-BPS Hessian matrix actually are ``flat
directions'' of $V_{BH}$ at the considered class of critical points. This
means that non-BPS attractors in $\mathcal{N}=2$, $d=4$ supergravity have a
related moduli space, spanned by those moduli which are not stabilized at
the BH horizon. However, it should be pointed out that such an emergence of
moduli spaces do not violate the Attractor Mechanism and/or the determinacy
of BH thermodinamical properties, because the non-BPS BH entropy simply does
not depend on the scalar degrees of freedom spanning the moduli space of the
considered class ($Z\neq 0$ or $Z=0$) of non-BPS critical points of $V_{BH}$%
. Such considerations hold also for $\mathcal{N}>2$-extended, $d=4$
supergravities (where also BPS attractors can have a related moduli space),
and in general in all theories with an homogeneous (not necessarily
symmetric) scalar manifold \cite{Ferrara-Marrani-1, CFM1, PASCOS07,AFMT1}.%
\setcounter{equation}0
\def\theequation{3.\arabic{subsection}.\arabic{equation}}
\subsection{\label{General-Formulation}$\mathcal{N}=2$, $d=4$ General
Formulation}

\setcounter{equation}0
\def\theequation{3.3.\arabic{subsubsection}.\arabic{equation}}
\subsubsection{\label{Derivation-SKG-identities}Special K\"{a}hler Geometry
Identities}

We will now derive some important identities of the SK geometry \cite
{FBC,K1,K2,BFM,AoB-book,K3} of the scalar manifold of $\mathcal{N}=2$, $d=4$
ungauged supergravity. Such identities extend the results obtained by
Ferrara and Kallosh in \cite{FK1}.

Let us start by considering the covariant antiholomorphic derivative of $%
\overline{Z}$; by recalling the definition (\ref{Z}) and using the second of
\textit{Ans\"{a}tze} (\ref{Ans1}), one gets
\begin{equation}
\overline{D}_{\overline{j}}\overline{Z}=q_{\Lambda }\overline{D}_{\overline{i%
}}\overline{L}^{\Lambda }-p^{\Lambda }\mathcal{N}_{\Lambda \Delta }\overline{%
D}_{\overline{j}}\overline{L}^{\Delta }.  \label{starting-expr}
\end{equation}
The contraction of both sides with $g^{i\overline{j}}D_{i}L^{\Sigma }$ then
yields
\begin{equation}
g^{i\overline{j}}\left( D_{i}L^{\Sigma }\right) \overline{D}_{\overline{j}}%
\overline{Z}=q_{\Lambda }g^{i\overline{j}}\left( D_{i}L^{\Sigma }\right)
\overline{D}_{\overline{j}}\overline{L}^{\Lambda }-p^{\Lambda }\mathcal{N}%
_{\Lambda \Delta }g^{i\overline{j}}\left( D_{i}L^{\Sigma }\right) \overline{D%
}_{\overline{j}}\overline{L}^{\Delta }.  \label{20apr1}
\end{equation}
By exploiting the symmetry of $\mathcal{N}_{\Lambda \Sigma }$ and its
inverse (see Eq. (\ref{N}) and Eq. (\ref{L}) further below, as well),
recalling the first of the Ans\"{a}tze (\ref{Ans1}), and using the result of
SK geometry (see \textit{e.g.} \cite{4})
\begin{equation}
g^{i\overline{j}}\left( D_{i}L^{\Lambda }\right) \overline{D}_{\overline{j}}%
\overline{L}^{\Sigma }=-\frac{1}{2}\left( Im\mathcal{N}\right) ^{-1\mid
\Lambda \Sigma }-\overline{L}^{\Lambda }L^{\Sigma },  \label{20apr2}
\end{equation}
Eq.(\ref{20apr1}) can be further elaborated as follows:
\begin{eqnarray}
&&
\begin{array}{l}
g^{i\overline{j}}\left( D_{i}L^{\Sigma }\right) \overline{D}_{\overline{j}}%
\overline{Z}=q_{\Lambda }\left[ -\frac{1}{2}\left( Im\mathcal{N}\right)
^{-1\mid \Sigma \Lambda }-\overline{L}^{\Sigma }L^{\Lambda }\right]
-p^{\Lambda }\mathcal{N}_{\Lambda \Delta }\left[ -\frac{1}{2}\left( Im%
\mathcal{N}\right) ^{-1\mid \Sigma \Delta }-\overline{L}^{\Sigma }L^{\Delta }%
\right] = \\
\\
=-\frac{1}{2}\left( Im\mathcal{N}\right) ^{-1\mid \Sigma \Lambda }q_{\Lambda
}-\overline{L}^{\Sigma }\left( L^{\Lambda }q_{\Lambda }-M_{\Lambda
}p^{\Lambda }\right) +\frac{1}{2}\left( Im\mathcal{N}\right) ^{-1\mid \Sigma
\Delta }\left( Re\mathcal{N}_{\Delta \Lambda }\right) p^{\Lambda }+\frac{i}{2%
}p^{\Sigma }= \\
\\
=\frac{i}{2}p^{\Sigma }-\overline{L}^{\Sigma }Z+\frac{1}{2}\left( Im\mathcal{%
N}\right) ^{-1\mid \Sigma \Delta }\left( Re\mathcal{N}_{\Delta \Lambda
}\right) p^{\Lambda }-\frac{1}{2}\left( Im\mathcal{N}\right) ^{-1\mid \Sigma
\Lambda }q_{\Lambda }.
\end{array}
\notag \\
&&  \label{SKG-p}
\end{eqnarray}
Now, by subtracting to the expression (\ref{SKG-p}) its complex conjugate,
one gets
\begin{equation}
p^{\Lambda }=2Re\left[ i\overline{Z}L^{\Lambda }+ig^{i\overline{j}}\left(
D_{i}Z\right) \overline{D}_{\overline{j}}\overline{L}^{\Lambda }\right] =-2Im%
\left[ \overline{Z}L^{\Lambda }+g^{i\overline{j}}\left( D_{i}Z\right)
\overline{D}_{\overline{j}}\overline{L}^{\Lambda }\right] .  \label{SKG-p2}
\end{equation}

On the other hand, by using the second of Ans\"{a}tze (\ref{Ans1}), the
contraction of both sides of Eq. (\ref{starting-expr}) with $g^{i\overline{j}%
}D_{j}M_{\Sigma }$ analogously yields
\begin{eqnarray}
&&
\begin{array}{l}
g^{i\overline{j}}\left( D_{i}M_{\Sigma }\right) \overline{D}_{\overline{j}}%
\overline{Z}=q_{\Lambda }g^{i\overline{j}}\left( D_{i}M_{\Sigma }\right)
\overline{D}_{\overline{j}}\overline{L}^{\Lambda }-p^{\Lambda }\mathcal{N}%
_{\Lambda \Delta }g^{i\overline{j}}\left( D_{i}M_{\Sigma }\right) \overline{D%
}_{\overline{j}}\overline{L}^{\Delta }= \\
\\
=q_{\Lambda }g^{i\overline{j}}\overline{\mathcal{N}}_{\Sigma \Delta }\left(
D_{i}L^{\Delta }\right) \overline{D}_{\overline{j}}\overline{L}^{\Lambda
}-p^{\Lambda }\mathcal{N}_{\Lambda \Delta }g^{i\overline{j}}\overline{%
\mathcal{N}}_{\Sigma \Xi }\left( D_{i}L^{\Xi }\right) \overline{D}_{%
\overline{j}}\overline{L}^{\Delta }.
\end{array}
\notag \\
&&  \label{20apr3}
\end{eqnarray}
Once again, by exploiting the symmetry of $\mathcal{N}_{\Lambda \Sigma }$
and its inverse, recalling the first of the Ans\"{a}tze (\ref{Ans1}), and
using Eq. (\ref{20apr2}), Eq. (\ref{20apr3}) can be further elaborated as
follows:
\begin{eqnarray}
&&
\begin{array}{l}
g^{i\overline{j}}\left( D_{i}M_{\Sigma }\right) \overline{D}_{\overline{j}}%
\overline{Z}=q_{\Lambda }\overline{\mathcal{N}}_{\Sigma \Delta }\left[ -%
\frac{1}{2}\left( Im\mathcal{N}\right) ^{-1\mid \Delta \Lambda }-\overline{L}%
^{\Delta }L^{\Lambda }\right] -p^{\Lambda }\mathcal{N}_{\Lambda \Delta }%
\overline{\mathcal{N}}_{\Sigma \Xi }\left[ -\frac{1}{2}\left( Im\mathcal{N}%
\right) ^{-1\mid \Xi \Delta }-\overline{L}^{\Xi }L^{\Delta }\right] = \\
\\
=-\frac{1}{2}\left( Im\mathcal{N}\right) ^{-1\mid \Delta \Lambda }\left( Re%
\mathcal{N}_{\Sigma \Delta }\right) q_{\Lambda }+\frac{i}{2}q_{\Sigma }-%
\overline{M}_{\Sigma }Z+\frac{1}{2}\left( Im\mathcal{N}\right) ^{-1\mid \Xi
\Delta }\left( Re\mathcal{N}_{\Sigma \Xi }\right) \left( Re\mathcal{N}%
_{\Lambda \Delta }\right) p^{\Lambda }+\frac{1}{2}\left( Im\mathcal{N}%
_{\Lambda \Sigma }\right) p^{\Lambda }.
\end{array}
\notag \\
&&  \label{SKG-q}
\end{eqnarray}
\medskip \medskip Thence, by subtracting to the expression (\ref{SKG-q}) its
complex conjugate, one gets
\begin{equation}
q_{\Lambda }=2Re\left[ i\overline{Z}M_{\Lambda }+ig^{i\overline{j}}\left(
D_{i}Z\right) \overline{D}_{\overline{j}}\overline{M}_{\Lambda }\right] =-2Im%
\left[ \overline{Z}M_{\Lambda }+g^{i\overline{j}}\left( D_{i}Z\right)
\overline{D}_{\overline{j}}\overline{M}_{\Lambda }\right] .  \label{SKG-q2}
\end{equation}

By expressing the identities (\ref{SKG-p2}) and (\ref{SKG-q2}) in a vector $%
Sp\left( 2n_{V}+2\right) $-covariant notation, one finally gets
\begin{eqnarray}
\left(
\begin{array}{c}
p^{\Lambda } \\
\\
q_{\Lambda }
\end{array}
\right)  &=&-2Im\left[ \overline{Z}\left(
\begin{array}{c}
L^{\Lambda } \\
\\
M_{\Lambda }
\end{array}
\right) +g^{i\overline{j}}D_{i}Z\left(
\begin{array}{c}
\overline{D}_{\overline{j}}\overline{L}^{\Lambda } \\
\\
\overline{D}_{\overline{j}}\overline{M}_{\Lambda }
\end{array}
\right) \right] ,  \notag \\
&&  \label{pre-SKG-final1}
\end{eqnarray}
or in compact form
\begin{equation}
Q^{T}=-2Im\left[ \overline{Z}V+g^{i\overline{j}}\left( D_{i}Z\right)
\overline{D}_{\overline{j}}\overline{V}\right] ,  \label{SKG-final1}
\end{equation}
where we recalled the definitions (\ref{Gamma-tilde}) and (\ref{PI}) of the $%
\left( 2n_{V}+2\right) \times 1$ vectors $Q^{T}$ and $V$, respectively.

It is worth pointing out that the vector identity (\ref{SKG-final1}) has
been obtained only by using the properties of the SK geometry. The relations
yielded by the identity (\ref{SKG-final1}) are $2n_{V}+2$ real ones, but
they have been obtained by starting from an expression for $\overline{D}_{%
\overline{i}}\overline{Z}$, corresponding to $n_{V}$ complex, and therefore $%
2n_{V}$ real, degrees of freedom. The two redundant real degrees of freedom
are encoded in the homogeneity (of degree 1) of the identity (\ref
{SKG-final1}) under complex rescalings of the symplectic BH charge vector $Q$%
; indeed, by recalling the definition (\ref{Z}) it is immediate to check
that the r.h.s. of identity (\ref{SKG-final1}) acquires an overall factor $%
\lambda $ under a global rescaling of $Q$ of the kind
\begin{equation}
Q\longrightarrow \lambda Q,\text{ \ }\lambda \in \mathbb{C}.
\label{complex-rescaling}
\end{equation}

The summation of the expressions (\ref{SKG-p}) and (\ref{SKG-q}) with their
complex conjugates respectively yields
\begin{eqnarray}
&&
\begin{array}{l}
\left( Im\mathcal{N}\right) ^{-1\mid \Delta \Lambda }\left( Re\mathcal{N}%
_{\Delta \Sigma }\right) p^{\Sigma }-\left( Im\mathcal{N}\right) ^{-1\mid
\Lambda \Sigma }q_{\Sigma }=2Re\left[ \overline{Z}L^{\Lambda }+g^{i\overline{%
j}}\left( D_{i}Z\right) \overline{D}_{\overline{j}}\overline{L}^{\Lambda }%
\right] ;
\end{array}
\notag \\
&&  \label{SKG-p3} \\
&&
\begin{array}{l}
\left[ Im\mathcal{N}_{\Lambda \Sigma }+\left( Im\mathcal{N}\right) ^{-1\mid
\Xi \Delta }\left( Re\mathcal{N}_{\Lambda \Xi }\right) Re\mathcal{N}_{\Sigma
\Delta }\right] p^{\Sigma }-\left( Im\mathcal{N}\right) ^{-1\mid \Delta
\Sigma }\left( Re\mathcal{N}_{\Lambda \Delta }\right) q_{\Sigma }= \\
\\
=2Re\left[ \overline{Z}M_{\Lambda }+g^{i\overline{j}}\left( D_{i}Z\right)
\overline{D}_{\overline{j}}\overline{M}_{\Lambda }\right] .
\end{array}
\notag \\
&&  \label{SKG-q3}
\end{eqnarray}

In order to elaborate a shorthand notation for the obtained SKG identities (%
\ref{SKG-p2}), (\ref{SKG-q2}) and (\ref{SKG-p3}), (\ref{SKG-q3}), let us now
reconsider the starting expressions (\ref{SKG-p}) and (\ref{SKG-q}),
respectively reading
\begin{eqnarray}
&&
\begin{array}{l}
\left[ \delta _{\Sigma }^{\Lambda }-i\left( Im\mathcal{N}\right) ^{-1\mid
\Lambda \Delta }Re\mathcal{N}_{\Delta \Sigma }\right] p^{\Sigma }+i\left( Im%
\mathcal{N}\right) ^{-1\mid \Lambda \Sigma }q_{\Sigma }=-2i\overline{L}%
^{\Lambda }Z-2ig^{i\overline{j}}\left( \overline{D}_{\overline{j}}\overline{Z%
}\right) D_{i}L^{\Lambda };
\end{array}
\notag \\
&&  \label{SKG-p-fund} \\
&&
\begin{array}{l}
-i\left[ \left( Im\mathcal{N}\right) ^{-1\mid \Xi \Delta }\left( Re\mathcal{N%
}_{\Lambda \Xi }\right) Re\mathcal{N}_{\Sigma \Delta }+Im\mathcal{N}%
_{\Lambda \Sigma }\right] p^{\Sigma }+\left[ \delta _{\Lambda }^{\Sigma
}+i\left( Im\mathcal{N}\right) ^{-1\mid \Delta \Sigma }Re\mathcal{N}%
_{\Lambda \Delta }\right] q_{\Sigma }= \\
\\
=-2i\overline{M}_{\Lambda }Z-2ig^{i\overline{j}}\left( \overline{D}_{%
\overline{j}}\overline{Z}\right) D_{i}M_{\Lambda }.
\end{array}
\notag \\
&&  \label{SKG-q-fund}
\end{eqnarray}
Thus, the identities (\ref{SKG-p-fund}) and (\ref{SKG-q-fund}) may be recast
as the following fundamental $\left( 2n_{V}+2\right) \times 1$ vector
identity, defining the geometric structure of SK manifolds \cite
{FBC,K1,K2,BFM,AoB-book,K3}:
\begin{equation}
Q^{T}-i\epsilon \mathcal{M}\left( \mathcal{N}\right) Q^{T}=-2i\overline{V}%
Z-2ig^{i\overline{j}}\left( \overline{D}_{\overline{j}}\overline{Z}\right)
D_{i}V.  \label{SKG-identities1}
\end{equation}
The $\left( 2n_{V}+2\right) \times \left( 2n_{V}+2\right) $ real symmetric
matrix $\mathcal{M}\left( \mathcal{N}\right) $ is defined as \cite{4,FK1,FK2}
\begin{equation}
\mathcal{M}\left( \mathcal{N}\right) =\mathcal{M}\left( Re\mathcal{N},Im%
\mathcal{N}\right) \equiv \left(
\begin{array}{ccc}
Im\mathcal{N}+\left( Re\mathcal{N}\right) \left( Im\mathcal{N}\right) ^{-1}Re%
\mathcal{N} &  & -\left( Re\mathcal{N}\right) \left( Im\mathcal{N}\right)
^{-1} \\
-\left( Im\mathcal{N}\right) ^{-1}Re\mathcal{N} &  & \left( Im\mathcal{N}%
\right) ^{-1}
\end{array}
\right) ,  \label{M}
\end{equation}
where $\mathcal{N}_{\Lambda \Sigma }$ is defined by Eq. (\ref{N}). It is
worth recalling that $\mathcal{M}\left( \mathcal{N}\right) $ is symplectic
with respect to the symplectic metric $\epsilon $, \textit{i.e.} it
satisfies ($\left( \mathcal{M}\left( \mathcal{N}\right) \right) ^{T}=%
\mathcal{M}\left( \mathcal{N}\right) $)
\begin{equation}
\mathcal{M}\left( \mathcal{N}\right) \epsilon \mathcal{M}\left( \mathcal{N}%
\right) =\epsilon .
\end{equation}

By using Eqs. (\ref{norm-PI}), (\ref{ortho-rels}), (\ref{ortho1}) and (\ref
{ortho2}), the identity (\ref{SKG-identities1}) implies the following
relations:
\begin{equation}
\left\{
\begin{array}{l}
\left\langle V,Q^{T}-i\epsilon \mathcal{M}\left( \mathcal{N}\right)
Q^{T}\right\rangle =-2Z; \\
\\
\left\langle \overline{V},Q^{T}-i\epsilon \mathcal{M}\left( \mathcal{N}%
\right) Q^{T}\right\rangle =0; \\
\\
\left\langle D_{i}V,Q^{T}-i\epsilon \mathcal{M}\left( \mathcal{N}\right)
Q^{T}\right\rangle =0; \\
\\
\left\langle \overline{D}_{\overline{i}}\overline{V},Q^{T}-i\epsilon
\mathcal{M}\left( \mathcal{N}\right) Q^{T}\right\rangle =-2\overline{D}_{%
\overline{i}}\overline{Z}.
\end{array}
\right.  \label{SKG-SKG-yielded}
\end{equation}

There are only $2n_{V}$ independent real relations out of the $4n_{V}+4$
real ones yielded by the $2n_{V}+2$ complex identities (\ref{SKG-identities1}%
). Indeed, by taking the real and imaginary part of the SKG vector identity (%
\ref{SKG-identities1}) one respectively obtains
\begin{eqnarray}
&&
\begin{array}{l}
Q^{T}=-2Re\left[ iZ\overline{V}+ig^{i\overline{j}}\left( \overline{D}_{%
\overline{j}}\overline{Z}\right) D_{i}V\right] =-2Im\left[ \overline{Z}V+g^{i%
\overline{j}}\left( D_{i}Z\right) \overline{D}_{\overline{j}}\overline{V}%
\right] ;
\end{array}
\label{Re-SKG-SKG} \\
&&  \notag \\
&&
\begin{array}{l}
\epsilon \mathcal{M}\left( \mathcal{N}\right) Q^{T}=2Im\left[ iZ\overline{V}%
+ig^{i\overline{j}}\left( \overline{D}_{\overline{j}}\overline{Z}\right)
D_{i}V\right] =2Re\left[ \overline{Z}V+g^{i\overline{j}}\left( D_{i}Z\right)
\overline{D}_{\overline{j}}\overline{V}\right] .
\end{array}
\label{Im-SKG-SKG}
\end{eqnarray}
Consequently, the imaginary and real parts of the SKG vector identity (\ref
{SKG-identities1}) are \textit{linearly dependent} one from the other, being
related by the $\left( 2n_{V}+2\right) \times \left( 2n_{V}+2\right) $ real
matrix
\begin{equation}
\epsilon \mathcal{M}\left( \mathcal{N}\right) =\left(
\begin{array}{ccc}
\left( Im\mathcal{N}\right) ^{-1}Re\mathcal{N} &  & -\left( Im\mathcal{N}%
\right) ^{-1} \\
Im\mathcal{N}+\left( Re\mathcal{N}\right) \left( Im\mathcal{N}\right) ^{-1}Re%
\mathcal{N} &  & -\left( Re\mathcal{N}\right) \left( Im\mathcal{N}\right)
^{-1}
\end{array}
\right) .  \label{epsilon-emme}
\end{equation}
Put another way, Eqs. (\ref{Re-SKG-SKG}) and (\ref{Im-SKG-SKG}) yield
\begin{equation}
Re\left[ Z\overline{V}+g^{i\overline{j}}\left( \overline{D}_{\overline{j}}%
\overline{Z}\right) D_{i}V\right] =\epsilon \mathcal{M}\left( \mathcal{N}%
\right) Im\left[ Z\overline{V}+g^{i\overline{j}}\left( \overline{D}_{%
\overline{j}}\overline{Z}\right) D_{i}V\right] ,  \label{rotation1}
\end{equation}
expressing the fact that the real and imaginary parts of the quantity $Z%
\overline{V}+g^{i\overline{j}}\left( \overline{D}_{\overline{j}}\overline{Z}%
\right) D_{i}V$ are simply related through a finite \textit{symplectic
rotation} given by the matrix $\epsilon \mathcal{M}\left( \mathcal{N}\right)
$ (see Eq. (\ref{S}) further below), whose simplecticity directly follows
from the symplectic nature of $\mathcal{M}\left( \mathcal{N}\right) $. Eq. (%
\ref{rotation1}) reduces the number of independent real relations implied by
the identity (\ref{SKG-identities1}) from $4n_{V}+4$ to $2n_{V}+2$. Two
additional real degrees of freedom are scaled out by the complex rescaling (%
\ref{complex-rescaling}).

This is clearly consistent with the fact that the $2n_{V}+2$\ complex
identities (\ref{SKG-identities1}) express nothing but a \textit{change of
basis} of the BH charge configurations, between the K\"{a}hler-invariant $%
1\times \left( 2n_{V}+2\right) $\ symplectic (magnetic/electric) basis
vector $Q$ defined by Eq. (\ref{Gamma-tilde}) and the complex,
moduli-dependent $1\times \left( n_{V}+1\right) $ \textit{matter charges}
vector (with K\"{a}hler weights $\left( 1,-1\right) $)
\begin{equation}
\mathcal{Z}\left( z,\overline{z}\right) \equiv \left( Z\left( z,\overline{z}%
\right) ,Z_{i}\left( z,\overline{z}\right) \right) _{i=1,...,n_{V}}.
\label{Z-call}
\end{equation}

It should be recalled that the BH charges are conserved due to the overall $%
\left( U(1)\right) ^{n_{V}+1}$ gauge-invariance of the system under
consideration, and $Q$ and $\mathcal{Z}\left( z,\overline{z}\right) $ are
two \textit{equivalent} basis for them. Their very equivalence relations are
given by the SKG identities (\ref{SKG-identities1}) themselves. By its very
definition (\ref{Gamma-tilde}), $Q$\ is \textit{moduli-independent} (at
least in a stationary, spherically symmetric and asymptotically flat
extremal BH background, as it is the case being treated here), whereas $Z$
is \textit{moduli-dependent}, since it refers to the eigenstates of the $%
\mathcal{N}=2$, $d=4$ supergravity multiplet and of the $n_{V}$\ Maxwell
vector supermultiplets. \setcounter{equation}0
\def\theequation{3.3.\arabic{subsubsection}.\arabic{equation}}
\subsubsection{\label{BH-New-Attractor}\textit{``New Attractor'' Approach}}

The evaluation of the (real part of the) fundamental SK geometrical
identities (\ref{pre-SKG-final1}) and (\ref{SKG-final1}) along the
constraints determining the various classes of critical points of $V_{BH}$
in $\mathcal{M}_{n_{V}}$ allows one to obtain a completely equivalent form
of the AEs for extremal (static, spherically symmetric, asymptotically flat)
BHs in $\mathcal{N}=2$, $d=4$ ungauged supergravity, which may be simpler in
some cases (see also \cite{BFMY} for the treatment of an explicit case).

I) \textit{Supersymmetric (}$\frac{1}{2}$\textit{-BPS) critical points}. By
evaluating the identities (\ref{pre-SKG-final1}) and (\ref{SKG-final1})
along the constraints (\ref{BPS-conds}), one obtains

\begin{equation}
\left(
\begin{array}{c}
p^{\Lambda } \\
\\
q_{\Lambda }
\end{array}
\right) =-2Im\left[ e^{K/2}\overline{Z}\left(
\begin{array}{c}
X^{\Lambda } \\
\\
F_{\Lambda }
\end{array}
\right) \right] _{\frac{1}{2}-BPS},  \label{alg-BPS-1}
\end{equation}
or in compact form
\begin{equation}
Q^{T}=-2Im\left[ e^{K/2}\overline{Z}\Pi \right] _{\frac{1}{2}-BPS}.  \notag
\label{alg-BPS-2}
\end{equation}

Eqs. (\ref{alg-BPS-1}) and (\ref{alg-BPS-2}) are equivalent, purely
algebraic forms of the $\frac{1}{2}$-BPS extremal BH AEs, given by the
(partly differential) conditions (\ref{BPS-conds})\textit{. }By inserting as
input the BH charge configuration $Q\equiv \left( p^{\Lambda },q_{\Lambda
}\right) $ and the covariantly holomorphic sections $L^{\Lambda }$ and $%
M_{\Lambda }$ of the $U(1)$-bundle over $\mathcal{M}_{n_{V}}$, Eqs. (\ref
{alg-BPS-1}) and (\ref{alg-BPS-2}) give as output (if any) the purely
charge-dependent $\frac{1}{2}$-BPS critical points $\left( z_{\frac{1}{2}%
-BPS}^{i}\left( p,q\right) ,\overline{z}_{\frac{1}{2}-BPS}^{\overline{i}%
}\left( p,q\right) \right) $ of $V_{BH}$.

By looking at Eqs. (\ref{alg-BPS-1}) and (\ref{alg-BPS-2}), it is easy to
realize that $\frac{1}{2}$-BPS critical points of $V_{BH}$ with $Z=0$ (which
are \textit{degenerate}, yielding $V_{BH,\frac{1}{2}-BPS}=0$) correspond to
the trivial case of \textit{all} vanishing magnetic and electric BH charges.
This means that (static, spherically symmetric, asymptotically flat)
extremal BHs with $\frac{1}{2}$-BPS attractor horizon scalar configurations
with $Z=0$ (\textit{i.e.} with no central extension of the $\mathcal{N}=2$, $%
d=4$ horizon supersymmetry algebra) cannot be described by the classical
extremal BH Attractor Mechanism encoded by Eqs. (\ref{alg-BPS-1}) and (\ref
{alg-BPS-2}). They are a particular case of the so-called \textit{''small''}
extremal BHs, which are \textit{classically degenerate}, acquiring a
non-vanishing, finite horizon area and entropy only taking into account
quantum/higher-derivative corrections.

It is worth pointing out that Eqs. (\ref{alg-BPS-1}) and (\ref{alg-BPS-2})
are purely algebraic ones, whereas Eqs. (\ref{BPS-conds}) are (partly)
differential, thus, in general, more complicated to be solved. Consequently,
at least in the $\frac{1}{2}$-BPS case, the \textit{``new attractor''
approach} is simpler of the \textit{``criticality conditions'' approach} to
the search of critical points of $V_{BH}$.

II) \textit{Non-BPS }$Z\neq 0$\textit{\ critical points}. By evaluating the
identities (\ref{pre-SKG-final1}) and (\ref{SKG-final1}) along the
constraints (\ref{non-BPS-Z<>0-conds}) and (\ref{non-BPS-Z<>0-fund}), one
obtains
\begin{eqnarray}
&&  \notag \\
&&
\begin{array}{l}
\left(
\begin{array}{c}
p^{\Lambda } \\
\\
q_{\Lambda }
\end{array}
\right) = \\
\\
=2Im\left\{ e^{K/2}\left[ Z\left(
\begin{array}{c}
\overline{X}^{\Lambda } \\
\\
\overline{F}_{\Lambda }
\end{array}
\right) +\frac{i}{2}\frac{\overline{Z}}{\left| Z\right| ^{2}}\overline{C}_{%
\overline{i}\overline{j}\overline{k}}g^{i\overline{i}}g^{j\overline{j}}g^{k%
\overline{k}}\left( D_{j}Z\right) \left( D_{k}Z\right) \left(
\begin{array}{c}
D_{i}X^{\Lambda } \\
\\
D_{i}F_{\Lambda }
\end{array}
\right) \right] \right\} _{non-BPS,Z\neq 0},
\end{array}
\notag \\
&&  \label{alg-non-BPS-Z<>0-1}
\end{eqnarray}
or in compact form
\begin{eqnarray}
&&Q^{T}=2Im\left\{ e^{K/2}\left[ Z\overline{\Pi }+\frac{i}{2}\frac{\overline{%
Z}}{\left| Z\right| ^{2}}\overline{C}_{\overline{i}\overline{j}\overline{k}%
}g^{i\overline{i}}g^{j\overline{j}}g^{k\overline{k}}\left( D_{j}Z\right)
\left( D_{k}Z\right) D_{i}\Pi \right] \right\} _{non-BPS,Z\neq 0}.  \notag \\
&&  \label{alg-non-BPS-Z<>0-2}
\end{eqnarray}

Eqs. (\ref{alg-non-BPS-Z<>0-1}) and (\ref{alg-non-BPS-Z<>0-2}) are
equivalent forms of the non-BPS $Z\neq 0$ extremal BH AEs, given by the
(partly differential) conditions (\ref{non-BPS-Z<>0-conds}) and (\ref
{non-BPS-Z<>0-fund})\textit{. }By inserting as input the BH charge
configuration $Q$, the covariantly holomorphic sections $L^{\Lambda }$ and $%
M_{\Lambda }$, the K\"{a}hler potential $K$ (and consequently the
contravariant metric tensor $g^{i\overline{j}}$) and the completely
symmetric, covariantly holomorphic rank-3 tensor $C_{ijk}$, Eqs. (\ref
{alg-non-BPS-Z<>0-1}) and (\ref{alg-non-BPS-Z<>0-2}) give as output (if any)
the purely charge-dependent non-BPS $Z\neq 0$ critical points $\left(
z_{non-BPS,Z\neq 0}^{i}\left( p,q\right) ,\overline{z}_{non-BPS,Z\neq 0}^{%
\overline{i}}\left( p,q\right) \right) $ of $V_{BH}$. Notice that,
differently from Eqs. (\ref{alg-BPS-1}) and (\ref{alg-BPS-2}), Eqs. (\ref
{alg-non-BPS-Z<>0-1}) and (\ref{alg-non-BPS-Z<>0-2}) are not purely
algebraic. Thus, in the non-BPS $Z\neq 0$ case the (computational)
simplification in the search of critical points of $V_{BH}$ obtained by
exploiting the \textit{``new attractor'' approach} rather than the \textit{%
``criticality conditions'' approach} is model-dependent.

It is interesting to point out that, as it is evident by looking for
instance at Eq. (\ref{alg-non-BPS-Z<>0-2}), at the non-BPS $Z\neq 0$
critical points of $V_{BH}$ the coefficients of $\overline{\Pi }$ and $%
D_{i}\Pi $ in the AEs have the same holomorphicity in the central charge $Z$%
, \textit{i.e.} they can be expressed only in terms of $Z$ and $D_{i}Z$,
without using $\overline{Z}$ and $\overline{D}_{\overline{i}}\overline{Z}$.
Such a fact does not happen in a generic point of $\mathcal{M}_{n_{V}}$, as
it is seen from the global identity (\ref{SKG-final1}). As it is evident,
the price to be paid in order to obtain the same holomorphicity in $Z$ at
the non-BPS $Z\neq 0$ critical points of $V_{BH}$ is the fact that the
coefficient of $D_{i}\Pi $ is not linear in some covariant derivative of $Z$
any more, also explicitly depending on the rank-3 covariantly
antiholomorphic tensor $\overline{C}_{\overline{i}\overline{j}\overline{k}}$.

III) \textit{Non-BPS }$Z=0$\textit{\ critical points}. By evaluating the
identities (\ref{pre-SKG-final1}) and (\ref{SKG-final1}) along the
constraints (\ref{non-BPS-Z=0-conds}) and (\ref{non-BPS-Z=0-fund}), one
obtains
\begin{eqnarray}
\left(
\begin{array}{c}
p^{\Lambda } \\
\\
q_{\Lambda }
\end{array}
\right) &=&2Im\left\{ e^{K/2}\left[ g^{i\overline{j}}\left( \overline{%
\partial }_{\overline{j}}\overline{Z}\right) \left(
\begin{array}{c}
D_{i}X^{\Lambda } \\
\\
D_{i}F_{\Lambda }
\end{array}
\right) \right] \right\} _{non-BPS,Z=0},  \label{alg-non-BPS-Z=0-1} \\
&&  \notag
\end{eqnarray}
or in compact form
\begin{equation}
Q^{T}=2Im\left\{ e^{K/2}\left[ g^{i\overline{j}}\left( \overline{\partial }_{%
\overline{j}}\overline{Z}\right) D_{i}\Pi \right] \right\} _{non-BPS,Z=0}.
\label{alg-non-BPS-Z=0-2}
\end{equation}

Eqs. (\ref{alg-non-BPS-Z=0-1}) and (\ref{alg-non-BPS-Z=0-2}) are equivalent
forms of the non-BPS $Z=0$ extremal BH AEs, given by the (partly
differential) conditions (\ref{non-BPS-Z=0-conds}) and (\ref
{non-BPS-Z=0-fund})\textit{. }By inserting as input the BH charge
configuration $Q$, the covariantly holomorphic sections $L^{\Lambda }$ and $%
M_{\Lambda }$, and the K\"{a}hler potential $K$ (and consequently the
contravariant metric tensor $g^{i\overline{j}}$), Eqs. (\ref
{alg-non-BPS-Z=0-1}) and (\ref{alg-non-BPS-Z=0-2}) give as output (if any)
the purely charge-dependent non-BPS $Z=0$ critical points $\left(
z_{non-BPS,Z=0}^{i}\left( p,q\right) ,\overline{z}_{non-BPS,Z=0}^{\overline{i%
}}\left( p,q\right) \right) $ of $V_{BH}$. Differently from Eqs. (\ref
{alg-BPS-1}) and (\ref{alg-BPS-2}), and similarly to Eqs. (\ref
{alg-non-BPS-Z<>0-1}) and (\ref{alg-non-BPS-Z<>0-2}), Eqs. (\ref
{alg-non-BPS-Z=0-1}) and (\ref{alg-non-BPS-Z=0-2}) are not purely algebraic.
Thus, in the non-BPS $Z=0$ case the (computational) simplification in the
search of critical points of $V_{BH}$ obtained by exploiting the \textit{%
``new attractor'' approach} rather than the \textit{``criticality
conditions'' approach} is model-dependent. \setcounter{equation}0
\def\theequation{3.\arabic{subsection}.\arabic{equation}}
\subsection{\label{Micro-BH}Type IIB Superstrings on $CY_{3}$}

\setcounter{equation}0
\def\theequation{3.4.\arabic{subsubsection}.\arabic{equation}}
\subsubsection{\label{Hodge-decomposition}Hodge Decomposition of $\mathcal{H}%
_{3}$}

We consider Type IIB superstring theory compactified on a Calabi-Yau
threefold ($CY_{3}$) \cite{Ferrara-Strominger, Candelas,4,Moore},
determining an effective $\mathcal{N}=2$, $d=4$ ungauged supergravity with a
number $n_{V}$ of Abelian vector multiplets. Within such a framework, the $%
CY_{3}$ has a complex structure (CS) moduli space (of complex dimension $%
n_{V}=h_{2,1}\equiv dim\left( H^{2,1}\left( CY_{3}\right) \right) $, where $%
H^{2,1}$ is the $\left( 2,1\right) $-cohomology group of the considered
manifold), which is a special K\"{a}hler (SK) manifold.

We introduce a\footnote{$b_{3}=2h_{2,1}+2$ is the so-called third Betti
number of the $CY_{3}$.} $b_{3}$-dim. real (manifestly symplectic-covariant)
basis of the third real\footnote{%
In the strict quantum regime, one should consider the third \textit{integer}
cohomology $H^{3}\left( CY_{3},\mathbb{Z}\right) $. The present
(semi)classical treatment deal with the \textit{large charges limit} and
thus consistently consider real, unquantized, rather than integer, quantized
quantities.} cohomology $H^{3}\left( CY_{3},\mathbb{R}\right) $, given by
the set of real 3-forms $\left\{ \alpha _{\Lambda },\beta ^{\Lambda
}\right\} $ ($\Lambda =0,1,...,h_{2,1}$ throughout) satisfying\footnote{%
Recall that the $\wedge $ (''wedge'') product among odd-forms is odd,
whereas the one among even-forms (and among odd- and even-forms) is even.}
\begin{equation}
\int_{CY_{3}}\alpha _{\Lambda }\wedge \alpha _{\Sigma
}=0,~~\int_{CY_{3}}\beta ^{\Lambda }\wedge \beta ^{\Sigma
}=0,~~\int_{CY_{3}}\alpha _{\Lambda }\wedge \beta ^{\Sigma }=\delta
_{\Lambda }^{\Sigma }.  \label{22apr1}
\end{equation}
By Poincar\`{e}-duality on $CY_{3}$, we may correspondingly introduce the $%
b_{3}$-dim. real (manifestly symplectic-covariant) basis of the third real
homology $H_{3}\left( CY_{3},\mathbb{R}\right) $, given by the set of real
3-cycles $\left\{ A^{\Lambda },B_{\Lambda }\right\} $ satisfying
\begin{equation}
\int_{A^{\Lambda }}\alpha _{\Sigma }=\delta _{\Sigma }^{\Lambda
},~~\int_{A^{\Lambda }}\beta ^{\Sigma }=0,~~\int_{B_{\Lambda }}\alpha
_{\Sigma }=0,~~\int_{B_{\Lambda }}\beta ^{\Sigma }=-\delta _{\Lambda
}^{\Sigma }.
\end{equation}

The $CY_{3}$ is endowed with a (\textit{nowhere-vanishing}) holomorphic
3-form
\begin{equation}
\Omega _{3}\left( z\right) \equiv X^{\Lambda }\left( z\right) \alpha
_{\Lambda }-F_{\Lambda }\left( z\right) \beta ^{\Lambda }\in H^{3,0}\left(
CY_{3}\right) ,  \label{Omega3}
\end{equation}
where ''$z$'' denotes the functional dependence on the CS moduli $\left\{
z^{i},\overline{z}^{\overline{i}}\right\} $ ($i=1,...,h_{2,1}$ throughout),
and $\left\{ X^{\Lambda },F_{\Lambda }\right\} $ stands for the basis of
symplectic holomorphic fundamental periods of $\Omega _{3}$ around the
3-cycles $\left\{ A^{\Lambda },B_{\Lambda }\right\} $, respectively:
\begin{equation}
X^{\Lambda }\left( z\right) \equiv \int_{A^{\Lambda }}\Omega _{3}\left(
z\right) ,~~F_{\Lambda }\left( z\right) \equiv \int_{B_{\Lambda }}\Omega
_{3}\left( z\right) .
\end{equation}
$\Omega _{3}$, as well as its fundamental periods, has K\"{a}hler weights $%
\left( 2,0\right) $:
\begin{equation}
\begin{array}{l}
D_{i}\Omega _{3}=\partial _{i}\Omega _{3}+\left( \partial _{i}K\right)
\Omega _{3}, \\
\\
\overline{D}_{\overline{i}}\Omega _{3}=\overline{\partial }_{\overline{i}%
}\Omega _{3}=0,
\end{array}
\end{equation}
where $K$ is the real K\"{a}hler potential in the $h_{2,1}$-dim. SK CS
moduli space of $CY_{3}$.

Type IIB compactified on $CY_{3}$ is characterized by a real 5-form
\begin{equation}
\mathcal{Z}\equiv \mathcal{F}^{\Lambda }\alpha _{\Lambda }-\mathcal{G}%
_{\Lambda }\beta ^{\Lambda },
\end{equation}
where $\mathcal{F}^{\Lambda }$ is the space-time 2-form given by the Abelian
field-strengths ($\Lambda =0$ pertains to the graviphoton, whereas $\Lambda
=i$ corresponds to the Maxwell vector supermultiplets), and $\mathcal{G}%
_{\Lambda }$ is the corresponding ''dual'' space-time 2-form, in the sense
of Legendre transform:
\begin{equation}
\mathcal{G}_{\Lambda }\equiv \frac{\delta \mathcal{L}}{\delta \mathcal{F}%
^{\Lambda }}=\left( ReN_{\Lambda \Sigma }\right) \mathcal{F}^{\Sigma }+\frac{%
1}{2}\left( ImN_{\Lambda \Sigma }\right) ^{\ast }\mathcal{F}^{\Sigma }.
\end{equation}
$^{\ast }\mathcal{F}^{\Sigma }$ denotes the Hodge $\ast $-dual of $\mathcal{F%
}^{\Lambda }$, defined in components as follows (the space-time indices $\mu
,\nu $ run $0,1,2,3$ throughout):
\begin{equation}
^{\ast }\mathcal{F}_{\mu \nu }^{\Lambda }\equiv \frac{1}{2}\epsilon _{\mu
\nu \rho \sigma }\mathcal{F}^{\Lambda \mid \rho \sigma }=\frac{1}{2}G^{\rho
\lambda }G^{\sigma \tau }\epsilon _{\mu \nu \rho \sigma }\mathcal{F}%
_{\lambda \tau }^{\Lambda },
\end{equation}
where $\epsilon _{\mu \nu \rho \sigma }$ is the $d=4$ completely
antisymmetric Ricci-Levi-Civita tensor, and $G^{\mu \nu }$ is the $d=4$
space-time completely contravariant metric tensor. $\mathcal{L}$ stands for
the (bosonic sector of the) $\mathcal{N}=2$, $d=4$ ungauged supergravity
Lagrangian density:
\begin{eqnarray}
\mathcal{L} &=&-\frac{R}{2}+g_{i\overline{j}}\left( \partial _{\mu
}z^{i}\right) \left( \partial _{\nu }\overline{z}^{\overline{j}}\right)
G^{\mu \nu }+\frac{1}{4}\left( ImN_{\Lambda \Sigma }\right) G^{\mu \lambda
}G^{\nu \rho }\mathcal{F}_{\mu \nu }^{\Lambda }\mathcal{F}_{\lambda \rho
}^{\Lambda }+\frac{1}{4}\left( ReN_{\Lambda \Sigma }\right) G^{\mu \lambda
}G^{\nu \rho }\mathcal{F}_{\mu \nu }^{\Lambda }{}^{\ast }\mathcal{F}%
_{\lambda \rho }^{\Lambda }.  \notag \\
&&  \label{L}
\end{eqnarray}

The Hodge $\ast $-duality acts as a symplectic $Sp\left( 2h_{1,2}+2,\mathbb{R%
}\right) $ rotation on the basis $\left\{ \alpha _{\Lambda },\beta ^{\Lambda
}\right\} $:
\begin{equation}
\left(
\begin{array}{c}
^{\ast }\alpha _{\Lambda } \\
^{\ast }\beta ^{\Lambda }
\end{array}
\right) =\mathcal{S}\left(
\begin{array}{c}
\alpha _{\Lambda } \\
\beta ^{\Lambda }
\end{array}
\right) ,\text{~~}\mathcal{S}\equiv -\epsilon \mathcal{M}\left( \mathcal{N}%
\right) ,~~\mathcal{S}^{T}\epsilon \mathcal{S}=\epsilon .  \label{S}
\end{equation}
where $\epsilon $ is the $\left( 2h_{1,2}+2\right) $-dim. symplectic metric
defined in Eq. (\ref{Omega})\textbf{, }$\mathcal{M}\left( \mathcal{N}\right)
$ is the real, symplectic matrix defined by Eq. (\ref{M}). Notice that $%
\mathcal{S}$ is nothing but the opposite of the matrix given by Eq. (\ref
{epsilon-emme}). It can be shown that $\mathcal{Z}$ is Hodge $\ast $%
-self-dual:
\begin{equation}
^{\ast }\mathcal{Z}=\mathcal{Z}.  \label{Z*}
\end{equation}

Whenever the relevant integrations over internal manifold $CY_{3}$ and over
space-time make sense, manifestly symplectic-covariant magnetic and electric
charges can be introduced as the asymptotical ''space-dressings'' of a
suitable contraction of $\mathcal{Z}$ over the symplectic 3-cycles of $%
CY_{3} $, \textit{i.e.} as the asymptotical fluxes of the space-time 2-forms
corresponding to the components of $\mathcal{Z}$ along the symplectic basis $%
\left\{ A^{\Lambda },B_{\Lambda }\right\} $ of $H_{3}\left( CY_{3},\mathbb{R}%
\right) $, respectively:
\begin{eqnarray}
&&
\begin{array}{c}
p^{\Lambda }\equiv \frac{1}{4\pi }\int_{A^{\Lambda }\times S_{\infty }^{2}}%
\mathcal{Z}=\frac{1}{4\pi }\int_{A^{\Lambda }\times S_{\infty }^{2}}\left(
\mathcal{F}^{\Sigma }\alpha _{\Sigma }-\mathcal{G}_{\Sigma }\beta ^{\Sigma
}\right) =\frac{1}{4\pi }\int_{S_{\infty }^{2}}\mathcal{F}^{\Lambda }; \\
\\
q_{\Lambda }\equiv \frac{1}{4\pi }\int_{B_{\Lambda }\times S_{\infty }^{2}}%
\mathcal{Z}=\frac{1}{4\pi }\int_{B_{\Lambda }\times S_{\infty }^{2}}\left(
\mathcal{F}^{\Sigma }\alpha _{\Sigma }-\mathcal{G}_{\Sigma }\beta ^{\Sigma
}\right) =\frac{1}{4\pi }\int_{S_{\infty }^{2}}\mathcal{G}^{\Lambda },
\end{array}
\notag \\
&&
\end{eqnarray}
where $S_{\infty }^{2}$ denotes the 2-sphere at spatial infinity\footnote{%
Consistently with (a proper subset of) the solutions of $\mathcal{N}=2$, $%
d=4 $ ungauged supergravity, the space-time metric is assumed to be static,
spherically symmetric, and asymptotically flat. In such a framework,
''spatial infinity'' corresponds to $r\rightarrow \infty $, where $r$ is the
radial coordinate.}.

$\left\{ p^{\Lambda },q_{\Lambda }\right\} $ can be seen as the components
(along the real symplectic basis $\left\{ \alpha _{\Lambda },\beta ^{\Lambda
}\right\} $ of $H^{3}\left( CY_{3},\mathbb{R}\right) $) of the real flux
3-form $\mathcal{H}_{3}$, defined as the asymptotical ''space-dressing'' of $%
\mathcal{Z}$:
\begin{equation}
\mathcal{H}_{3}\equiv \frac{1}{4\pi }\int_{S_{\infty }^{2}}\mathcal{Z}%
=p^{\Lambda }\alpha _{\Lambda }-q_{\Lambda }\beta ^{\Lambda }\in H^{3}\left(
CY_{3},\mathbb{R}\right) .
\end{equation}
$\left\{ p^{\Lambda },q_{\Lambda }\right\} $ are the physical charges, and
they are conserved, due to the overall $\left( U(1)\right) ^{h_{2,1}+1}$
gauge symmetry of the considered framework. They respectively are the
magnetic and electric charges of the $\left( U(1)\right) ^{h_{2,1}+1}$ gauge
group of the (symplectic) real parameterization of $H^{3}\left( CY_{3},%
\mathbb{R}\right) $, which however is not the only possible one.

Indeed, in general the third real cohomology $H^{3}\left( CY_{3},\mathbb{R}%
\right) $ can be Hodge-decomposed along the third Dalbeault cohomogy of $%
CY_{3}$ as follows:
\begin{equation}
H^{3}\left( CY_{3},\mathbb{R}\right) =H^{3,0}\left( CY_{3}\right) \oplus
_{s}H^{2,1}\left( CY_{3}\right) \oplus _{s}H^{1,2}\left( CY_{3}\right)
\oplus _{s}H^{0,3}\left( CY_{3}\right) ,  \label{decomp1}
\end{equation}
corresponding to perform a change of basis from the symplectic real basis to
the Dalbeault basis:
\begin{equation}
\left\{ \alpha _{\Lambda },\beta ^{\Lambda }\right\} \longrightarrow \left\{
\Omega _{3},D_{i}\Omega _{3},\overline{D}_{\overline{i}}\overline{\Omega }%
_{3},\overline{\Omega }_{3}\right\} .  \label{change1}
\end{equation}
The subscript ''$s$'' in in Eq. (\ref{decomp1}) stands for the semidirect
cohomological sum, due to the fact that (some of the) cohomologies in the
r.h.s. of the Hodge decomposition (\ref{decomp1}) have non-vanishing
intersections. Indeed, as it can be checked by recalling Eqs. (\ref
{ortho-rels}), (\ref{ortho1}) and (\ref{22apr1}), the following results
hold:
\begin{eqnarray}
&&
\begin{array}{l}
\int_{CY_{3}}\Omega _{3}\wedge \Omega _{3}=0,~~\int_{CY_{3}}\Omega
_{3}\wedge D_{i}\Omega _{3}=0,~~\int_{CY_{3}}\Omega _{3}\wedge \overline{D}_{%
\overline{i}}\overline{\Omega }_{3}=0, \\
\\
\int_{CY_{3}}\Omega _{3}\wedge \overline{\Omega }_{3}=-ie^{-K}%
\Leftrightarrow K=-ln\left( i\int_{CY_{3}}\Omega _{3}\wedge \overline{\Omega
}_{3}\right) , \\
\\
\int_{CY_{3}}\left( D_{i}\Omega _{3}\right) \wedge D_{j}\Omega _{3}=0, \\
\\
\\
\begin{array}{c}
\int_{CY_{3}}\left( D_{i}\Omega _{3}\right) \wedge \overline{D}_{\overline{j}%
}\overline{\Omega }_{3}=\left[ \overline{\partial }_{\overline{j}}\partial
_{i}ln\left( i\int_{CY_{3}}\Omega _{3}\wedge \overline{\Omega }_{3}\right)
\right] \int_{CY_{3}}\Omega _{3}\wedge \overline{\Omega }_{3}=-\overline{%
\partial }_{\overline{j}}\partial _{i}K\int_{CY_{3}}\Omega _{3}\wedge
\overline{\Omega }_{3}=ie^{-K}g_{i\overline{j}} \\
\Updownarrow \\
g_{i\overline{j}}=-\overline{\partial }_{\overline{j}}\partial _{i}ln\left(
i\int_{CY_{3}}\Omega _{3}\wedge \overline{\Omega }_{3}\right) =-\frac{%
\int_{CY_{3}}\left( D_{i}\Omega _{3}\right) \wedge \overline{D}_{\overline{j}%
}\overline{\Omega }_{3}}{\int_{CY_{3}}\Omega _{3}\wedge \overline{\Omega }%
_{3}}.
\end{array}
\end{array}
\notag \\
&&  \label{N=2-intersections}
\end{eqnarray}
In particular, the second line of Eq. (\ref{N=2-intersections}) allows one
to write the covariant derivatives of $\Omega _{3}$ (which are the basis of $%
H^{2,1}\left( CY_{3}\right) $) as follows:
\begin{equation}
D_{i}\Omega _{3}=\left( \partial _{i}-\frac{\int_{CY_{3}}\left( \partial
_{i}\Omega _{3}\right) \wedge \overline{\Omega }_{3}}{\int_{CY_{3}}\Omega
_{3}\wedge \overline{\Omega }_{3}}\right) \Omega _{3}.
\end{equation}
It is worth pointing out that the $2h_{2,1}+2$ 3-forms $\left\{ \Omega
_{3},D_{i}\Omega _{3},\overline{D}_{\overline{i}}\overline{\Omega }_{3},%
\overline{\Omega }_{3}\right\} _{i=1,...,h_{2,1}}$ are \textit{all} the
possible ($\left( 2,0\right) $ and $\left( 0,2\right) $)-K\"{a}hler-weighted
independent 3-forms which can be defined on $CY_{3}$ in the considered
framework. This is due to the two fundamental relations
\begin{eqnarray}
\overline{D}_{\overline{j}}D_{i}\Omega _{3} &=&g_{i\overline{j}}\Omega _{3};
\label{1} \\
D_{i}D_{j}\Omega _{3} &=&iC_{ijk}g^{k\overline{l}}\overline{D}_{\overline{l}}%
\overline{\Omega }_{3}=D_{(i}D_{j)}\Omega _{3},  \label{2}
\end{eqnarray}
which are the translation, in the language of forms on $CY_{3}$, of the
third and second of Eqs. (\ref{SKG-rels1}), respectively. Notice that the
third of Eqs. (\ref{SKG-rels1}) and Eq. (\ref{1}) hold in a generic
K\"{a}hler framework, whereas the second of Eqs. (\ref{SKG-rels1}) and Eq. (%
\ref{2}) in general hold only in SK geometry. Due to Eq. (\ref{2}), the
completely symmetric, covariantly holomorphic tensor $C_{ijk}$ of SK
geometry can be obtained by intersecting the elements of the basis of $%
H^{2,1}\left( CY_{3}\right) $ with their covariant derivatives (and
normalizing with respect to the intersection of $H^{3,0}\left( CY_{3}\right)
$ and $H^{0,3}\left( CY_{3}\right) $):
\begin{equation}
C_{ijk}=C_{\left( ijk\right) }=-i\frac{\int_{CY_{3}}\left( D_{i}D_{j}\Omega
_{3}\right) \wedge D_{k}\Omega _{3}}{\int_{CY_{3}}\Omega _{3}\wedge
\overline{\Omega }_{3}}=e^{K}\int_{CY_{3}}\left( D_{i}D_{j}\Omega
_{3}\right) \wedge D_{k}\Omega _{3}.  \label{CC}
\end{equation}

According to the Hodge-decomposition (\ref{decomp1}) implemented through the
change of basis (\ref{change1}), the charges undergo the following change of
basis:
\begin{equation}
\left\{ p^{\Lambda },q_{\Lambda }\right\} \longrightarrow \left\{
Z^{3,0}\left( z,p,q\right) ,Z_{i}^{2,1}\left( z,\overline{z},p,q\right) ,Z_{%
\overline{i}}^{1,2}\left( z,\overline{z},p,q\right) ,Z^{0,3}\left( \overline{%
z},p,q\right) \right\} ,
\end{equation}
where the complex, (CS) moduli-dependent quantities on the r.h.s. are
defined as follows:
\begin{eqnarray}
&&
\begin{array}{l}
Z^{3,0}\left( z;p,q\right) \equiv \int_{CY_{3}}\mathcal{H}_{3}\wedge \Omega
_{3}\left( z\right) =\frac{1}{4\pi }\int_{CY_{3}}\left( \int_{S_{\infty
}^{2}}\mathcal{Z}\right) \wedge \Omega _{3}\left( z\right) =X^{\Lambda
}\left( z\right) q_{\Lambda }-F_{\Lambda }\left( z\right) p^{\Lambda
}=W\left( z;p,q\right) ;
\end{array}
\notag \\
&&  \label{Z3,0} \\
&&
\begin{array}{l}
Z_{i}^{2,1}\left( z,\overline{z};p,q\right) \equiv \int_{CY_{3}}\mathcal{H}%
_{3}\wedge \left( D_{i}\Omega _{3}\right) \left( z,\overline{z}\right)
=\int_{CY_{3}}\left( \int_{S_{\infty }^{2}}\mathcal{Z}\right) \wedge \left(
D_{i}\Omega _{3}\right) \left( z,\overline{z}\right) = \\
\\
=\left( D_{i}X^{\Lambda }\right) \left( z,\overline{z}\right) q_{\Lambda
}-\left( D_{i}F_{\Lambda }\right) \left( z,\overline{z}\right) p^{\Lambda
}=\left( D_{i}W\right) \left( z,\overline{z};p,q\right) ;
\end{array}
\notag \\
&&  \label{Z2,1} \\
&&
\begin{array}{l}
Z_{\overline{i}}^{1,2}\left( z,\overline{z};p,q\right) \equiv \overline{%
Z_{i}^{2,1}\left( z,\overline{z};p,q\right) };
\end{array}
\label{Z1,2} \\
&&  \notag \\
&&
\begin{array}{l}
Z^{0,3}\left( \overline{z};p,q\right) \equiv \overline{Z^{3,0}\left(
z;p,q\right) }.
\end{array}
\label{Z0,3}
\end{eqnarray}
As it can be seen, $Z^{3,0}\left( z;p,q\right) $ and $Z_{i}^{2,1}\left( z,%
\overline{z};p,q\right) $ are respectively nothing but the $\mathcal{N}=2$, $%
d=4$ holomorphic central charge function $W\left( z;p,q\right) $, also named
$\mathcal{N}=2$ superpotential (see Eq. (\ref{Z}) and comments below), and
its covariant derivatives, introduced \textit{\`{a} la Gukov-Vafa-Witten}
(GVW) \cite{Gukov:1999ya} also in the considered $\mathcal{N}=2$, $d=4$
framework. In other words, Eq. (\ref{Z3,0}) defines the holomorphic central
extension of the $\mathcal{N}=2$, $d=4$ local supersymmetry algebra, whereas
Eq. (\ref{Z3,0}) defines in a geometrical way the charges of the other field
strength vectors, orthogonal to the graviphoton. $\left\{
Z^{3,0},Z_{i}^{2,1},Z_{\overline{i}}^{1,2},Z^{0,3}\right\} $ correspond to
electric and magnetic charges of the $\left( U(1)\right) ^{h_{2,1}+1}$ gauge
group of the complex Dalbeault parameterization of $H^{3}\left( CY_{3},%
\mathbb{R}\right) $. Their dependence on moduli can be understood by taking
into account that they refer to the supermultiplet eigenstates, which are
moduli-dependent (as already pointed out below Eq. (\ref{rotation1})). They
satisfy the following model-independent sum rules \cite{FK1}:
\begin{eqnarray}
&&
\begin{array}{l}
\left( \left| Z^{3,0}\right| ^{2}+g^{i\overline{j}}Z_{i}^{2,1}Z_{\overline{j}%
}^{1,2}\right) e^{K}=-\frac{1}{2}\left( p^{\Lambda },q_{\Lambda }\right)
\mathcal{M}\left( \mathcal{N}\right) \left(
\begin{array}{c}
p^{\Sigma } \\
q_{\Sigma }
\end{array}
\right) =I_{1}\left( z,\overline{z};p,q\right) =V_{BH}\left( z,\overline{z}%
;p,q\right) \geqslant 0;
\end{array}
\notag \\
&& \\
&&
\begin{array}{l}
\left( \left| Z^{3,0}\right| ^{2}-g^{i\overline{j}}Z_{i}^{2,1}Z_{\overline{j}%
}^{1,2}\right) e^{K}=-\frac{1}{2}\left( p^{\Lambda },q_{\Lambda }\right)
\mathcal{M}\left( \mathcal{F}\right) \left(
\begin{array}{c}
p^{\Sigma } \\
q_{\Sigma }
\end{array}
\right) =I_{2}\left( z,\overline{z};p,q\right) \gtreqless 0,
\end{array}
\end{eqnarray}
where $\mathcal{M}\left( \mathcal{N}\right) $ is the real, symplectic matrix
defined by Eq. (\ref{M}), $\mathcal{F}\equiv F_{\Lambda \Sigma }=\partial
_{\Sigma }F_{\Lambda }$, $\mathcal{M}\left( \mathcal{F}\right) =\mathcal{M}%
\left( \mathcal{N}\rightarrow \mathcal{F}\right) $. $I_{1}$ and $I_{2}$ are
the first and second lowest-order (quadratic in charges) invariants of SK
geometry. As far as the metric $g_{i\overline{j}}$ of the SK CS moduli space
is regular, $I_{1}$ has positive signature and it is nothing but the ``BH
effective potential'' $V_{BH}$, whereas $I_{2}$ has signature $\left(
1,h_{2,1}\right) $. Since the considered extremal BH background is static
(and spherically symmetric), the undressed charges $p^{\Lambda }$ and $%
q_{\Lambda }$ are conserved in time, and so are the dressed charges $\left\{
Z^{3,0},Z_{i}^{2,1},Z_{\overline{i}}^{1,2},Z^{0,3}\right\} $ (which however,
through their dependence on scalars, do depend on radial coordinate).

The real, K\"{a}hler gauge-invariant 3-form $\mathcal{H}_{3}$ can be thus
Hodge-decomposed as follows ($\gamma _{1},\gamma _{2},\gamma _{3},\gamma
_{4}\in \mathbb{C}$)
\begin{eqnarray}
&&
\begin{array}{l}
\mathcal{H}_{3}=e^{K}\left[
\begin{array}{l}
\gamma _{1}\left( \int_{CY_{3}}\mathcal{H}_{3}\wedge \Omega _{3}\right)
\overline{\Omega }_{3}+\gamma _{2}g^{i\overline{j}}\left( \int_{CY_{3}}%
\mathcal{H}_{3}\wedge \left( D_{i}\Omega _{3}\right) \right) \overline{D}_{%
\overline{j}}\overline{\Omega }_{3}+ \\
\\
+\gamma _{3}g^{j\overline{i}}\left( \int_{CY_{3}}\mathcal{H}_{3}\wedge
\overline{D}_{\overline{i}}\overline{\Omega }_{3}\right) D_{j}\Omega
_{3}+\gamma _{4}\left( \int_{CY_{3}}\mathcal{H}_{3}\wedge \overline{\Omega }%
_{3}\right) \Omega _{3}
\end{array}
\right] = \\
\\
\\
=e^{K}\left[
\begin{array}{l}
\gamma _{1}W\overline{\Omega }_{3}+\gamma _{2}g^{i\overline{j}}\left(
D_{i}W\right) \overline{D}_{\overline{j}}\overline{\Omega }_{3}+ \\
\\
+\gamma _{3}g^{j\overline{i}}\left( \overline{D}_{\overline{i}}\overline{W}%
\right) D_{j}\Omega _{3}+\gamma _{4}\overline{W}\Omega _{3}
\end{array}
\right] ,
\end{array}
\notag \\
&&  \label{decomp2}
\end{eqnarray}
\textbf{\ }where Eqs. (\ref{Z3,0})-(\ref{Z0,3}) were used. The r.h.s. of the
Hodge-decomposition (\ref{decomp2}) is the most general K\"{a}hler
gauge-invariant combination of \textit{all} the possible ($\left( 2,0\right)
$ and $\left( 0,2\right) $)-K\"{a}hler-weighted independent 3-forms for Type
IIB on $CY_{3}$s. The overall factor $e^{K}$ (with K\"{a}hler weights $%
\left( -2,-2\right) $) is necessary to make the r.h.s. of the identity (\ref
{decomp2}) K\"{a}hler gauge-invariant. The reality condition $\overline{%
\mathcal{H}}_{3}=$ $\mathcal{H}_{3}$ implies $\gamma _{3}=\overline{\gamma
_{2}}$ and $\gamma _{4}=\overline{\gamma _{1}}$. The complex coefficients $%
\gamma _{1}$ and $\gamma _{2}$ can be determined by computing $\int_{CY_{3}}%
\mathcal{H}_{3}\wedge \Omega _{3}$ and $\int_{CY_{3}}\mathcal{H}_{3}\wedge
D_{l}\Omega _{3}$, using the identity (\ref{decomp2}) and recalling Eqs. (%
\ref{Z3,0})-(\ref{Z0,3}) and the intersections (\ref{N=2-intersections}). By
doing so, one obtains:
\begin{eqnarray}
&&
\begin{array}{l}
W=\int_{CY_{3}}\mathcal{H}_{3}\wedge \Omega _{3}=e^{K}\gamma
_{1}W\int_{CY_{3}}\overline{\Omega }_{3}\wedge \Omega _{3}=i\gamma
_{1}W\Leftrightarrow \gamma _{1}=-i; \\
\\
D_{l}W=\int_{CY_{3}}\mathcal{H}_{3}\wedge D_{l}\Omega _{3}=e^{K}\gamma
_{2}g^{i\overline{j}}\left( D_{i}W\right) \int_{CY_{3}}\overline{D}_{%
\overline{j}}\overline{\Omega }_{3}\wedge D_{l}\Omega _{3}=-i\gamma
_{2}D_{l}W\Leftrightarrow \gamma _{2}=i.
\end{array}
\notag \\
&&
\end{eqnarray}
Thus, the complete Hodge-decomposition of the real flux 3-form $\mathcal{H}%
_{3}$ of Type IIB on $CY_{3}$s reads
\begin{eqnarray}
&&
\begin{array}{l}
\mathcal{H}_{3}=-ie^{K}\left[ W\overline{\Omega }_{3}-g^{i\overline{j}%
}\left( D_{i}W\right) \overline{D}_{\overline{j}}\overline{\Omega }_{3}+g^{j%
\overline{i}}\left( \overline{D}_{\overline{i}}\overline{W}\right)
D_{j}\Omega _{3}-\overline{W}\Omega _{3}\right] = \\
\\
=-2Im\left[ \overline{Z}\hat{\Omega}_{3}-g^{j\overline{i}}\left( \overline{D}%
_{\overline{i}}\overline{Z}\right) D_{j}\hat{\Omega}_{3}\right] ,
\end{array}
\notag \\
&&  \label{decomp3}
\end{eqnarray}
where in the second line we recalled the definition of the $\mathcal{N}=2$, $%
d=4$ covariantly holomorphic central charge function $Z\left( z,\overline{z}%
,p,q\right) $ (with K\"{a}hler weights $\left( 1,-1\right) $) given by Eq. (%
\ref{Z}) (see also Eq. (\ref{DiZ})), and introduced the covariantly
holomorphic $\left( 3,0\right) $-form $\hat{\Omega}_{3}$ (with K\"{a}hler
weights $\left( 1,-1\right) $) on $CY_{3}$:
\begin{eqnarray}
&&
\begin{array}{l}
Z\left( z,\overline{z};p,q\right) \equiv e^{\frac{K\left( z,\overline{z}%
\right) }{2}}W\left( z;p,q\right) , \\
\\
D_{i}Z=e^{\frac{K}{2}}D_{i}W,~\overline{D}_{\overline{i}}Z=0;
\end{array}
\\
&&  \notag \\
&&
\begin{array}{l}
\hat{\Omega}_{3}\left( z,\overline{z}\right) \equiv \frac{\Omega _{3}}{\sqrt{%
i\int_{CY_{3}}\Omega _{3}\wedge \overline{\Omega }_{3}}}=e^{\frac{K\left( z,%
\overline{z}\right) }{2}}\Omega _{3}\left( z\right) , \\
\\
D_{i}\hat{\Omega}_{3}=e^{\frac{K}{2}}D_{i}\Omega _{3},~\overline{D}_{%
\overline{i}}\hat{\Omega}_{3}=0.
\end{array}
\label{Omega3-hat}
\end{eqnarray}

Let us now compare the Hodge-decomposition identity (\ref{decomp3}) with the
real part (\ref{pre-SKG-final1}) and (\ref{SKG-final1}) of the SK
geometrical identities (\ref{SKG-identities1}). It is immediate to realize
that the identity (\ref{decomp3}) is nothing but the translation, in the
language of forms of Type IIB on $CY_{3}$ (\textit{i.e.} in a particular
stringy framework) of the identity (\ref{pre-SKG-final1})-(\ref{SKG-final1}%
), which is the real part of the fundamental identities (\ref
{SKG-identities1}), holding for any SK geometry, irrespectively of its
microscopic/stringy origin. \setcounter{equation}0
\def\theequation{3.4.\arabic{subsubsection}.\arabic{equation}}
\subsubsection{\label{Stringy-BH-New-Attractor}\textit{``New Attractor''
Approach}}

The evaluation of the Hodge-decomposition identity (\ref{decomp3}) along the
constraints determining the various classes of critical points of $V_{BH}$
in $\mathcal{M}_{n_{V}}$ (which in the considered stringy framework is
nothing but the CS moduli space of $CY_{3}$) allows one to obtain a
completely equivalent form of the AEs for extremal (static, spherically
symmetric, asymptotically flat) BHs in the particular framework in which $%
\mathcal{N}=2$, $d=4$ ungauged supergravity is obtained by compactifying
Type IIB on $CY_{3}$. As already pointed out in the treatment at macroscopic
level, in some cases such equivalent forms of AEs may be simpler to solve
than the AEs obtained by exploiting the \textit{``criticality conditions''
approach} (see Sect. \ref{Sect2}).

I) \textit{Supersymmetric (}$\frac{1}{2}$\textit{-BPS) critical points}. By
evaluating the Hodge-decomposition identity (\ref{decomp3}) along the
constraints (\ref{BPS-conds}), one obtains

\begin{eqnarray}
&&
\begin{array}{l}
\mathcal{H}_{3}=-i\left[ e^{K}\left( W\overline{\Omega }_{3}-\overline{W}%
\Omega _{3}\right) \right] _{\frac{1}{2}-BPS}= \\
\\
=-2Im\left( \overline{Z}\hat{\Omega}_{3}\right) _{\frac{1}{2}-BPS}.
\end{array}
\notag \\
&&  \label{form-alg-BPS}
\end{eqnarray}

Eq. (\ref{form-alg-BPS}) is the translation, for Type IIB on $CY_{3}$, of
Eqs. (\ref{alg-BPS-1}) and (\ref{alg-BPS-2}), which in turn are equivalent,
purely algebraic forms of the $\frac{1}{2}$-BPS extremal BH AEs, given by
the (partly differential) conditions (\ref{BPS-conds})\textit{. }By
recalling Eqs. (\ref{decomp1}) and (\ref{change1}), Eq. (\ref{form-alg-BPS})
implies that at $\frac{1}{2}$-BPS critical points of $V_{BH}$ the real flux
3-form $\mathcal{H}_{3}$ of Type IIB on $CY_{3}$ has vanishing components
along the Dalbeault third cohomologies $H^{2,1}\left( CY_{3}\right) $ and $%
H^{1,2}\left( CY_{3}\right) $. This can be understood easily by recalling
Eq. (\ref{Z2,1}):
\begin{eqnarray}
D_{i}W &=&0,~\forall i\Longleftrightarrow \left\{
\begin{array}{c}
\int_{CY_{3}}\mathcal{H}_{3}\wedge D_{i}\Omega _{3}=0,~\forall i; \\
\Updownarrow \\
\int_{CY_{3}}\mathcal{H}_{3}\wedge \overline{D}_{\overline{i}}\overline{%
\Omega }_{3}=0,~\forall \overline{i}.
\end{array}
\right.  \notag \\
&&  \label{form-ortho-BPS-1}
\end{eqnarray}
Thus, at $\frac{1}{2}$-BPS critical points of $V_{BH}$ $\mathcal{H}_{3}$ is
\textit{''orthogonal'' }(in the sense of Eq. (\ref{form-ortho-BPS-1}), as
understood below, as well) to \textit{all} the 3-forms which are basis
elements of $H^{2,1}\left( CY_{3}\right) $ and $H^{1,2}\left( CY_{3}\right) $%
.

Consequently, the complete supersymmetry breaking at the horizon of (static,
spherically symmetric, asymptotically flat) extremal BHs in $\mathcal{N}=2$,
$d=4$ supergravity as low-energy, effective theory of Type IIB on $CY_{3}$
can be traced back to the non-vanishing \textit{``intersections''} (defined
by Eqs. (\ref{Z2,1}) and (\ref{Z1,2})) of $\mathcal{H}_{3}$ with $%
H^{2,1}\left( CY_{3}\right) $ and $H^{1,2}\left( CY_{3}\right) $. Moreover,
in light of Eq. (\ref{Z3,0}), the $\frac{1}{2}$-BPS non-degeneracy condition
$W_{\frac{1}{2}-BPS}\neq 0$ corresponds to a condition of non(-complete)-%
\textit{''orthogonality''} between $\mathcal{H}_{3}$ and $\Omega _{3}$,
basis of $H^{3,0}\left( CY_{3}\right) $:
\begin{eqnarray}
W &\neq &0\Longleftrightarrow \left\{
\begin{array}{c}
\int_{CY_{3}}\mathcal{H}_{3}\wedge \Omega _{3}\neq 0; \\
\Updownarrow \\
\int_{CY_{3}}\mathcal{H}_{3}\wedge \overline{\Omega }_{3}\neq 0.
\end{array}
\right.  \notag \\
&&  \label{form-orto-BPS-2}
\end{eqnarray}

II) \textit{Non-BPS }$Z\neq 0$\textit{\ critical points}. By evaluating the
Hodge-decomposition identity (\ref{decomp3}) along the constraints (\ref
{non-BPS-Z<>0-conds}) and (\ref{non-BPS-Z<>0-fund}), one obtains
\begin{eqnarray}
&&\mathcal{H}_{3}=2Im\left[ Z\overline{\hat{\Omega}}_{3}+\frac{i}{2}\frac{Z}{%
\left| Z\right| ^{2}}C_{ikl}g^{i\overline{j}}g^{k\overline{k}}g^{l\overline{l%
}}\left( \overline{D}_{\overline{k}}\overline{Z}\right) \left( \overline{D}_{%
\overline{l}}\overline{Z}\right) \overline{D}_{\overline{j}}\overline{\hat{%
\Omega}}_{3}\right] _{non-BPS,Z\neq 0},  \notag \\
&&  \label{form-alg-non-BPS-Z<>0}
\end{eqnarray}

Eq. (\ref{form-alg-non-BPS-Z<>0}) is the translation, for Type IIB on $%
CY_{3} $, of Eqs. (\ref{alg-non-BPS-Z<>0-1}) and (\ref{alg-non-BPS-Z<>0-2}),
which in turn are equivalent forms of the non-BPS $Z\neq 0$ extremal BH AEs,
given by the (partly differential) conditions (\ref{non-BPS-Z<>0-conds}) and
(\ref{non-BPS-Z<>0-fund})\textit{.}

By recalling Eqs. (\ref{decomp1}) and (\ref{change1}), Eq. (\ref
{form-alg-non-BPS-Z<>0}) implies that at the non-BPS $Z\neq 0$ critical
points of $V_{BH}$ the real flux 3-form $\mathcal{H}_{3}$ of type IIB on $%
CY_{3}$ has components along $H^{0,3}\left( CY_{3}\right) $ and $%
H^{2,1}\left( CY_{3}\right) $ with the same holomorphicity in the
holomorphic central charge $Z$. In other words, such components can be
expressed only in terms of $Z$ and $D_{i}Z$, without using $\overline{Z}$
and $\overline{D}_{\overline{i}}\overline{Z}$. Such a fact does not happen
in a generic point of the CS moduli space of $CY_{3}$, as it is seen from
the global Hodge-decomposition identity (\ref{decomp3}). As it is evident,
the price to be paid in order to obtain the same holomorphicity in $Z$ at
the non-BPS $Z\neq 0$ critical points of $V_{BH}$ is the fact that the
component of $\mathcal{H}_{3}$ along $H^{2,1}\left( CY_{3}\right) $ is not
linear in some covariant derivative of $Z$ any more, also explicitly
depending on the rank-3 covariantly antiholomorphic tensor $\overline{C}_{%
\overline{i}\overline{j}\overline{k}}$. By recalling Eqs. (\ref{Z3,0})-(\ref
{Z0,3}), this can be understood by considering the translation of Eq. (\ref
{non-BPS-Z<>0-fund}) in the language of (3-)forms of Type IIB on $CY_{3}$:
\begin{equation}
\int_{CY_{3}}\mathcal{H}_{3}\wedge \overline{D}_{\overline{i}}\overline{%
\Omega }_{3}=\frac{i}{2\int_{CY_{3}}\mathcal{H}_{3}\wedge \Omega _{3}}%
\overline{C}_{\overline{i}\overline{j}\overline{k}}g^{l\overline{j}}g^{m%
\overline{k}}\left( \int_{CY_{3}}\mathcal{H}_{3}\wedge D_{l}\Omega
_{3}\right) \int_{CY_{3}}\mathcal{H}_{3}\wedge D_{m}\Omega _{3},~~\forall
\overline{i}=\overline{1},...,\overline{n_{V}}.
\label{form-ortho-non-BPS-Z<>0-1}
\end{equation}
Eq. (\ref{form-ortho-non-BPS-Z<>0-1}), holding at non-BPS $Z\neq 0$ critical
points of $V_{BH}$, expresses the \textit{``intersections''} of $\mathcal{H}%
_{3}$ with $H^{1,2}\left( CY_{3}\right) $ (\textit{i.e.} the components of $%
\mathcal{H}_{3}$ along $H^{2,1}\left( CY_{3}\right) $; see Eq. (\ref{decomp2}%
)) non-linearly in terms of \textit{``intersections''} of $\mathcal{H}_{3}$
with $H^{3,0}\left( CY_{3}\right) $ and $H^{2,1}\left( CY_{3}\right) $,
which can be all expressed only in terms of $Z$ and $D_{i}Z$, without using $%
\overline{Z}$ and $\overline{D}_{\overline{i}}\overline{Z}$.

III) \textit{Non-BPS }$Z=0$\textit{\ critical points}. By evaluating the
Hodge-decomposition identity (\ref{decomp3}) along the constraints (\ref
{non-BPS-Z=0-conds}) and (\ref{non-BPS-Z=0-fund}), one obtains
\begin{eqnarray}
&&
\begin{array}{l}
\mathcal{H}_{3}=-ie^{K}\left[ -g^{i\overline{j}}\left( \partial _{i}W\right)
\overline{D}_{\overline{j}}\overline{\Omega }_{3}+g^{j\overline{i}}\left(
\overline{\partial }_{\overline{i}}\overline{W}\right) D_{j}\Omega _{3}%
\right] = \\
\\
=2Im\left[ g^{j\overline{i}}\left( \overline{\partial }_{\overline{i}}%
\overline{Z}\right) D_{j}\hat{\Omega}_{3}\right] ,
\end{array}
\notag \\
&&  \label{form-alg-non-BPS-Z=0}
\end{eqnarray}
Eq. (\ref{form-alg-non-BPS-Z=0}) is the translation, for Type IIB on $CY_{3}$%
, of Eqs. (\ref{alg-non-BPS-Z=0-1}) and (\ref{alg-non-BPS-Z=0-2}), which in
turn are equivalent forms of the non-BPS $Z=0$ extremal BH AEs, given by the
(partly differential) conditions (\ref{non-BPS-Z=0-conds}) and (\ref
{non-BPS-Z=0-fund})\textit{. }By recalling Eqs. (\ref{decomp1}) and (\ref
{change1}), Eq. (\ref{form-alg-non-BPS-Z=0}) implies that at non-BPS $Z=0$
critical points of $V_{BH}$, in an opposite fashion with respect to the case
of $\frac{1}{2}$-BPS critical points of $V_{BH}$, the real flux 3-form $%
\mathcal{H}_{3}$ of Type IIB on $CY_{3}$ has vanishing components along the
Dalbeault third cohomologies $H^{3,0}\left( CY_{3}\right) $ and $%
H^{0,3}\left( CY_{3}\right) $. This can be understood easily by recalling
Eq. (\ref{Z3,0}):
\begin{eqnarray}
W &=&0\Longleftrightarrow \left\{
\begin{array}{c}
\int_{CY_{3}}\mathcal{H}_{3}\wedge \Omega _{3}=0; \\
\Updownarrow \\
\int_{CY_{3}}\mathcal{H}_{3}\wedge \overline{\Omega }_{3}=0.
\end{array}
\right.  \notag \\
&&  \label{form-ortho-non-BPS-Z=0-1}
\end{eqnarray}
Thus, at non-BPS $Z=0$ critical points of $V_{BH}$ $\mathcal{H}_{3}$ is
\textit{''orthogonal'' }(in the sense of Eq. (\ref{form-ortho-non-BPS-Z=0-1}%
), as understood below, as well) to $\Omega _{3}$ and $\overline{\Omega }%
_{3} $, basis of $H^{3,0}\left( CY_{3}\right) $ and $H^{0,3}\left(
CY_{3}\right) $, respectively. Moreover, in light of Eqs. (\ref{Z2,1}) and (%
\ref{Z1,2}), the non-BPS $Z=0$ non-degeneracy condition (\textit{at least}
for strictly positive-definite $g_{i\overline{j}}$ at the considered
critical points of $V_{BH}$)
\begin{equation}
\left( D_{i}W\right) _{non-BPS,Z=0}\neq 0,\text{\textit{at least} for some }%
i\in \left\{ 1,...,n_{V}\right\}
\end{equation}
corresponds to a condition of non(-complete)-\textit{''orthogonality''}
between $\mathcal{H}_{3}$ and the $D_{i}\Omega _{3}$s, basis elements of $%
H^{2,1}\left( CY_{3}\right) $:
\begin{eqnarray}
D_{i}W &\neq &0,\text{\textit{at least} for some }i\in \left\{
1,...,n_{V}\right\} \Longleftrightarrow \left\{
\begin{array}{c}
\int_{CY_{3}}\mathcal{H}_{3}\wedge D_{i}\Omega _{3}\neq 0,,\text{\textit{at
least} for some }i\in \left\{ 1,...,n_{V}\right\} ; \\
\Updownarrow \\
\int_{CY_{3}}\mathcal{H}_{3}\wedge \overline{D}_{\overline{i}}\overline{%
\Omega }_{3}\neq 0,\text{\textit{at least} for some }\overline{i}\in \left\{
\overline{1},...,\overline{n_{V}}\right\} .
\end{array}
\right.  \notag \\
&&  \label{form-ortho-non-BPS-Z=0-2}
\end{eqnarray}
\setcounter{equation}0

\section{\label{Flux-Vacua-Attractors}Flux Vacua Attractor Equations\newline
in $\mathcal{N}=1$, $d=4$ Supergravity from Type IIB on $\frac{CY_{3}\times
T^{2}}{\mathbb{Z}_{2}}$}

\setcounter{equation}0
\def\theequation{4.\arabic{subsection}.\arabic{equation}}
\subsection{\label{CY3-Orientifolds}$CY_{3}$ Orientifolds}

We consider Type IIB superstring theory compactified on a $CY_{3}$
orientifold with O3/O7-planes (as in the GPK-KKLT model \cite
{GPK,Kachru:2003aw}), determining an $\mathcal{N}=1$, $d=4$ supergravity as
effective, low-energy theory. Within such a framework, we will derive FV AEs%
\footnote{%
In \cite{Dall'Agata} the FV Attractor Mechanism has been shown to act also
in the landscape of non-K\"{a}hler vacua emerging in the flux
compactifications of heterotic superstrings.}, similarly to what done in
Subsect. \ref{Micro-BH} for extremal BH AEs in effective $\mathcal{N}=2$, $%
d=4$ supergravity from Type IIB on $CY_{3}$. We will mainly follow \cite{K1}%
, \cite{Denef-Douglas-1} and \cite{Soroush:2007ed}. In our treatment, the
relevant moduli moduli space $M$ of the $CY_{3}$ orientifold is the one
composed by the (direct) product of the CS moduli space (of complex
dimension $h_{2,1}\equiv dim\left( H^{2,1}\left( CY_{3}\right) \right) $,
which is a SK manifold, and the $1$-dim. K\"{a}hler manifold parameterized
by the universal axion-dilaton. We will denote the CS moduli by $\left(
x^{i},\overline{x}^{\overline{i}}\right) _{i=1,...,h_{2,1}}\equiv \left(
t^{i},\overline{t}^{\overline{i}}\right) _{i=1,...,h_{2,1}}$ (not to be
confused with the projective coordinates in the SK CS moduli space) and the
axion-dilaton by $\tau \equiv t^{0}$:
\begin{equation}
M=\mathcal{M}_{t^{0}}\otimes \mathcal{M}_{CS}.  \label{MM}
\end{equation}
No K\"{a}hler structure (KS) moduli will be considered in our treatment of
the classical FV Attractor Mechanism; indeed, in the considered framework
the stabilization of KS moduli requires quantum perturbative or
non-perturbative mechanisms, such as worldsheet instantons and gaugino
condensation (see \textit{e.g.} \cite{Kachru:2003aw})\textbf{.}
\setcounter{equation}0
\def\theequation{4.1.\arabic{subsubsection}.\arabic{equation}}
\subsubsection{\label{Bein and Metric}Vielbein and Metric Tensor in the
Moduli Space}

We start by defining the structure of the $\left( h_{2,1}+1\right) $-dim. K%
\"{a}hler manifold spanned by the CS moduli and the axion-dilaton. Its K\"{a}%
hler potential can be written as follows ($\Lambda =1,...,h_{2,1},h_{2,1}+1$
throughout\footnote{%
Notice the different range of the symplectic (capital Greek) indices in the
present treatment of Type IIB on $CY_{3}$ orientifold with O3/O7-planes with
respect to the range $0,1,...,h_{2,1}$ of the previous treatment of Type IIB
on $CY_{3}$. In general, the reference to the graviphoton degree of freedom
``$0$'' is lost, due to the orientifolding truncation of the low-energy,
effective supergravity.}):
\begin{eqnarray}
&&
\begin{array}{l}
K\left( t,\overline{t}\right) =-ln\left[ -i\left( t^{0}-\overline{t}^{%
\overline{0}}\right) \right] -ln\left[ i\int_{CY_{3}}\Omega _{3}\left(
x\right) \wedge \overline{\Omega }_{3}\left( \overline{x}\right) \right] =
\\
\\
=-ln\left[ \int_{CY_{3}}\left[ t^{0}\Omega _{3}\left( x\right) \wedge
\overline{\Omega }_{3}\left( \overline{x}\right) -\Omega _{3}\left( x\right)
\wedge \overline{t}^{\overline{0}}\overline{\Omega }_{3}\left( \overline{x}%
\right) \right] \right] = \\
\\
=-ln\left[ \left( t^{0}-\overline{t}^{\overline{0}}\right) \left( \overline{X%
}^{\Lambda }\left( \overline{x}\right) F_{\Lambda }\left( x\right)
-X^{\Lambda }\left( x\right) \overline{F}_{\Lambda }\left( \overline{x}%
\right) \right) \right] ,
\end{array}
\notag \\
&&  \label{FV-K}
\end{eqnarray}
where $\Omega _{3}$ is the holomorphic $\left( 3,0\right) $-form defined on $%
CY_{3}$. Thus, one can write:
\begin{eqnarray}
&&\left\{
\begin{array}{l}
K\left( t,\overline{t}\right) =K_{1}\left( t^{0},\overline{t}^{\overline{0}%
}\right) +K_{3}\left( x,\overline{x}\right) ; \\
\\
K_{1}\left( t^{0},\overline{t}^{\overline{0}}\right) \equiv -ln\left[
-i\left( t^{0}-\overline{t}^{\overline{0}}\right) \right] ; \\
\\
K_{3}\left( x,\overline{x}\right) \equiv -ln\left[ i\left( \overline{X}%
^{\Lambda }\left( \overline{x}\right) F_{\Lambda }\left( x\right)
-X^{\Lambda }\left( x\right) \overline{F}_{\Lambda }\left( \overline{x}%
\right) \right) \right] .
\end{array}
\right.   \notag \\
&&
\end{eqnarray}
The reality condition on $K_{1}$ and $K_{3}$ yields the conditions
\begin{equation}
Imt^{0}>0,~Im\left( X^{\Lambda }\left( x\right) \overline{F}_{\Lambda
}\left( \overline{x}\right) \right) >0.
\end{equation}
The metric of the whole moduli space is given by ($a=0,1,...,h_{2,1}$
throughout)
\begin{equation}
g_{a\overline{b}}\left( t,\overline{t}\right) =\overline{\partial }_{%
\overline{b}}\partial _{a}K\left( t,\overline{t}\right) =\overline{\partial }%
_{\overline{b}}\partial _{a}\left[ K_{1}\left( t^{0},\overline{t}^{\overline{%
0}}\right) +K_{3}\left( x,\overline{x}\right) \right] ,
\end{equation}
yielding
\begin{eqnarray}
&&\left\{
\begin{array}{l}
g_{0\overline{0}}=-\left( \overline{t}^{\overline{0}}-t^{0}\right)
^{-2}=e^{2K_{1}\left( t^{0},\overline{t}^{\overline{0}}\right) }; \\
\\
g_{0\overline{i}}=0=g_{i\overline{0}}; \\
\\
g_{i\overline{j}}=\overline{\partial }_{\overline{j}}\partial
_{i}K_{3}\left( x,\overline{x}\right) .
\end{array}
\right.   \notag \\
&&  \label{g-M}
\end{eqnarray}
In our treatment we will make extensive use of the local ``flat''
coordinates in $M$ (denoted by capital indices $A=\underline{0},\underline{1}%
,...,\underline{h_{2,1}}$ throughout), defined as usual by ($g_{a\overline{b}%
}g^{a\overline{c}}=\delta _{\overline{b}}^{\overline{c}}$, $g_{a\overline{b}%
}g^{c\overline{b}}=\delta _{a}^{c}$)
\begin{equation}
g_{a\overline{b}}\left( t,\overline{t}\right) \equiv e_{a}^{A}\left( t,%
\overline{t}\right) \overline{e}_{\overline{b}}^{\overline{B}}\left( t,%
\overline{t}\right) \delta _{A\overline{B}}\Leftrightarrow g^{a\overline{b}%
}\left( t,\overline{t}\right) \equiv e_{A}^{a}\left( t,\overline{t}\right)
\overline{e}_{\overline{B}}^{\overline{b}}\left( t,\overline{t}\right)
\delta ^{A\overline{B}},  \label{M-bein}
\end{equation}
where $e_{a}^{A}\left( t,\overline{t}\right) $ is the local vielbein in $M$,
and $e_{A}^{a}\left( t,\overline{t}\right) $ is its inverse ($%
e_{a}^{A}e_{A}^{b}=\delta _{a}^{b}$, $e_{a}^{A}e_{B}^{a}=\delta _{B}^{A}$).
Due to Eqs. (\ref{g-M}), the $h_{2,1}^{2}+2h_{2,1}+1$ components of the
vielbein $e_{a}^{A}=\left\{ e_{0}^{\underline{0}},e_{i}^{\underline{0}%
},e_{0}^{I},e_{i}^{I}\right\} $ ($I=\underline{1},...,\underline{h_{2,1}}$
throughout), defined by Eq. (\ref{M-bein}), satisfy the following set of
Eqs.:
\begin{eqnarray}
&&\left\{
\begin{array}{l}
\left| e_{0}^{\underline{0}}\left( t,\overline{t}\right) \right|
^{2}+e_{0}^{I}\left( t,\overline{t}\right) \overline{e}_{\overline{0}}^{%
\overline{J}}\left( t,\overline{t}\right) \delta _{I\overline{J}}=-\left(
\overline{t}^{\overline{0}}-t^{0}\right) ^{-2}; \\
\\
e_{0}^{\underline{0}}\left( t,\overline{t}\right) \overline{e}_{\overline{i}%
}^{\overline{\underline{0}}}\left( t,\overline{t}\right) +e_{0}^{I}\left( t,%
\overline{t}\right) \overline{e}_{\overline{i}}^{\overline{J}}\left( t,%
\overline{t}\right) \delta _{I\overline{J}}=0; \\
\\
e_{i}^{\underline{0}}\left( t,\overline{t}\right) \overline{e}_{\overline{j}%
}^{\overline{\underline{0}}}\left( t,\overline{t}\right) +e_{i}^{I}\left( t,%
\overline{t}\right) \overline{e}_{\overline{j}}^{\overline{J}}\left( t,%
\overline{t}\right) \delta _{I\overline{J}}=\overline{\partial }_{\overline{j%
}}\partial _{i}K_{3}\left( x,\overline{x}\right) ,
\end{array}
\right.   \notag \\
&&  \label{M-bein-2}
\end{eqnarray}
admitting as a solution\footnote{%
Notice that the solutions given by Eqs. (\ref{M-bein-2-sol}) and (\ref
{M-bein-inv-sol}) are clearly not unique. Indeed, for a given metric, one
can always transform the vielbein and its inverse by a Lorentz
transformation, which however will not affect the metric itself.}:
\begin{eqnarray}
&&\left\{
\begin{array}{l}
\left| e_{0}^{\underline{0}}\left( t,\overline{t}\right) \right|
^{2}=-\left( \overline{t}^{\overline{0}}-t^{0}\right) ^{-2}\Leftarrow e_{0}^{%
\underline{0}}\left( t,\overline{t}\right) =\left( \overline{t}^{\overline{0}%
}-t^{0}\right) ^{-1}=ie^{K_{1}\left( t^{0},\overline{t}^{\overline{0}%
}\right) }=e_{0}^{\underline{0}}\left( t^{0},\overline{t}^{\overline{0}%
}\right) ; \\
\\
e_{0}^{I}\left( t,\overline{t}\right) =0,~\forall I=\underline{1},...,%
\underline{h_{2,1}}; \\
\\
e_{i}^{\underline{0}}\left( t,\overline{t}\right) =0,~\forall
i=1,...,h_{2,1}; \\
\\
e_{i}^{I}\left( t,\overline{t}\right) \overline{e}_{\overline{j}}^{\overline{%
J}}\left( t,\overline{t}\right) \delta _{I\overline{J}}=\overline{\partial }%
_{\overline{j}}\partial _{i}K_{3}\left( x,\overline{x}\right) .
\end{array}
\right.   \notag \\
&&  \label{M-bein-2-sol}
\end{eqnarray}

By inverting Eq. (\ref{M-bein}) one gets
\begin{equation}
\delta _{A\overline{B}}=e_{A}^{a}\left( t,\overline{t}\right) \overline{e}_{%
\overline{B}}^{\overline{b}}\left( t,\overline{t}\right) g_{a\overline{b}%
}\left( t,\overline{t}\right) \Leftrightarrow \delta ^{A\overline{B}%
}=e_{a}^{A}\left( t,\overline{t}\right) \overline{e}_{\overline{b}}^{%
\overline{B}}\left( t,\overline{t}\right) g^{a\overline{b}}\left( t,%
\overline{t}\right) ,  \label{M-bein-inv}
\end{equation}
which by Eq. (\ref{g-M}) implies that the $h_{2,1}^{2}+2h_{2,1}+1$
components of the inverse vielbein $e_{A}^{a}=\left\{ e_{\underline{0}%
}^{0},e_{\underline{0}}^{i},e_{I}^{0},e_{I}^{i}\right\} $, defined by Eq. (%
\ref{M-bein}), satisfy the set following set of Eqs.:
\begin{eqnarray}
&&\left\{
\begin{array}{l}
-\left( \overline{t}^{\overline{0}}-t^{0}\right) ^{-2}\left| e_{\underline{0}%
}^{0}\left( t,\overline{t}\right) \right| ^{2}+e_{\underline{0}}^{i}\left( t,%
\overline{t}\right) \overline{e}_{\overline{\underline{0}}}^{\overline{j}%
}\left( t,\overline{t}\right) \overline{\partial }_{\overline{j}}\partial
_{i}K_{3}\left( x,\overline{x}\right) =1; \\
\\
-\left( \overline{t}^{\overline{0}}-t^{0}\right) ^{-1}\overline{e}_{%
\overline{I}}^{\overline{0}}\left( t,\overline{t}\right) +e_{\underline{0}%
}^{i}\left( t,\overline{t}\right) \overline{e}_{\overline{I}}^{\overline{j}%
}\left( t,\overline{t}\right) \overline{\partial }_{\overline{j}}\partial
_{i}K_{3}\left( x,\overline{x}\right) =0; \\
\\
-\left( \overline{t}^{\overline{0}}-t^{0}\right) ^{-2}e_{I}^{0}\left( t,%
\overline{t}\right) \overline{e}_{\overline{J}}^{\overline{0}}\left( t,%
\overline{t}\right) +e_{I}^{i}\left( t,\overline{t}\right) \overline{e}_{%
\overline{J}}^{\overline{j}}\left( t,\overline{t}\right) \overline{\partial }%
_{\overline{j}}\partial _{i}K_{3}\left( x,\overline{x}\right) =\delta _{I%
\overline{J}},
\end{array}
\right.  \notag \\
&&  \label{M-bein-inv-2}
\end{eqnarray}
admitting as a\textbf{\ }solution:
\begin{eqnarray}
&&\left\{
\begin{array}{l}
\begin{array}{c}
\left| e_{\underline{0}}^{0}\left( t,\overline{t}\right) \right|
^{2}=-\left( \overline{t}^{\overline{0}}-t^{0}\right) ^{2}=\left| e_{0}^{%
\underline{0}}\left( t,\overline{t}\right) \right| ^{-2}; \\
\Uparrow \\
e_{\underline{0}}^{0}\left( t,\overline{t}\right) =\left( \overline{t}^{%
\overline{0}}-t^{0}\right) =-ie^{-K_{1}\left( t^{0},\overline{t}^{\overline{0%
}}\right) }=\left[ e_{0}^{\underline{0}}\left( t^{0},\overline{t}^{\overline{%
0}}\right) \right] ^{-1}=e_{\underline{0}}^{0}\left( t^{0},\overline{t}^{%
\overline{0}}\right) ;
\end{array}
\\
\\
\\
e_{\underline{0}}^{i}\left( t,\overline{t}\right) =0,~\forall
i=1,...,h_{2,1}; \\
\\
e_{I}^{0}\left( t,\overline{t}\right) =0,~\forall I=\underline{1},...,%
\underline{h_{2,1}}; \\
\\
e_{I}^{i}\left( t,\overline{t}\right) \overline{e}_{\overline{J}}^{\overline{%
j}}\left( t,\overline{t}\right) \overline{\partial }_{\overline{j}}\partial
_{i}K_{3}\left( x,\overline{x}\right) =\delta _{I\overline{J}},
\end{array}
\right.  \notag \\
&&  \label{M-bein-inv-sol}
\end{eqnarray}
implying, by Eq. (\ref{M-bein}), that the components of the inverse metric
tensor of $M$ read as follows:
\begin{eqnarray}
&&\left\{
\begin{array}{l}
g^{0\overline{0}}=-\left( \overline{t}^{\overline{0}}-t^{0}\right)
^{2}=e^{-2K_{1}\left( t^{0},\overline{t}^{\overline{0}}\right) }=\left( g_{0%
\overline{0}}\right) ^{-1}; \\
\\
g^{0\overline{i}}=0=g^{i\overline{0}}; \\
\\
g^{i\overline{j}}:g^{i\overline{j}}\overline{\partial }_{\overline{j}%
}\partial _{k}K_{3}\left( x,\overline{x}\right) =\delta _{k}^{i},~g^{i%
\overline{j}}\overline{\partial }_{\overline{k}}\partial _{i}K_{3}\left( x,%
\overline{x}\right) =\delta _{\overline{k}}^{\overline{j}}.
\end{array}
\right.  \notag \\
&&  \label{g-M-inv}
\end{eqnarray}
Moreover, it should be noticed that actually $e_{i}^{I}=e_{i}^{I}\left( x,%
\overline{x}\right) $ and $e_{I}^{i}=e_{I}^{i}\left( x,\overline{x}\right) $%
, as obtained by differentiating with respect to the axion-dilaton $t^{0}$
the fourth Eq. of the systems of solutions (\ref{M-bein-2-sol}) and (\ref
{M-bein-inv-sol}), respectively:
\begin{eqnarray}
&&
\begin{array}{l}
\begin{array}{c}
\left\{ \left[ \partial _{0}e_{i}^{I}\left( t,\overline{t}\right) \right]
\overline{e}_{\overline{j}}^{\overline{J}}\left( t,\overline{t}\right)
+e_{i}^{I}\left( t,\overline{t}\right) \partial _{0}\overline{e}_{\overline{j%
}}^{\overline{J}}\left( t,\overline{t}\right) \right\} \delta _{I\overline{J}%
}=0; \\
\Updownarrow \\
\left\{
\begin{array}{l}
\partial _{0}e_{i}^{I}\left( t,\overline{t}\right) =0, \\
\\
\partial _{0}\overline{e}_{\overline{i}}^{\overline{I}}\left( t,\overline{t}%
\right) =0\Leftrightarrow \overline{\partial }_{\overline{0}}e_{i}^{I}\left(
t,\overline{t}\right) =0;
\end{array}
\right. \\
\Updownarrow \\
e_{i}^{I}=e_{i}^{I}\left( x,\overline{x}\right) ;
\end{array}
\\
\\
\\
\begin{array}{c}
\left\{ \left[ \partial _{0}e_{I}^{i}\left( t,\overline{t}\right) \right]
\overline{e}_{\overline{J}}^{\overline{j}}\left( t,\overline{t}\right)
+e_{I}^{i}\left( t,\overline{t}\right) \partial _{0}\overline{e}_{\overline{J%
}}^{\overline{j}}\left( t,\overline{t}\right) \right\} \overline{\partial }_{%
\overline{j}}\partial _{i}K_{3}\left( x,\overline{x}\right) =0; \\
\Updownarrow \\
\left\{
\begin{array}{l}
\partial _{0}e_{I}^{i}\left( t,\overline{t}\right) =0, \\
\\
\partial _{0}\overline{e}_{\overline{I}}^{\overline{i}}\left( t,\overline{t}%
\right) \Leftrightarrow \overline{\partial }_{\overline{0}}e_{I}^{i}\left( t,%
\overline{t}\right) =0;
\end{array}
\right. \\
\Updownarrow \\
e_{I}^{i}=e_{I}^{i}\left( x,\overline{x}\right) .
\end{array}
\end{array}
\notag \\
&&  \label{funct-dep1}
\end{eqnarray}

In the following treatment, we will use the solutions (\ref{M-bein-2-sol})
and (\ref{M-bein-inv-sol}) of the systems of Eqs. (\ref{M-bein-2}) and (\ref
{M-bein-inv-2}), respectively, \textit{i.e.} we will assume that a system of
local ``flat'' coordinates in $M$ defined by Eqs. (\ref{M-bein}) and (\ref
{M-bein-inv}) always exists such that the corresponding vielbein and its
inverse are given by Eqs. (\ref{M-bein-2-sol}) and (\ref{M-bein-inv-sol})
(implemented by Eqs. (\ref{funct-dep1})), in turn consistent with the
covariant and contravariant metric tensor of $M$ given by Eqs. (\ref{g-M})
and (\ref{g-M-inv}), respectively. \setcounter{equation}0
\def\theequation{4.1.\arabic{subsubsection}.\arabic{equation}}
\subsubsection{\label{1-3-4-Forms}1-, 3- and 4-Forms on $\frac{CY_{3}\times
T^{2}}{\mathbb{Z}_{2}}$}

Next, we introduce the Ramond-Ramond (RR) and Neveu-Schwarz-Neveu-Schwarz
(NSNS) flux 3-forms of Type IIB on $CY_{3}$ orientifold (with O3/O7-planes)
as follows:
\begin{equation}
\begin{array}{l}
RR:\frak{F}_{3}\equiv p_{f}^{\Lambda }\alpha _{\Lambda }-q_{f\mid \Lambda
}\beta ^{\Lambda }\in H^{3}\left( CY_{3},\mathbb{R}\right) ; \\
\\
NSNS:\frak{H}_{3}\equiv p_{h}^{\Lambda }\alpha _{\Lambda }-q_{h\mid \Lambda
}\beta ^{\Lambda }\in H^{3}\left( CY_{3},\mathbb{R}\right) ,
\end{array}
\label{RR-NS-1}
\end{equation}
where we introduced the $1\times \left( 2h_{2,1}+2\right) $ symplectic
vector of RR and NSNS fluxes (charges), respectively:
\begin{equation}
\begin{array}{l}
Q_{RR}\equiv \left( p_{f}^{\Lambda },q_{f\mid \Lambda }\right) ; \\
\\
Q_{NSNS}\equiv \left( p_{h}^{\Lambda },q_{h\mid \Lambda }\right) ,
\end{array}
\label{RR-NS-charges}
\end{equation}
and $\left\{ \alpha _{\Lambda },\beta ^{\Lambda }\right\} $ is the $b_{3}$%
-dim. real (manifestly symplectic-covariant) basis of the third real
cohomology $H^{3}\left( CY_{3},\mathbb{R}\right) $, satisfying Eq. (\ref
{22apr1}). In the considered framework, the flux 3-forms defined by Eq. (\ref
{RR-NS-1}) can be unified in the $t^{0}$-dependent, complex flux 3-form
\begin{equation}
\frak{G}_{3}\left( t^{0}\right) \equiv \frak{F}_{3}-t^{0}\frak{H}_{3}=\left(
p_{f}^{\Lambda }-t^{0}p_{h}^{\Lambda }\right) \alpha _{\Lambda }-\left(
q_{f\mid \Lambda }-t^{0}q_{h\mid \Lambda }\right) \beta ^{\Lambda }\in
H^{3}\left( CY_{3},\mathbb{C};t^{0}\right) ,  \label{G3}
\end{equation}
thus determining the GVW $\mathcal{N}=1$, $d=4$ holomorphic superpotential
as follows:
\begin{eqnarray}
&&
\begin{array}{l}
W\left( t\right) \equiv \int_{CY_{3}}\frak{G}_{3}\left( t^{0}\right) \wedge
\Omega _{3}\left( x\right) =\int_{CY_{3}}\frak{F}_{3}\wedge \Omega
_{3}\left( x\right) -t^{0}\int_{CY_{3}}\frak{H}_{3}\wedge \Omega _{3}\left(
x\right) = \\
\\
=q_{f\mid \Lambda }X^{\Lambda }\left( x\right) -p_{f}^{\Lambda }F_{\Lambda
}\left( x\right) +q_{h\mid \Lambda }\left( -t^{0}X^{\Lambda }\left( x\right)
\right) -p_{h}^{\Lambda }\left( -t^{0}F_{\Lambda }\left( x\right) \right) .
\end{array}
\notag \\
&&  \label{W-FV-1}
\end{eqnarray}
The second line of Eq. (\ref{W-FV-1}) suggests to redefine the holomorphic $%
\left( 3,0\right) $-form in the \textit{''NSNS sector''} as follows:
\begin{equation}
\Omega _{3,NS}\left( t\right) \equiv -t^{0}\Omega _{3,RR}\left( x\right)
=-t^{0}\Omega _{3}\left( x\right) .
\end{equation}
Since in Type IIB on the considered $CY_{3}$ orientifold the flux 3-forms $%
\frak{F}_{3}$ and $\frak{H}_{3}$ form the $SL\left( 2,H^{3}\left( CY_{3},%
\mathbb{R}\right) \right) $-doublet
\begin{equation}
\widehat{F}\equiv \left(
\begin{array}{c}
\frak{F}_{3} \\
\\
\frak{H}_{3}
\end{array}
\right) \in SL\left( 2,H^{3}\left( CY_{3},\mathbb{R}\right) \right) ,
\end{equation}
correspondingly, one can introduce the $SL\left( 2,H^{3,0}\left(
CY_{3};t\right) \right) $-doublet
\begin{equation}
\Xi \left( t\right) \equiv \left(
\begin{array}{l}
\Xi _{1}\left( x\right) \equiv \Omega _{3}\left( x\right)  \\
\\
\Xi _{2}\left( t\right) \equiv -t^{0}\Omega _{3}\left( x\right)
\end{array}
\right) \in SL\left( 2,H^{3,0}\left( CY_{3};t\right) \right) .
\end{equation}
By exploiting such a manifest $SL(2)$-covariance, Eqs. (\ref{FV-K}) and (\ref
{W-FV-1}) can be rewritten as follows:
\begin{eqnarray}
&&
\begin{array}{l}
K\left( t,\overline{t}\right) =-ln\left[ \int_{CY_{3}}\left[ \Xi _{1}\left(
x\right) \wedge \overline{\Xi }_{2}\left( \overline{t}\right) -\Xi
_{2}\left( t\right) \wedge \overline{\Xi }_{1}\left( \overline{x}\right)
\right] \right] ;
\end{array}
\label{FV-K-2} \\
&&  \notag \\
&&
\begin{array}{l}
W\left( t\right) =\int_{CY_{3}}\left[ \frak{F}_{3}\wedge \Xi _{1}\left(
x\right) +\frak{H}_{3}\wedge \Xi _{2}\left( t\right) \right] =\int_{CY_{3}}%
\widehat{F}^{T}\wedge \Xi \left( t\right) .
\end{array}
\label{W-FV-2}
\end{eqnarray}
Thus, the $\mathcal{N}=1$, $d=4$ covariantly holomorphic central charge
function of Type IIB on $CY_{3}$ orientifold with O3/O7-planes can be
introduced:
\begin{eqnarray}
Z\left( t,\overline{t}\right)  &\equiv &e^{\frac{1}{2}K\left( t,\overline{t}%
\right) }W\left( t\right) =e^{\frac{1}{2}K\left( t,\overline{t}\right)
}\int_{CY_{3}}\frak{G}_{3}\left( t^{0}\right) \wedge \Omega _{3}\left(
x\right) = \\
&=&\frac{\int_{CY_{3}}\widehat{F}^{T}\wedge \Xi \left( t\right) }{\sqrt{%
\int_{CY_{3}}\left[ \Xi _{1}\left( x\right) \wedge \overline{\Xi }_{2}\left(
\overline{t}\right) -\Xi _{2}\left( t\right) \wedge \overline{\Xi }%
_{1}\left( \overline{x}\right) \right] }}=  \label{Z-FV-1} \\
&=&\frac{\left( q_{f\mid \Lambda }-t^{0}q_{h\mid \Lambda }\right) X^{\Lambda
}\left( x\right) -\left( p_{f}^{\Lambda }-t^{0}p_{h}^{\Lambda }\right)
F_{\Lambda }\left( x\right) }{\sqrt{\left( t^{0}-\overline{t}^{\overline{0}%
}\right) \left( \overline{X}^{\Lambda }\left( \overline{x}\right) F_{\Lambda
}\left( x\right) -X^{\Lambda }\left( x\right) \overline{F}_{\Lambda }\left(
\overline{x}\right) \right) }},  \label{Z-FV-2}
\end{eqnarray}
with K\"{a}hler weights $\left( 1,-1\right) $ with respect to $K\left( t,%
\overline{t}\right) $:
\begin{eqnarray}
&&
\begin{array}{l}
D_{a}Z\left( t,\overline{t}\right) =\partial _{a}Z\left( t,\overline{t}%
\right) +\frac{1}{2}\left( \partial _{a}K\left( t,\overline{t}\right)
\right) Z\left( t,\overline{t}\right) ; \\
\\
\overline{D}_{\overline{a}}Z\left( t,\overline{t}\right) =\overline{\partial
}_{\overline{a}}Z\left( t,\overline{t}\right) -\frac{1}{2}\left( \overline{%
\partial }_{\overline{a}}K\left( t,\overline{t}\right) \right) Z\left( t,%
\overline{t}\right) =0.
\end{array}
\notag \\
&&
\end{eqnarray}

Now, we can perform an unifying simplification of notation, by using the
language of 4-forms on Calabi-Yau 4-folds ($CY_{4}$); in such a framework,
Type IIB on $CY_{3}$ orientifold with O3/O7-planes can be described by
4-forms defined on $CY_{4}=\frac{CY_{3}\times T^{2}}{\mathbb{Z}_{2}}$, where
$T^{2}$ denotes the ``auxiliary'' $2$-torus, whose complex modulus is the
universal-axion dilaton $\tau \equiv t^{0}$. Thus, beside the $b_{3}$-dim.
real (manifestly symplectic-covariant) basis $\left\{ \alpha _{\Lambda
},\beta ^{\Lambda }\right\} $ of $H^{3}\left( CY_{3},\mathbb{R}\right) $
(satisfying Eq. (\ref{22apr1})), one can introduce the $2$-dim. basis $%
\left\{ \alpha ,\beta \right\} $ of $H^{1}\left( T^{2},\mathbb{R}\right) $,
satisfying
\begin{equation}
\int_{T^{2}}\alpha \wedge \alpha =0=\int_{T^{2}}\beta \wedge \beta
,~~\int_{T^{2}}\alpha \wedge \beta =1,  \label{24apr1}
\end{equation}
and the holomorphic $\left( 1,0\right) $-form $\Omega _{1}\left(
t^{0}\right) $ on $T^{2}$:
\begin{equation}
\Omega _{1}\left( t^{0}\right) \equiv -t^{0}\alpha +\beta \in H^{1,0}\left(
T^{2}\right) .  \label{Omega1}
\end{equation}
By recalling Eq. (\ref{Omega3}), it is thus possible to define an
holomorphic $\left( 4,0\right) $-form on $CY_{4}$($=\frac{CY_{3}\times T^{2}%
}{\mathbb{Z}_{2}}$, as always understood in treatment below) as follows:
\begin{eqnarray}
\Omega _{4}\left( t\right) &\equiv &\Omega _{1}\left( t^{0}\right) \wedge
\Omega _{3}\left( x\right) =X^{\Lambda }\left( x\right) \beta \wedge \alpha
_{\Lambda }-t^{0}X^{\Lambda }\left( x\right) \alpha \wedge \alpha _{\Lambda
}-F_{\Lambda }\left( x\right) \beta \wedge \beta ^{\Lambda }+t^{0}F_{\Lambda
}\left( x\right) \alpha \wedge \beta ^{\Lambda }.  \notag \\
&&  \label{Omega4}
\end{eqnarray}
Instead of using the complex, $t^{0}$-dependent flux 3-form $\frak{G}%
_{3}\left( t^{0}\right) \in H^{3}\left( CY_{3},\mathbb{C};t^{0}\right) $
defined by Eq. (\ref{G3}), the RR and NSNS flux 3-forms can be unified
elegantly by introducing the real flux 4-form
\begin{eqnarray}
\frak{F}_{4} &\equiv &-\alpha \wedge \frak{F}_{3}+\beta \wedge \frak{H}%
_{3}=\left( p_{h}^{\Lambda }\beta -p_{f}^{\Lambda }\alpha \right) \wedge
\alpha _{\Lambda }-\left( q_{h\mid \Lambda }\beta -q_{f\mid \Lambda }\alpha
\right) \wedge \beta ^{\Lambda }\in H^{4}\left( CY_{4},\mathbb{R}\right) .
\notag \\
&&  \label{F4}
\end{eqnarray}

By using Eqs. (\ref{22apr1}), (\ref{Omega3}), (\ref{24apr1}), (\ref{Omega1}%
), (\ref{Omega4}) and (\ref{F4}), Eqs. (\ref{FV-K-2})-(\ref{Z-FV-1}) can be
elegantly rewritten as follows:
\begin{eqnarray}
&&
\begin{array}{l}
K\left( t,\overline{t}\right) =-ln\left( \int_{CY_{4}}\Omega _{4}\left(
t\right) \wedge \overline{\Omega }_{4}\left( \overline{t}\right) \right) ;
\end{array}
\label{FV-KK} \\
&&  \notag \\
&&
\begin{array}{l}
W\left( t\right) =\int_{CY_{4}}\frak{F}_{4}\wedge \Omega _{4}\left( t\right)
;
\end{array}
\label{FV-WW} \\
&&  \notag \\
&&
\begin{array}{l}
Z\left( t,\overline{t}\right) =e^{\frac{1}{2}K\left( t,\overline{t}\right)
}\int_{CY_{4}}\frak{F}_{4}\wedge \Omega _{4}\left( t\right) =\frac{%
\int_{CY_{4}}\frak{F}_{4}\wedge \Omega _{4}\left( t\right) }{\sqrt{%
\int_{CY_{4}}\Omega _{4}\left( t\right) \wedge \overline{\Omega }_{4}\left(
\overline{t}\right) }}=\int_{CY_{4}}\frak{F}_{4}\wedge \hat{\Omega}%
_{4}\left( t,\overline{t}\right) ,
\end{array}
\label{FV-ZZ}
\end{eqnarray}
where in Eq. (\ref{FV-ZZ}) we defined the covariantly holomorphic 4-form on $%
CY_{4}$:
\begin{eqnarray}
&&
\begin{array}{l}
\hat{\Omega}_{4}\left( t,\overline{t}\right) \equiv e^{\frac{1}{2}K\left( t,%
\overline{t}\right) }\Omega _{4}\left( t\right) =e^{\frac{1}{2}K_{1}\left(
t^{0},\overline{t}^{\overline{0}}\right) }e^{\frac{1}{2}K_{3}\left( x,%
\overline{x}\right) }\Omega _{1}\left( t^{0}\right) \wedge \Omega _{3}\left(
x\right) =\hat{\Omega}_{1}\left( t^{0},\overline{t}^{\overline{0}}\right)
\wedge \hat{\Omega}_{3}\left( x,\overline{x}\right) ; \\
\\
\hat{\Omega}_{1}\left( t^{0},\overline{t}^{\overline{0}}\right) \equiv e^{%
\frac{1}{2}K_{1}\left( t^{0},\overline{t}^{\overline{0}}\right) }\Omega
_{1}\left( t^{0}\right) ;
\end{array}
\notag \\
&&  \label{Omega4-hat}
\end{eqnarray}
$\hat{\Omega}_{3}\left( x,\overline{x}\right) $ is the covariantly
holomorphic 3-form on $CY_{3}$, defined by Eq. (\ref{Omega3-hat}); it has
K\"{a}hler weights $\left( 1,-1\right) $ with respect to the K\"{a}hler
potential $K_{3}\left( x,\overline{x}\right) $ of the SK CS moduli space $%
\mathcal{M}_{CS}$ of $CY_{3}$:
\begin{eqnarray}
&&
\begin{array}{l}
D_{i}\hat{\Omega}_{3}=\left\{
\begin{array}{l}
\partial _{i}\hat{\Omega}_{3}+\frac{1}{2}\left( \partial _{i}K_{3}\right)
\hat{\Omega}_{3}=e^{\frac{1}{2}K_{3}}D_{i}\Omega _{3}= \\
\\
=\frac{1}{\sqrt{i\left( \overline{X}^{\Delta }F_{\Delta }-X^{\Delta }%
\overline{F}_{\Delta }\right) }}\left\{
\begin{array}{l}
\left[ \partial _{i}X^{\Lambda }-\frac{\left( \overline{X}^{\Sigma }\partial
_{i}F_{\Sigma }-\left( \partial _{i}X^{\Sigma }\right) \overline{F}_{\Sigma
}\right) }{\overline{X}^{\Xi }F_{\Xi }-X^{\Xi }\overline{F}_{\Xi }}%
X^{\Lambda }\right] \alpha _{\Lambda }+ \\
\\
-\left[ \partial _{i}F_{\Lambda }-\frac{\left( \overline{X}^{\Sigma
}\partial _{i}F_{\Sigma }-\left( \partial _{i}X^{\Sigma }\right) \overline{F}%
_{\Sigma }\right) }{\overline{X}^{\Xi }F_{\Xi }-X^{\Xi }\overline{F}_{\Xi }}%
F_{\Lambda }\right] \beta ^{\Lambda }
\end{array}
\right\} ;
\end{array}
\right. \\
\\
\\
\overline{D}_{\overline{i}}\hat{\Omega}_{3}=\overline{\partial }_{\overline{i%
}}\hat{\Omega}_{3}-\frac{1}{2}\left( \overline{\partial }_{\overline{i}%
}K_{3}\right) \hat{\Omega}_{3}=0; \\
\\
D_{i}D_{j}\hat{\Omega}_{3}=iC_{ijk}g^{k\overline{l}}\overline{D}_{\overline{l%
}}\overline{\hat{\Omega}}_{3};~~\overline{D}_{\overline{i}}D_{j}\hat{\Omega}%
_{3}=g_{j\overline{i}}\hat{\Omega}_{3}; \\
\\
D_{0}\hat{\Omega}_{3}=0;~~\overline{D}_{\overline{0}}\hat{\Omega}_{3}=0.
\end{array}
\notag \\
&&
\end{eqnarray}
On the other hand, $\hat{\Omega}_{1}\left( t^{0},\overline{t}^{\overline{0}%
}\right) $ is the covariantly holomorphic 1-form on $T^{2}$, defined by the
second line of Eq. (\ref{Omega4-hat}); it has K\"{a}hler weights $\left(
1,-1\right) $ with respect to the K\"{a}hler potential $K_{1}\left( t^{0},%
\overline{t}^{\overline{0}}\right) $ of the K\"{a}hler $1$-dim. moduli space
$\mathcal{M}_{t^{0}}$ of $T^{2}$, spanned by the universal axion-dilaton $%
\tau \equiv t^{0}$:
\begin{eqnarray}
&&
\begin{array}{l}
\begin{array}{c}
D_{0}\hat{\Omega}_{1}=\partial _{0}\hat{\Omega}_{1}+\frac{1}{2}\left(
\partial _{0}K_{1}\right) \hat{\Omega}_{1}=e^{\frac{1}{2}K_{1}}D_{0}\Omega
_{1}=ie^{K_{1}}\overline{\hat{\Omega}}_{1}=\left( \overline{t}^{\overline{0}%
}-t^{0}\right) ^{-1}\overline{\hat{\Omega}}_{1}; \\
\Updownarrow \\
\overline{\hat{\Omega}}_{1}=-ie^{-K_{1}}D_{0}\hat{\Omega}_{1}\Leftrightarrow
\hat{\Omega}_{1}=ie^{-K_{1}}\overline{D}_{\overline{0}}\overline{\hat{\Omega}%
}_{1}\Leftrightarrow \overline{D}_{\overline{0}}\overline{\hat{\Omega}}%
_{1}=-ie^{K_{1}}\hat{\Omega}_{1};
\end{array}
\\
\\
\\
\overline{D}_{\overline{0}}\hat{\Omega}_{1}=\overline{\partial }_{\overline{0%
}}\hat{\Omega}_{1}-\frac{1}{2}\left( \overline{\partial }_{\overline{0}%
}K_{1}\right) \hat{\Omega}_{1}=0; \\
\\
D_{0}D_{0}\hat{\Omega}_{1}=0;~~\overline{D}_{\overline{0}}D_{0}\hat{\Omega}%
_{1}=g_{0\overline{0}}\hat{\Omega}_{1}=e^{2K_{1}}\hat{\Omega}_{1}=-\left(
\overline{t}^{\overline{0}}-t^{0}\right) ^{-2}\hat{\Omega}_{1}; \\
\\
D_{i}\hat{\Omega}_{1}=0;~~\overline{D}_{\overline{i}}\hat{\Omega}_{1}=0.
\end{array}
\notag \\
&&
\end{eqnarray}
Resultingly, the covariantly holomorphic 4-form $\hat{\Omega}_{4}\left( t,%
\overline{t}\right) $ on $CY_{4}=\frac{CY_{3}\times T^{2}}{\mathbb{Z}_{2}}$,
defined by the first line of Eq. (\ref{Omega4-hat}), has K\"{a}hler weights $%
\left( 1,-1\right) $ with respect to the whole K\"{a}hler potential $K\left(
t,\overline{t}\right) =K_{1}\left( t^{0},\overline{t}^{\overline{0}}\right)
+K_{3}\left( x,\overline{x}\right) $ of the $\left( h_{2,1}+1\right) $-dim.
moduli space $M=\mathcal{M}_{t^{0}}\otimes \mathcal{M}_{CS}.$ of $CY_{4}$
(recall Eq. (\ref{MM})):
\begin{eqnarray}
&&
\begin{array}{l}
D_{a}\hat{\Omega}_{4}\left( t,\overline{t}\right) =\partial _{a}\hat{\Omega}%
_{4}\left( t,\overline{t}\right) +\frac{1}{2}\left( \partial _{a}K\left( t,%
\overline{t}\right) \right) \hat{\Omega}_{4}\left( t,\overline{t}\right) ;
\\
\\
\overline{D}_{\overline{a}}\hat{\Omega}_{4}\left( t,\overline{t}\right) =%
\overline{\partial }_{\overline{a}}\hat{\Omega}_{4}\left( t,\overline{t}%
\right) -\frac{1}{2}\left( \overline{\partial }_{\overline{a}}K\left( t,%
\overline{t}\right) \right) \hat{\Omega}_{4}\left( t,\overline{t}\right) =0,
\end{array}
\notag \\
&&  \label{25apr1}
\end{eqnarray}
implying that
\begin{equation}
D_{b}\overline{D}_{\overline{a}}\hat{\Omega}_{4}\left( t,\overline{t}\right)
=g_{b\overline{a}}\hat{\Omega}_{4}\left( t,\overline{t}\right) .
\label{25apr2}
\end{equation}
\setcounter{equation}0
\def\theequation{4.1.\arabic{subsubsection}.\arabic{equation}}
\subsubsection{\label{FV-Hodge-Decomposition}Hodge Decomposition of $\frak{F}%
_{4}$}

Now, in order to derive the Hodge-decomposition\footnote{%
For an elegant and detailed derivation of the Hodge-decomposition of $\frak{F%
}_{4}$ using methods of algebraic geometry, see \textit{e.g.} Sect. 2 of
\cite{Soroush:2007ed}.} of the real flux 4-form $\frak{F}_{4}$, we have to
determine all the possible independent 4-forms on $CY_{4}$($=\frac{%
CY_{3}\times T^{2}}{\mathbb{Z}_{2}}$, as always understood throughout). Due
to Eqs. (\ref{25apr1}) and (\ref{25apr2}), it is easy to realize that, up to
the third order of covariant differentiation included, the possible
independent 4-forms ($\left( 1,-1\right) $-K\"{a}hler weighted with respect
to $K$) on $CY_{4}$ are $\hat{\Omega}_{4}$, $D_{a}\hat{\Omega}_{4}$, $%
D_{a}D_{b}\hat{\Omega}_{4}$, $D_{a}D_{b}D_{c}\hat{\Omega}_{4}$ and $%
\overline{D}_{\overline{a}}D_{b}D_{c}\hat{\Omega}_{4}$.

As it can be realized by considering Eqs. (\ref{DaOmega4})-(\ref
{DabDbDcOmega4}) of Appendix I, $D_{a}D_{b}\hat{\Omega}_{4}$ cannot be
expressed in terms of $D_{a}\hat{\Omega}_{4}$ (as instead it happens in the
extremal BH case, see Eq. (\ref{2})), and all the independent, $\left(
1,-1\right) $-K\"{a}hler-weighted 4-forms on the considered $CY_{4}$ are
given by the $2h_{2,1}+2$ forms
\begin{equation}
\hat{\Omega}_{4},~D_{0}\hat{\Omega}_{4},~D_{i}\hat{\Omega}_{4},~D_{0}D_{i}%
\hat{\Omega}_{4}.  \label{25apr-1952}
\end{equation}
The third covariant derivatives of $\hat{\Omega}_{4}$ do not add any other
independent 4-form, and so do all the other higher order covariant
derivatives of $\hat{\Omega}_{4}$. Thus, the possible candidates along which
one might decompose the real flux 4-form $\frak{F}_{4}$ are the 4-forms
given by Eq. (\ref{25apr-1952}) and their complex conjugated $\overline{\hat{%
\Omega}}_{4},~\overline{D}_{\overline{0}}\overline{\hat{\Omega}}_{4},~%
\overline{D}_{\overline{i}}\overline{\hat{\Omega}}_{4},~\overline{D}_{%
\overline{0}}\overline{D}_{\overline{i}}\overline{\hat{\Omega}}_{4}$.

The \textit{``intersections''} among the elements of the set of 4-forms $%
\hat{\Omega}_{4}$,$~D_{0}\hat{\Omega}_{4}$,$~D_{i}\hat{\Omega}_{4}$,$%
~D_{0}D_{i}\hat{\Omega}_{4}$, $\overline{\hat{\Omega}}_{4}$,$~\overline{D}_{%
\overline{0}}\overline{\hat{\Omega}}_{4}$,$~\overline{D}_{\overline{i}}%
\overline{\hat{\Omega}}_{4}$ and$~\overline{D}_{\overline{0}}\overline{D}_{%
\overline{i}}\overline{\hat{\Omega}}_{4}$ in generic local ``curved'' and in
local ``flat'' coordinates of $M$ are given in Appendix II. By using such
results, the real, K\"{a}hler gauge-invariant 4-form $\frak{F}_{4}$ can be
thus Hodge-decomposed as follows ($\eta _{1},...,\eta _{6}\in \mathbb{C}$)
\begin{eqnarray}
\frak{F}_{4} &=&\left[
\begin{array}{l}
\eta _{1}\left( \int_{CY_{4}}\frak{F}_{4}\wedge \hat{\Omega}_{4}\right)
\overline{\hat{\Omega}}_{4}+\eta _{2}\delta ^{A\overline{B}}\left(
\int_{CY_{4}}\frak{F}_{4}\wedge \left( D_{A}\hat{\Omega}_{4}\right) \right)
\overline{D}_{\overline{B}}\overline{\hat{\Omega}}_{4}+ \\
\\
+\eta _{3}\delta ^{A\overline{B}}\left( \int_{CY_{4}}\frak{F}_{4}\wedge D_{%
\underline{0}}D_{A}\hat{\Omega}_{4}\right) \overline{D}_{\underline{%
\overline{0}}}\overline{D}_{\overline{B}}\overline{\hat{\Omega}}_{4}+\eta
_{4}\delta ^{B\overline{A}}\left( \int_{CY_{4}}\frak{F}_{4}\wedge \overline{D%
}_{\underline{\overline{0}}}\overline{D}_{\overline{A}}\overline{\hat{\Omega}%
}_{4}\right) D_{\underline{0}}D_{B}\hat{\Omega}_{4}+ \\
\\
+\eta _{5}\delta ^{B\overline{A}}\left( \int_{CY_{4}}\frak{F}_{4}\wedge
\left( \overline{D}_{\overline{A}}\overline{\hat{\Omega}}_{4}\right) \right)
D_{B}\hat{\Omega}_{4}+\eta _{6}\left( \int_{CY_{4}}\frak{F}_{4}\wedge
\overline{\hat{\Omega}}_{4}\right) \hat{\Omega}_{4}
\end{array}
\right] =  \notag \\
&&  \notag \\
&& \\
&=&\left[
\begin{array}{l}
\eta _{1}Z\overline{\hat{\Omega}}_{4}+\eta _{2}\delta ^{A\overline{B}}\left(
D_{A}Z\right) \overline{D}_{\overline{B}}\overline{\hat{\Omega}}_{4}+\eta
_{3}\delta ^{A\overline{B}}\left( D_{\underline{0}}D_{A}Z\right) \overline{D}%
_{\underline{\overline{0}}}\overline{D}_{\overline{B}}\overline{\hat{\Omega}}%
_{4}+ \\
\\
+\eta _{4}\delta ^{B\overline{A}}\left( \overline{D}_{\underline{\overline{0}%
}}\overline{D}_{\overline{A}}\overline{Z}\right) D_{\underline{0}}D_{B}\hat{%
\Omega}_{4}+\eta _{5}\delta ^{B\overline{A}}\left( \overline{D}_{\overline{A}%
}\overline{Z}\right) D_{B}\hat{\Omega}_{4}+\eta _{6}\overline{Z}\hat{\Omega}%
_{4}
\end{array}
\right] ,  \label{FV-decomp1}
\end{eqnarray}
\textbf{\ }where Eq. (\ref{FV-ZZ}) was used, also implying:
\begin{eqnarray}
&&
\begin{array}{l}
\int_{CY_{4}}\frak{F}_{4}\wedge D_{a}\hat{\Omega}_{4}=D_{a}Z,~~\int_{CY_{4}}%
\frak{F}_{4}\wedge D_{a}D_{b}\hat{\Omega}_{4}=D_{a}D_{b}Z; \\
\\
\int_{CY_{4}}\frak{F}_{4}\wedge D_{A}\hat{\Omega}_{4}=D_{A}Z,~~\int_{CY_{4}}%
\frak{F}_{4}\wedge D_{A}D_{B}\hat{\Omega}_{4}=D_{A}D_{B}Z.
\end{array}
\notag \\
&&  \label{FV-ZZZ}
\end{eqnarray}
The r.h.s. of the Hodge-decomposition (\ref{FV-decomp1}) is the most general
K\"{a}hler gauge-invariant combination of \textit{all} the possible ($\left(
1,-1\right) $ and $\left( -1,1\right) $)-K\"{a}hler-weighted independent
4-forms for Type IIB on $CY_{4}=\frac{CY_{3}\times T^{2}}{\mathbb{Z}_{2}}$.
The reality condition $\overline{\frak{F}}_{4}=$ $\frak{F}_{4}$ implies $%
\eta _{4}=\overline{\eta _{3}}$, $\eta _{5}=\overline{\eta _{2}}$ and $\eta
_{6}=\overline{\eta _{1}}$. The (a priori) complex coefficients $\eta _{1}$,
$\eta _{2}$ and $\eta _{3}$ can be determined by computing $\int_{CY_{4}}%
\frak{F}_{4}\wedge \hat{\Omega}_{4}$, $\int_{CY_{4}}\frak{F}_{4}\wedge D_{A}%
\hat{\Omega}_{4}$ and $\int_{CY_{4}}\frak{F}_{4}\wedge D_{\underline{0}}D_{A}%
\hat{\Omega}_{4}$, and using the identity (\ref{FV-decomp1}) and recalling
Eqs. (\ref{FV-ZZ}), (\ref{FV-ZZZ}) and the \textit{``intersections''} in
local ``flat'' coordinates (\ref{flat-intersect1})-(\ref{flat-intersect4}).
By doing so, one obtains:
\begin{eqnarray}
&&
\begin{array}{l}
Z=\int_{CY_{4}}\frak{F}_{4}\wedge \hat{\Omega}_{4}=\eta _{1}Z\int_{CY_{4}}%
\overline{\hat{\Omega}}_{4}\wedge \hat{\Omega}_{4}=\eta _{1}Z\Leftrightarrow
\eta _{1}=1; \\
\\
\\
\begin{array}{l}
D_{C}Z=\int_{CY_{4}}\frak{F}_{4}\wedge D_{C}\hat{\Omega}_{4}=\eta _{2}\delta
^{A\overline{B}}\left( D_{A}Z\right) \int_{CY_{4}}\left( \overline{D}_{%
\overline{B}}\overline{\hat{\Omega}}_{4}\right) \wedge D_{C}\hat{\Omega}_{4}=
\\
\\
=-\eta _{2}\delta ^{A\overline{B}}D_{A}Z\delta _{C\overline{B}}=-\eta
_{2}D_{C}Z\Leftrightarrow \eta _{2}=-1;
\end{array}
\\
\\
\\
\begin{array}{l}
D_{\underline{0}}D_{C}Z=\int_{CY_{4}}\frak{F}_{4}\wedge D_{\underline{0}%
}D_{C}\hat{\Omega}_{4}=\eta _{3}\delta ^{A\overline{B}}\left( D_{\underline{0%
}}D_{A}Z\right) \int_{CY_{4}}\left( \overline{D}_{\underline{\overline{0}}}%
\overline{D}_{\overline{B}}\overline{\hat{\Omega}}_{4}\right) \wedge D_{%
\underline{0}}D_{C}\hat{\Omega}_{4}= \\
\\
=\eta _{3}\delta ^{A\overline{B}}\delta _{C\overline{B}}D_{\underline{0}%
}D_{A}Z=\eta _{3}D_{\underline{0}}D_{C}Z\Leftrightarrow \eta _{3}=1.
\end{array}
\end{array}
\notag \\
&&
\end{eqnarray}
Thus, the complete Hodge-decomposition of the real, K\"{a}hler
gauge-invariant 4-form $\frak{F}_{4}$ of Type IIB on $\frac{CY_{3}\times
T^{2}}{\mathbb{Z}_{2}}$ in generic local ``flat'' coordinates\footnote{%
For the analogous expression in generic local ``curved'' coordinates in $M$,
see Eqs. (\ref{curved-FV-decomp2})-(\ref{curved-FV-decomp5}).} in $M$ reads
\begin{eqnarray}
\frak{F}_{4} &=&2Re\left[ \overline{Z}\hat{\Omega}_{4}-\left( \overline{D}%
^{A}\overline{Z}\right) D_{A}\hat{\Omega}_{4}+\left( \overline{D}^{%
\underline{0}}\overline{D}^{A}\overline{Z}\right) D_{\underline{0}}D_{A}\hat{%
\Omega}_{4}\right] =  \label{FV-decomp2} \\
&&  \notag \\
&=&2Re\left[
\begin{array}{l}
\overline{Z}\hat{\Omega}_{1}\wedge \hat{\Omega}_{3}-\left( \overline{D}_{%
\overline{\underline{0}}}\overline{Z}\right) \overline{\hat{\Omega}}%
_{1}\wedge \hat{\Omega}_{3}-\left( \overline{D}^{I}\overline{Z}\right) \hat{%
\Omega}_{1}\wedge D_{I}\hat{\Omega}_{3}+ \\
\\
+\left( \overline{D}^{\underline{0}}\overline{D}^{I}\overline{Z}\right)
\overline{\hat{\Omega}}_{1}\wedge D_{I}\hat{\Omega}_{3}
\end{array}
\right] .  \label{FV-decomp3}
\end{eqnarray}
\setcounter{equation}0
\def\theequation{4.\arabic{subsection}.\arabic{equation}}
\subsection{\label{FV-Criticality-Conditions-approach}$\mathcal{N}=1$, $d=4$
Effective Potential and ``\textit{Criticality Conditions'' Approach}}

The potential of $\mathcal{N}=1$, $d=4$ supergravity (from Type IIB on $%
\frac{CY_{3}\times T^{2}}{\mathbb{Z}_{2}}$), which acts as effective
potential for the FV attractors, is given by Eq. (\ref{14-Nov-1}), which we
repeat here \cite{Bagger-Witten,Cremmer-et-al}:
\begin{eqnarray}
&&
\begin{array}{l}
V_{\mathcal{N}=1}=e^{K}\left[ -3\left| W\right| ^{2}+g^{a\overline{b}}D_{a}W%
\overline{D}_{\overline{b}}\overline{W}\right] =-3\left| Z\right| ^{2}+g^{a%
\overline{b}}D_{a}Z\overline{D}_{\overline{b}}\overline{Z}= \\
\\
=-3\left| Z\right| ^{2}-\left( \overline{t}^{\overline{0}}-t^{0}\right)
^{2}\left| D_{0}Z\right| ^{2}+g^{i\overline{j}}\left( t^{k},\overline{t}^{%
\overline{k}}\right) D_{i}Z\overline{D}_{\overline{j}}\overline{Z}\gtreqless
0.
\end{array}
\notag \\
&&  \label{V-N=1}
\end{eqnarray}
At a glance, the first difference between the ``BH effective potential'' $%
V_{BH}$ given by Eq. (\ref{VBH1}) and the ``FV effective potential'' $V_{%
\mathcal{N}=1}$ given by Eq. (\ref{V-N=1}) concerns their sign. Indeed, $%
V_{BH}$ is positive-definite and it can be recognized as the first,
quadratic invariant of SK geometry; through the Bekenstein-Hawking
entropy-area formula, it is related to the classical entropy and to the area
of the event horizon of the considered extremal (static, spherically
symmetric, asymptotically flat) BH. On the other hand, $V_{\mathcal{N}=1}$
does not have a definite sign, and critical points of $V_{\mathcal{N}=1}$
can exist with $V_{\mathcal{N}=1}\gtreqless 0$:

1) $\left. V_{\mathcal{N}=1}\right| _{\partial V_{\mathcal{N}=1}=0}>0$
corresponds to De Sitter (dS) vacua;

2) $\left. V_{\mathcal{N}=1}\right| _{\partial V_{\mathcal{N}=1}=0}=0$
determines Minkowski vacua;

3) $\left. V_{\mathcal{N}=1}\right| _{\partial V_{\mathcal{N}=1}=0}<0$
corresponds to anti De Sitter (AdS) vacua.

By differentiating Eq. (\ref{V-N=1}) with respect to the moduli and
recalling Eqs. (\ref{FV-WW}) and (\ref{25apr2}), one obtains the general
criticality conditions of $V_{\mathcal{N}=1}$ ($\forall a=0,1,...,h_{2,1}$):
\begin{eqnarray}
&&
\begin{array}{c}
D_{a}V_{\mathcal{N}=1}=\partial _{a}V_{\mathcal{N}=1}=0; \\
\\
\Updownarrow \\
\\
\begin{array}{l}
e^{K}\left[ -3\overline{W}D_{a}W+g^{b\overline{c}}\left( D_{a}D_{b}W\right)
\overline{D}_{\overline{c}}\overline{W}+g^{b\overline{c}}\left(
D_{b}W\right) D_{a}\overline{D}_{\overline{c}}\overline{W}\right] = \\
\\
=e^{K}\left[ -2\overline{W}D_{a}W+g^{b\overline{c}}\left( D_{a}D_{b}W\right)
\overline{D}_{\overline{c}}\overline{W}\right] =0;
\end{array}
\\
\\
\Updownarrow \\
\\
-2\overline{W}D_{a}W+g^{b\overline{c}}\left( D_{a}D_{b}W\right) \overline{D}%
_{\overline{c}}\overline{W}=0,
\end{array}
\notag \\
&&  \label{FV-AEs1}
\end{eqnarray}
where, as in the case of extremal BH attractors in $\mathcal{N}=2$, $d=4$
supergravity, we assumed the K\"{a}hler potential to be regular, \textit{i.e.%
} that $\left| K\right| <\infty $ globally in $M$ (or \textit{at least} at
the critical points of $V_{\mathcal{N}=1}$). Eqs. (\ref{FV-AEs1}) are the
what one should rigorously refer to as the $\mathcal{N}=1$, $d=4$ FV AEs (in
Type IIB on $\frac{CY_{3}\times T^{2}}{\mathbb{Z}_{2}}$). By recalling Eq. (%
\ref{g-M-inv}), they can finally be rewritten as
\begin{eqnarray}
D_{a}V_{\mathcal{N}=1} &=&\partial _{a}V_{\mathcal{N}=1}=0\Leftrightarrow -2%
\overline{W}D_{a}W+g^{0\overline{0}}\left( D_{a}D_{0}W\right) \overline{D}_{%
\overline{0}}\overline{W}+g^{i\overline{j}}\left( D_{a}D_{i}W\right)
\overline{D}_{\overline{j}}\overline{W}=0,\forall a=0,1,...,h_{2,1}.  \notag
\\
&&  \label{FV-AEs2}
\end{eqnarray}
Let us specify such FV AEs for the two classes of (local ``curved'')
indices; with some elaborations, one obtains:
\begin{eqnarray}
&&
\begin{array}{l}
a=0\text{~(axion-dilaton direction in }M\text{)}: \\
\\
D_{0}V_{\mathcal{N}=1}=\partial _{0}V_{\mathcal{N}=1}=0\Leftrightarrow -2%
\overline{W}D_{0}W+g^{j\overline{k}}\left( D_{0}D_{j}W\right) \overline{D}_{%
\overline{k}}\overline{W}=0;
\end{array}
\label{FV-AEs2-0} \\
&&  \notag \\
&&  \notag \\
&&
\begin{array}{l}
a=i\text{~}\in \left\{ 1,...,h_{2,1}\right\} ~\text{(CS directions in }M%
\text{)}: \\
\\
\begin{array}{c}
D_{i}V_{\mathcal{N}=1}=\partial _{i}V_{\mathcal{N}=1}=0 \\
\\
\Updownarrow \\
\\
-2\overline{W}D_{i}W+g^{0\overline{0}}\left( D_{i}D_{0}W\right) \overline{D}%
_{\overline{0}}\overline{W}+g^{j\overline{k}}\left( D_{i}D_{j}W\right)
\overline{D}_{\overline{k}}\overline{W}=0; \\
\\
\Updownarrow \\
\\
-2\overline{W}D_{i}W+e^{-2K_{1}}\left( D_{0}D_{i}W\right) \overline{D}_{%
\overline{0}}\overline{W}-e^{-K_{1}}g^{j\overline{k}}C_{ijl}g^{l\overline{m}%
}\left( \overline{D}_{\overline{0}}\overline{D}_{\overline{m}}\overline{W}%
\right) \overline{D}_{\overline{k}}\overline{W}=0.
\end{array}
\end{array}
\notag \\
&&  \label{FV-AEs2-i}
\end{eqnarray}
Thus, despite the presence of the universal axion-dilaton direction in the $%
\left( h_{2,1}+1\right) $-dim. K\"{a}hler moduli space $M$, Eqs. (\ref
{FV-AEs2-i}) yields that the tensor $C_{ijk}$, defined in the $h_{2,1}$-dim.
SK CS moduli space $\mathcal{M}_{CS}\subsetneq M$, still plays a key role.
The FV AEs (\ref{FV-AEs2-0}) and (\ref{FV-AEs2-i}) of $\mathcal{N}=1$, $d=4$
supergravity from Type IIB on $\frac{CY_{3}\times T^{2}}{\mathbb{Z}_{2}}$
relate, at the critical points of the ``FV effective potential'' $V_{%
\mathcal{N}=1}$ (given by Eq. (\ref{V-N=1})), the $\mathcal{N}=1$, $d=4$
holomorphic superpotential $W$, the \textit{supersymmetry order parameters }$%
D_{i}Z=e^{\frac{1}{2}K}D_{i}W$ and the \textit{axino-dilatino-CS modulino
mixings }$D_{0}D_{i}Z=e^{\frac{1}{2}K}D_{0}D_{i}W$, which is part of the $%
\left( h_{2,1}+1\right) \times \left( h_{2,1}+1\right) $ \textit{modulino
mass matrix} $\Lambda _{ab}\equiv D_{a}D_{b}Z=e^{\frac{1}{2}K}D_{a}D_{b}W$
(note that in the considered $\mathcal{N}=1$, $d=4$ framework the
axino-dilatino and the $h_{2,1}$ CS modulinos play the role of the $%
n_{V}=h_{2,1}$ CS modulinos in the context of $\mathcal{N}=2$, $d=4$
supergravity from Type IIB on $CY_{3}$). It is worth pointing out that $%
\Lambda _{ab}$ is part of the holomorphic/anti-holomorphic form of the $%
\left( 2h_{2,1}+2\right) \times \left( 2h_{2,1}+2\right) $ covariant Hessian
of $Z$, which is nothing but the holomorphic/anti-holomorphic form of the
scalar (axion-dilaton + CS moduli, in the stringy description as Type IIB on
$\frac{CY_{3}\times T^{2}}{\mathbb{Z}_{2}}$) mass matrix.

The structure of the criticality conditions (\ref{FV-AEs2}) and (\ref
{FV-AEs2-0})-(\ref{FV-AEs2-i}) suggests the classification of the critical
points of $V_{\mathcal{N}=1}$ in two general classes:

I) The \textit{supersymmetric (SUSY)} critical points of $V_{\mathcal{N}=1}$%
, determined by the differential constraints
\begin{equation}
D_{a}W=0,\forall a=0,1,...,h_{2,1},  \label{FV-SUSY}
\end{equation}
which directly solve the conditions (\ref{FV-AEs2}) and (\ref{FV-AEs2-0})-(%
\ref{FV-AEs2-i}). By substituting the SUSY FV constraints (\ref{FV-SUSY})
into the expression (\ref{V-N=1}) of $V_{\mathcal{N}=1}$, one obtains that
SUSY dS critical points of $V_{\mathcal{N}=1}$ (\textit{i.e.}, independently
on the stability, SUSY dS FV described by a classical FV Attractor Mechanism
encoded by - the criticality conditions of - the potential $V_{\mathcal{N}%
=1} $) cannot exist, because
\begin{equation}
V_{\mathcal{N}=1,SUSY}=-3\left( e^{K}\left| W\right| ^{2}\right)
_{SUSY}=-3\left| Z\right| _{SUSY}^{2}\leqslant 0.
\end{equation}

II) The \textit{non-supersymmetric (non-SUSY)} critical points of $V_{%
\mathcal{N}=1}$, determined by the differential constraints
\begin{equation}
\left\{
\begin{array}{l}
D_{a}W\neq 0,\text{ (\textit{at least}) for some }a\in \left\{
0,1,...,h_{2,1}\right\} ; \\
\\
\partial _{a}V_{\mathcal{N}=1}=0,\forall a=0,1,...,h_{2,1}.
\end{array}
\right.  \label{FV-non-SUSY}
\end{equation}
The expression (\ref{V-N=1}) of $V_{\mathcal{N}=1}$ suggests that \textit{a
priori} such critical points of $V_{\mathcal{N}=1}$ are of all possible
species (dS, Minkowski, AdS). \setcounter{equation}0
\def\theequation{4.\arabic{subsection}.\arabic{equation}}
\subsection{\label{SUSY-Flux-Vacua-Attractors}Supersymmetric Flux Vacua
Attractor Equations}

In the present Subsection we will concentrate on the SUSY critical points of
$V_{\mathcal{N}=1}$, determining the supersymmetric FV AEs in $\mathcal{N}=1$%
, $d=4$ supergravity from Type IIB on $\frac{CY_{3}\times T^{2}}{\mathbb{Z}%
_{2}}$. This can be achieved respectively by evaluating the Hodge identities
(\ref{FV-decomp2})-(\ref{FV-decomp3}) and (\ref{curved-FV-decomp2})-(\ref
{curved-FV-decomp5}) at the SUSY FV constraints (\ref{FV-SUSY}).

The evaluation of the identities (\ref{FV-decomp2})-(\ref{FV-decomp3}) and (%
\ref{curved-FV-decomp2})-(\ref{curved-FV-decomp5}) along the constraints (%
\ref{FV-SUSY}) respectively yields the supersymmetric FV AEs in $\mathcal{N}%
=1$, $d=4$ supergravity from Type IIB on $\frac{CY_{3}\times T^{2}}{\mathbb{Z%
}_{2}}$ in local ``flat'' coordinates\footnote{%
For the analogous expression in generic local ``curved'' coordinates in $M$,
see Eq. (\ref{SUSY-FV-AEs2}).}:
\begin{eqnarray}
&&
\begin{array}{l}
\frak{F}_{4}=2Re\left[ \overline{Z}\hat{\Omega}_{4}+\delta ^{A\overline{B}%
}\left( \overline{D}_{\underline{\overline{0}}}\overline{D}_{\overline{B}}%
\overline{Z}\right) D_{\underline{0}}D_{A}\hat{\Omega}_{4}\right] _{SUSY}=
\\
\\
=2Re\left[ \overline{Z}\hat{\Omega}_{1}\wedge \hat{\Omega}_{3}+\delta ^{I%
\overline{J}}\left( \overline{D}_{\underline{\overline{0}}}\overline{D}_{%
\overline{J}}\overline{Z}\right) \overline{\hat{\Omega}}_{1}\wedge D_{I}\hat{%
\Omega}_{3}\right] _{SUSY}= \\
\\
=2e^{K_{1}+K_{3}}Re\left[ \overline{W}\Omega _{1}\wedge \Omega _{3}+\delta
^{I\overline{J}}\left( \overline{D}_{\underline{\overline{0}}}\overline{D}_{%
\overline{J}}\overline{W}\right) \overline{\Omega }_{1}\wedge D_{I}\Omega
_{3}\right] _{SUSY}.
\end{array}
\notag \\
&&  \label{SUSY-FV-AEs1}
\end{eqnarray}

Notice that, as in Eqs. (\ref{SUSY-FV-AEs1}) as well as in the treatment
below, the subscript ``$SUSY$'' denotes the evaluation at the SUSY FV
constraints (\ref{FV-SUSY}).

Furthermore, Eqs. (\ref{SUSY-FV-AEs1}) can be further elaborated by
computing that
\begin{eqnarray}
&&
\begin{array}{l}
\left( D_{\underline{0}}D_{J}W\right) _{SUSY}=\left( e_{J}^{j}\partial
_{j}D_{\underline{0}}W\right) _{SUSY}=\left( e_{J}^{j}e_{\underline{0}%
}^{0}\partial _{j}D_{0}W\right) _{SUSY}=\left( e_{J}^{j}e_{\underline{0}%
}^{0}\partial _{0}D_{j}W\right) _{SUSY}= \\
\\
=\left\{ e_{J}^{j}\left( \overline{\tau }-\tau \right) \left[ \partial
_{j}\partial _{0}W+\frac{1}{2}\left( \partial _{j}K_{3}\right) \partial _{0}W%
\right] \right\} _{SUSY}= \\
\\
=\left\{ e_{J}^{j}\left( \overline{\tau }-\tau \right) \left[ -q_{h\mid
\Lambda }\partial _{j}X^{\Lambda }+p_{h}^{\Lambda }\partial _{j}F_{\Lambda }+%
\frac{1}{2}\frac{\left( \overline{X}^{\Sigma }\partial _{j}F_{\Sigma }-%
\overline{F}_{\Sigma }\partial _{j}X^{\Sigma }\right) }{\overline{X}^{\Delta
}F_{\Delta }-\overline{F}_{\Delta }X^{\Delta }}\left( q_{h\mid \Xi }X^{\Xi
}-p_{h}^{\Xi }F_{\Xi }\right) \right] \right\} _{SUSY}.
\end{array}
\notag \\
&&  \label{D0DJW-SUSY}
\end{eqnarray}

The structure of the SUSY FV AEs (\ref{SUSY-FV-AEs1}) suggests the
classification of the SUSY critical points of $V_{\mathcal{N}=1}$ in three
general classes:

I) \textit{Type ``(3,0)''} SUSY FV, determined by the constraints (\ref
{FV-SUSY}) and by the further conditions
\begin{equation}
\left\{
\begin{array}{l}
W_{SUSY}\neq 0; \\
\\
\begin{array}{c}
\left[ g^{i\overline{j}}\left( \overline{D}_{\overline{0}}\overline{D}_{%
\overline{j}}\overline{W}\right) \overline{\Omega }_{1}\wedge D_{i}\Omega
_{3}\right] _{SUSY}=0; \\
\\
\Updownarrow \\
\\
\forall \Lambda =1,...,h_{2,1}+1:\left\{
\begin{array}{l}
\left[ g^{i\overline{j}}\left( \overline{D}_{\overline{0}}\overline{D}_{%
\overline{j}}\overline{W}\right) D_{i}X^{\Lambda }\right] _{SUSY}=0; \\
\\
\left[ g^{i\overline{j}}\left( \overline{D}_{\overline{0}}\overline{D}_{%
\overline{j}}\overline{W}\right) D_{i}F_{\Lambda }\right] _{SUSY}=0.
\end{array}
\right.
\end{array}
\end{array}
\right.  \label{FV-SUSY-1}
\end{equation}

Because of
\begin{equation}
V_{\mathcal{N}=1,SUSY,(3,0)}=-3\left( e^{K}\left| W\right| ^{2}\right)
_{SUSY,(3,0)}<0,
\end{equation}
the class \textit{``(3,0)''} of SUSY FV is composed only by AdS FV. In this
case, the SUSY FV AEs read as follows:
\begin{eqnarray}
&&
\begin{array}{l}
\frak{F}_{4}=2Re\left[ \overline{Z}\hat{\Omega}_{4}\right] _{SUSY,(3,0)}= \\
\\
=2Re\left[ \overline{Z}\hat{\Omega}_{1}\wedge \hat{\Omega}_{3}\right]
_{SUSY,(3,0)}=2\left( e^{K_{1}+K_{3}}\right) _{SUSY,(3,0)}Re\left[ \overline{%
W}\Omega _{1}\wedge \Omega _{3}\right] _{SUSY,(3,0)};
\end{array}
\notag \\
&&  \label{SUSY-FV-AEs-(3,0)}
\end{eqnarray}

Notice that for such a class of SUSY FV the condition of consistence of Eqs.
(\ref{SUSY-FV-AEs-(3,0)}) is $W_{SUSY}\neq 0$. In other words, Minkowski ($%
V_{\mathcal{N}=1}=0$) SUSY FV satisfying the constraints (\ref{FV-SUSY}) and
\begin{equation}
\left\{
\begin{array}{l}
W_{SUSY}=0; \\
\\
\begin{array}{c}
\left[ g^{i\overline{j}}\left( \overline{D}_{\overline{0}}\overline{D}_{%
\overline{j}}\overline{W}\right) \overline{\Omega }_{1}\wedge D_{i}\Omega
_{3}\right] _{SUSY}=0; \\
\\
\Updownarrow \\
\\
\forall \Lambda =1,...,h_{2,1}+1:\left\{
\begin{array}{l}
\left[ g^{i\overline{j}}\left( \overline{D}_{\overline{0}}\overline{D}_{%
\overline{j}}\overline{W}\right) D_{i}X^{\Lambda }\right] _{SUSY}=0; \\
\\
\left[ g^{i\overline{j}}\left( \overline{D}_{\overline{0}}\overline{D}_{%
\overline{j}}\overline{W}\right) D_{i}F_{\Lambda }\right] _{SUSY}=0.
\end{array}
\right.
\end{array}
\end{array}
\right.  \label{SUSY-FV-(3,0)-degenerate}
\end{equation}
are not described by the classical FV Attractor Mechanism encoded by the
SUSY FV AEs (\ref{SUSY-FV-AEs1}). Indeed, such Eqs., when evaluated along
the constraints (\ref{SUSY-FV-(3,0)-degenerate}) simply return \textit{all}
(RR and NSNS) vanishing fluxes.

It is worth pointing out that, beside the substitution of ``$Im$'' with ``$%
Re $'' and the doubling of the vector dimension due to the $SL(2,\mathbb{R})$%
-doublet of RR and NSNS (3-form) fluxes, the \textit{``(3,0)''} SUSY FV AEs (%
\ref{SUSY-FV-AEs-(3,0)}) are very close to the SUSY extremal BH AEs in $%
\mathcal{N}=2$, $d=4$ supergravity, given by Eqs. (\ref{alg-BPS-1})-(\ref
{alg-BPS-2}).

II) \textit{Type ``(2,1)''} SUSY FV, determined by the constraints (\ref
{FV-SUSY}) and by the further conditions
\begin{equation}
\left\{
\begin{array}{l}
W_{SUSY}=0; \\
\\
\begin{array}{c}
\left[ g^{i\overline{j}}\left( \overline{D}_{\overline{0}}\overline{D}_{%
\overline{j}}\overline{W}\right) \overline{\Omega }_{1}\wedge D_{i}\Omega
_{3}\right] _{SUSY}\neq 0; \\
\\
\Updownarrow \\
\\
\text{(\textit{at least}) for some }\Lambda \in \left\{
1,...,h_{2,1}+1\right\} :\left\{
\begin{array}{l}
\left[ g^{i\overline{j}}\left( \overline{D}_{\overline{0}}\overline{D}_{%
\overline{j}}\overline{W}\right) D_{i}X^{\Lambda }\right] _{SUSY}\neq 0; \\
\text{and/or} \\
\left[ g^{i\overline{j}}\left( \overline{D}_{\overline{0}}\overline{D}_{%
\overline{j}}\overline{W}\right) D_{i}F_{\Lambda }\right] _{SUSY}\neq 0.
\end{array}
\right.
\end{array}
\end{array}
\right.  \label{SUSY-FV-(2,1)}
\end{equation}

Because of
\begin{equation}
V_{\mathcal{N}=1,SUSY,(2,1)}=-3\left( e^{K}\left| W\right| ^{2}\right)
_{SUSY,(2,1)}=0,
\end{equation}
the class \textit{``(2,1)''} of SUSY FV is composed only by Minkowski FV. In
this case, the SUSY FV AEs read as follows:
\begin{eqnarray}
&&
\begin{array}{l}
\frak{F}_{4}=\left[ \delta ^{A\overline{B}}\left( D_{\underline{0}%
}D_{A}Z\right) \overline{D}_{\underline{\overline{0}}}\overline{D}_{%
\overline{B}}\overline{\hat{\Omega}}_{4}+\delta ^{B\overline{A}}\left(
\overline{D}_{\underline{\overline{0}}}\overline{D}_{\overline{A}}\overline{Z%
}\right) D_{\underline{0}}D_{B}\hat{\Omega}_{4}\right] _{SUSY,(2,1)}= \\
\\
=2Re\left[ \left( t^{0}-\overline{t}^{\overline{0}}\right) g^{i\overline{j}%
}\left( \overline{D}_{\overline{0}}\overline{D}_{\overline{j}}\overline{Z}%
\right) \overline{\hat{\Omega}}_{1}\wedge D_{i}\hat{\Omega}_{3}\right]
_{SUSY,(2,1)}= \\
\\
=2Re\left[ \overline{e}_{\overline{\underline{0}}}^{\overline{0}}e_{I}^{i}%
\overline{e}_{\overline{J}}^{\overline{j}}\delta ^{I\overline{J}}\left(
\overline{D}_{\overline{0}}\overline{D}_{\overline{j}}\overline{Z}\right)
\overline{\hat{\Omega}}_{1}\wedge D_{i}\hat{\Omega}_{3}\right] _{SUSY,(2,1)}=
\\
\\
=2\left( e^{K_{1}+K_{3}}\right) _{SUSY,(2,1)}Re\left[ \overline{e}_{%
\overline{\underline{0}}}^{\overline{0}}e_{I}^{i}\overline{e}_{\overline{J}%
}^{\overline{j}}\delta ^{I\overline{J}}\left( \overline{D}_{\overline{0}}%
\overline{D}_{\overline{j}}\overline{W}\right) \overline{\Omega }_{1}\wedge
D_{i}\Omega _{3}\right] _{SUSY,(2,1)}.
\end{array}
\notag \\
&&  \label{SUSY-FV-AEs-(2,1)}
\end{eqnarray}

It is worth observing that the SUSY FV constraints (\ref{FV-SUSY}) and the
condition $W_{SUSY,(2,1)}=0$ imply
\begin{equation}
\begin{array}{c}
\left( \partial _{a}W\right) _{SUSY,(2,1)}=0,\forall a=0,1,...,h_{2,1}; \\
\\
\Updownarrow \\
\\
\multicolumn{1}{l}{\left\{
\begin{array}{l}
\left( \partial _{0}W\right) _{SUSY,(2,1)}=\left( -q_{h\mid \Lambda
}X^{\Lambda }+p_{h}^{\Lambda }F_{\Lambda }\right) _{SUSY,(2,1)}=0; \\
\\
\\
\forall i=1,...,h_{2,1}:\left\{
\begin{array}{c}
\left( \partial _{i}W\right) _{SUSY,(2,1)}=\left[ q_{f\mid \Lambda }\partial
_{i}X^{\Lambda }-p_{f}^{\Lambda }\partial _{i}F_{\Lambda }-\tau \left(
q_{h\mid \Lambda }\partial _{i}X^{\Lambda }-p_{h}^{\Lambda }\partial
_{i}F_{\Lambda }\right) \right] _{SUSY,(2,1)}=0; \\
\\
\Updownarrow \\
\\
\left( q_{h\mid \Lambda }\partial _{i}X^{\Lambda }-p_{h}^{\Lambda }\partial
_{i}F_{\Lambda }\right) _{SUSY,(2,1)}=\left[ \frac{1}{\tau }\left( q_{f\mid
\Lambda }\partial _{i}X^{\Lambda }-p_{f}^{\Lambda }\partial _{i}F_{\Lambda
}\right) \right] _{SUSY,(2,1)},
\end{array}
\right.
\end{array}
\right.}
\end{array}
\label{SUSY-(2,1)-elaboration}
\end{equation}
where we used the fact that $\tau \neq 0$ is a necessary (but not
sufficient) condition for the (assumed) regularity of $K_{1}$ in $\mathcal{M}%
_{t^{0}\equiv \tau }\subsetneq M$ (or at least in the considered critical
points of $V_{\mathcal{N}=1}$). Thus, Eq. (\ref{D0DJW-SUSY}) can be further
elaborated as follows:
\begin{eqnarray}
&&
\begin{array}{l}
\left( D_{\underline{0}}D_{J}W\right) _{SUSY,(2,1)}=\left( e_{J}^{j}\partial
_{j}D_{\underline{0}}W\right) _{SUSY,(2,1)}= \\
\\
=\left( e_{J}^{j}e_{\underline{0}}^{0}\partial _{j}\partial _{0}W\right)
_{SUSY,(2,1)}=\left( e_{J}^{j}e_{\underline{0}}^{0}\partial _{0}\partial
_{j}W\right) _{SUSY,(2,1)}= \\
\\
=\left[ e_{J}^{j}\left( \overline{\tau }-\tau \right) \partial _{j}\partial
_{0}W\right] _{SUSY,(2,1)}=\left\{ e_{J}^{j}\left( \overline{\tau }-\tau
\right) \left( -q_{h\mid \Lambda }\partial _{j}X^{\Lambda }+p_{h}^{\Lambda
}\partial _{j}F_{\Lambda }\right) \right\} _{SUSY,(2,1)}= \\
\\
=\left\{ -e_{J}^{j}\frac{\left( \overline{\tau }-\tau \right) }{\tau }\left(
q_{f\mid \Lambda }\partial _{j}X^{\Lambda }-p_{f}^{\Lambda }\partial
_{j}F_{\Lambda }\right) \right\} _{SUSY,(2,1)},
\end{array}
\notag \\
&&  \label{D0DJW-SUSY-(2,1)}
\end{eqnarray}
where in the last line we used Eq. (\ref{SUSY-(2,1)-elaboration}).
Furthermore, by using Eq. (\ref{SUSY-(2,1)-elaboration}) with some
elaborations, one obtains that
\begin{equation}
\left.
\begin{array}{l}
W_{SUSY,(2,1)}=0 \\
\\
\left( \partial _{0}W\right) _{SUSY,(2,1)}=0
\end{array}
\right\} \Rightarrow \left( q_{f\mid \Lambda }X^{\Lambda }-p_{f}^{\Lambda
}F_{\Lambda }\right) _{SUSY,(2,1)}=0,
\end{equation}
and therefore at the class \textit{``(2,1)''} of SUSY critical points of $V_{%
\mathcal{N}=1}$ the \textit{''RR sector''} $q_{f\mid \Lambda }X^{\Lambda
}-p_{f}^{\Lambda }F_{\Lambda }$ and the ``\textit{NSNS sector''} $-\left(
q_{h\mid \Lambda }X^{\Lambda }-p_{h}^{\Lambda }F_{\Lambda }\right) $ of the
holomorphic superpotential $W$ vanish separately.

By looking at the \textit{``(2,1)''} SUSY FV AEs (\ref{SUSY-FV-AEs-(2,1)}),
it is interesting to note that \textit{``(2,1)''} SUSY FV do not have a
counterpart in the theory of extremal BH attractors in $\mathcal{N}=2$, $d=4$
supergravity. Indeed, as implied by the SUSY extremal BH AEs (\ref{alg-BPS-1}%
)-(\ref{alg-BPS-2}), the classical extremal BH Attractor Mechanism in $%
\mathcal{N}=2$, $d=4$ supergravity is not consistent with SUSY critical
points of $V_{BH}$ also having $W=0$, and thus determining $V_{BH}=0$. In
such a case, the SUSY extremal BH AEs (\ref{alg-BPS-1})-(\ref{alg-BPS-2})
simply yield \textit{all} (magnetic and electric) BH charges vanishing.

Contrarily to the extremal BH attractors in $\mathcal{N}=2$, $d=4$
supergravity, and as yielded by the \textit{``(2,1)''} SUSY FV AEs (\ref
{SUSY-FV-AEs-(2,1)}), the classical FV Attractor Mechanism allows for
stabilization of (axion-dilaton + CS) moduli in the SUSY case with vanishing
\textit{gravitino mass} $Z_{USY}=\left( e^{K}W\right) _{SUSY}=0$.

III) \textit{Type ``(3,0)+(2,1)''} SUSY FV, determined by the constraints (%
\ref{FV-SUSY}) and by the further conditions
\begin{equation}
\left\{
\begin{array}{l}
W_{SUSY}\neq 0; \\
\\
\begin{array}{c}
\left[ g^{i\overline{j}}\left( \overline{D}_{\overline{0}}\overline{D}_{%
\overline{j}}\overline{W}\right) \overline{\Omega }_{1}\wedge D_{i}\Omega
_{3}\right] _{SUSY}\neq 0; \\
\\
\Updownarrow \\
\\
\text{(\textit{at least}) for some }\Lambda \in \left\{
1,...,h_{2,1}+1\right\} :\left\{
\begin{array}{l}
\left[ g^{i\overline{j}}\left( \overline{D}_{\overline{0}}\overline{D}_{%
\overline{j}}\overline{W}\right) D_{i}X^{\Lambda }\right] _{SUSY}\neq 0; \\
\text{and/or} \\
\left[ g^{i\overline{j}}\left( \overline{D}_{\overline{0}}\overline{D}_{%
\overline{j}}\overline{W}\right) D_{i}F_{\Lambda }\right] _{SUSY}\neq 0.
\end{array}
\right.
\end{array}
\end{array}
\right.  \label{SUSY-FV-(3,0)+(2,1)}
\end{equation}

Because of
\begin{equation}
V_{\mathcal{N}=1,SUSY,\left( 3,0\right) +(2,1)}=-3\left( e^{K}\left|
W\right| ^{2}\right) _{SUSY,\left( 3,0\right) +(2,1)}<0,
\end{equation}
the class \textit{``(3,0)+(2,1)''} of SUSY FV is composed only by AdS FV.
For such a class of SUSY FV the FV AEs are simply given by Eqs. (\ref
{SUSY-FV-AEs1}) (which can be further elaborated by considering Eq. (\ref
{D0DJW-SUSY})), constrained by Eqs. (\ref{SUSY-FV-(3,0)+(2,1)}).

In \cite{G} examples of SUSY FV of all the classes considered above (\textit{%
``(3,0)''}, \textit{``(2,1)''}, and \textit{``(3,0)+(2,1)''}) have been
explicitly checked to satisfy the corresponding SUSY FV AEs of $\mathcal{N}%
=1 $, $d=4$ supergravity from Type IIB on $\frac{CY_{3}\times T_{2}}{\mathbb{%
Z}_{2}} $, in a model with $h_{2,1}=1$, where $CY_{3}$ is the so-called
Fermat \textit{sixtic} hypersurface. \setcounter{equation}0
\def\theequation{\arabic{section}.\arabic{equation}}
\section{Some Recent Developments on Extremal Black Hole Attractors}

In these lectures we have described the general theory of attractors for a
generic $\mathcal{N}=2$, $d=4$ SK geometry, both in the supergravity
language and in terms of Type IIB superstrings compactified on Calabi-Yau
threefolds. We have then described, in a similar way, the Attractor
Mechanism arising in $\mathcal{N}=1$, $d=4$ flux vacua, focussing on the
case of (the F-theory limit of) compactifications of Type IIB on Calabi-Yau
orientifolds (see also \cite{Dall'Agata} for an extension to the landscape
of non-K\"{a}hler vacua emerging in the flux compactifications of heterotic
superstrings).

In the last years, more results have been obtained for the non-BPS extremal $%
d=4$ BH attractors, especially with regard to symmetric $\mathcal{N}=2$ SK
geometries and to $\mathcal{N}>2$ extended theories.

The classification of the charge orbits of the $U$-duality \cite{HT}
groups supporting attractors with non-vanishing entropy was
performed in \cite{FG} (see also
\cite{FKlast}) and \cite{BFGM1}, respectively for $\mathcal{N}=8$ and $%
\mathcal{N}=2$ symmetric supergravities, whereas the corresponding moduli
spaces were found and studied respectively in \cite{fm07} and \cite
{Ferrara-Marrani-1}. Furthermore, the classification of attractors for $%
\mathcal{N}=3$, $4$ (along with the corresponding maximal compact
symmetries) was performed in \cite{ADFT}. Notice that the $\mathcal{N}=6$
theory has the same attractors, orbits and related moduli spaces of the
quaternionic magic $\mathcal{N}=2$ model \cite{BFGM1,Ferrara-Gimon}.

For the sake of completeness we report here the charge orbits and the moduli
space\footnote{%
The scalar manifolds of $3\leqslant \mathcal{N}\leqslant 8$, $d=4$
supergravities can be found \textit{e.g.} in \cite{ADFT}.} of
attractors for all $3\leqslant \mathcal{N}\leqslant 8$, $d=4$
supergravities (for the treatment of extremal BHs in such theories,
see \textit{e.g.}
\cite{Duff-Khuri,Duff-stu,Cvetic1,Cvetic2,Cvetic3}), including the
cases $\mathcal{N}=3$, $4$, $5$, not exhaustively discussed in
literature.

All $d=4$ theories with $\mathcal{N}$ even can be uplifted to $d=5$, and
their $U$-duality group admits a unique quartic invariant (see \textit{e.g.}
\cite{ADF-U-duality-d=4}). All such supergravities have a non-BPS attractor
solution whose moduli space coincide with the $d=5$ real scalar manifold.
This is the non-BPS solution with non-vanishing central charge matrix $%
Z_{AB} $ ($A$, $B=1,...,\mathcal{N}$), which breaks the $d=4$ $\mathcal{R}$%
-symmetry to the $d=5$ $\mathcal{R}$-symmetry. Since the cases $\mathcal{N}%
=6 $, $8$ have been treated in \cite
{FKlast,BFGM1,ADFT,Ferrara-Marrani-1,fm07}, let us now consider the case $%
\mathcal{N}=4$; as previously mentioned, its attractors with non-vanishing
entropy (and the corresponding maximal compact symmetries) have been
classified in \cite{ADFT}. The non-BPS attractor with $Z_{AB}\neq 0$ breaks
the $d=4$ $\mathcal{R}$-symmetry $SU\left( 4\right) \sim SO\left( 6\right) $
down to the $d=5$ $\mathcal{R}$-symmetry $USp\left( 4\right) \sim SO\left(
5\right) $, and the maximal compact symmetry exhibited by the solution is $%
USp\left( 4\right) \otimes SO\left( n-1\right) $, where $n$ denotes the
number of matter multiplets coupled to the supergravity one. The other
non-BPS attractor solution of $\mathcal{N}=4$, $d=4$ supergravity has $%
Z_{AB}=0$; thus, the $d=4$ $\mathcal{R}$-symmetry $SU\left( 4\right) $ is
unbroken, and the corresponding maximal compact symmetry is $SU\left(
4\right) \otimes SO\left( n-2\right) $.

On the other hand, the $d=4$ theories with $\mathcal{N}$ odd ($=3$, $5$)
cannot be uplifted to $d=5$, and their $U$-duality group admits a unique
quadratic invariant (see \textit{e.g.} \cite{ADF-U-duality-d=4}). The case $%
\mathcal{N}=3$ admits only $\frac{1}{3}$-BPS and non-BPS $Z_{AB}=0$
attractors with non-vanishing entropy; notice that such a result is similar
to the one obtained for the $\mathcal{N}=2$ symmetric sequence of SK
manifolds based on quadratic holomorphic prepotential (see \cite{BFGM1},
\cite{fm07} and Refs. therein), and it is ultimately due to the
aforementioned fact that the $\mathcal{N}=3$, $d=4$ $U$-duality group $%
SU\left( 3,n\right) $ has a unique quadratic (rather than quartic)
invariant. $\mathcal{N}=5$ is peculiar, as discussed in \cite{ADFT}, in such
a case only the $\frac{1}{5}$-BPS attractor has non-vanishing entropy (this
solution splits in BPS and non-BPS $Z=0$ ones when performing the $\mathcal{N%
}=5\rightarrow \mathcal{N}=2$ truncation of the theory \cite{ADFT}).

By knowing the real, symplectic representation $R$ (with $dim_{\mathbb{R}}=%
\frak{r}$) of the $U$-duality group $G$ in which the charge vector $Q$ sits,
the orbits of $R$ supporting attractors with non-vanishing entropy can be
computed; their dimension is always $\frak{r}-1$, because they are defined
by a fixed, non-vanishing value of the unique $U$-invariant of the theory.
For ($\mathcal{N}=2$ symmetric and) $3\leqslant \mathcal{N}\leqslant 8$, $%
d=4 $ supergravities such orbits are homogeneous symmetric manifolds of the
form $\frac{G}{\frak{H}}$ ($\frak{H}=\mathcal{H},\widehat{\mathcal{H}},%
\widetilde{\mathcal{H}}$ respectively for BPS, non-BPS $Z_{AB}\neq 0$ and
non-BPS $Z_{AB}=0$); the corresponding moduli space is given by the
symmetric manifold $\frac{\frak{H}}{h}$, where $h$ ($=\frak{h},\widehat{%
\frak{h}},\widetilde{\frak{h}}$, respectively) is the maximal compact
subgroup of $\frak{H}$. It is worth remarking that the ($\frac{1}{\mathcal{N}%
}$-)BPS moduli spaces of $3\leqslant \mathcal{N}\leqslant 8$, $d=4$
supergravities all are quaternionic K\"{a}hler manifolds; such a geometrical
property can be understood by noticing that in the supersymmetry reduction
down to $\mathcal{N}=2$ such spaces are spanned by the hypermultiplets'
scalar degrees of freedom \cite{ADF-U-duality-d=4,Ferrara-Marrani-1}.

Following \cite{ADF-U-duality-d=4} and \cite{ADFT}, the relation among the
signs of the $U$-invariant $I_{2}$ (quadratic in charges) or $I_{4}$
(quartic in charges) of the considered supergravity and the various BH
charge orbits is (for the $\mathcal{N}=8$ case see also \cite
{Ortin,Ferrara-Maldacena,FG,KL,FKlast}):
\begin{equation}
\begin{array}{ll}
\mathcal{N}=3: &
\begin{array}{l}
\left\{
\begin{array}{l}
\frac{1}{3}-BPS:I_{2}>0; \\
non-BPS,Z_{AB}=0:I_{2}<0;
\end{array}
\right. \\
~
\end{array}
\\
\mathcal{N}=4: &
\begin{array}{l}
\left\{
\begin{array}{l}
\frac{1}{4}-BPS:I_{4}>0; \\
non-BPS,Z_{AB}\neq 0:I_{4}<0; \\
non-BPS,Z_{AB}=0:I_{4}>0;
\end{array}
\right. \\
~
\end{array}
\\
\mathcal{N}=5: &
\begin{array}{l}
\frac{1}{5}-BPS:I_{2}\gtrless 0\text{\textit{(sign does not matter)}}; \\
~
\end{array}
\\
\mathcal{N}=6: &
\begin{array}{l}
\left\{
\begin{array}{l}
\frac{1}{6}-BPS:I_{4}>0; \\
non-BPS,Z_{AB}\neq 0:I_{4}<0; \\
non-BPS,Z_{AB}=0:I_{4}>0;
\end{array}
\right. \\
~
\end{array}
\\
\mathcal{N}=8: & \left\{
\begin{array}{l}
\frac{1}{8}-BPS:I_{4}>0; \\
non-BPS,Z_{AB}\neq 0:I_{4}<0.
\end{array}
\right.
\end{array}
\end{equation}

In Tables 1 and 2 we respectively list all charge orbits supporting extremal
BH attractors with non-vanishing classical Bekenstein-Hawking \cite{BH1}
entropy (\textit{i.e.} corresponding to the so-called ``large'' BHs) and
their corresponding moduli spaces for $3\leqslant \mathcal{N}\leqslant 8$, $%
d=4$ supergravities.

Some of the above results hold also for a generic (non-symmetric nor
eventually homogeneous) $\mathcal{N}=2$, $d=4$ SK geometry based on a cubic
holomorphic prepotential (usually named SK $d$-geometry \cite{dWVVP}). For
instance, for any SK $d$-geometry of $\mathcal{N}=2$, $d=4$ supergravity
coupled to $n$ Abelian vector multiplets, the so-called $D0$-$D6$ BH charge
configuration supports only non-BPS $Z\neq 0$ attractors, whose moduli space
is $\left( n-1\right) $%
-dimensional, and it is given by the corresponding $\mathcal{N}=2$, $d=5$
scalar manifold, endowed with real special geometry \cite{TT2}. It is worth
pointing out here that the existence of $n-1$ massless modes of the $%
2n\times 2n$ (real from of the) Hessian matrix of the BH effective potential
$V_{BH}$ at its non-BPS $Z\neq 0$ critical points was shown in \cite{TT} to
hold in any SK $d$-geometry of $\mathcal{N}=2$, $d=4$ supergravity coupled
to $n$ Abelian vector multiplets. However, the issue of the stability of the
non-BPS $Z\neq 0$ critical points of $V_{BH}$ (as well as of the non-BPS $%
Z=0 $ ones) in non-homogeneous SK $d$-geometry has not been
thoroughly
investigated so far\footnote{%
The case of homogeneous non-symmetric SK $d$-geometry has been studied in
\cite{DFT07-1}.}.

Let us finally remark that it is also possible to relate the \textit{flat}
directions of non-BPS attractor solutions in $\mathcal{N}=2$, $d=4$
symmetric supergravities with the \textit{flat} directions of ($\frac{1}{%
\mathcal{N}}$-)BPS and non-BPS attractors in $\mathcal{N}>2$, $d=4$ theories
\cite{Ferrara-Marrani-1}. Moreover, the moduli spaces of extremal BH
attractors with non-vanishing entropy in supergravity theories in $d=5$ and $%
d=6$ have been found and their relations with the corresponding Attractor
Eqs. in $d=4$ studied in \cite{CFM1} and \cite{AFMT1}.
\begin{table}[p]
\begin{center}
\begin{tabular}{|c||c|c|c|}
\hline
& $
\begin{array}{c}
\\
\frac{1}{\mathcal{N}}\text{-BPS orbits }~\frac{G}{\mathcal{H}} \\
~
\end{array}
$ & $
\begin{array}{c}
\\
\text{non-BPS, }Z_{AB}\neq 0\text{ orbits}~\frac{G}{\widehat{\mathcal{H}}}
\\
~
\end{array}
$ & $
\begin{array}{c}
\\
\text{non-BPS, }Z_{AB}=0\text{ orbits }\frac{G}{\widetilde{\mathcal{H}}} \\
~
\end{array}
$ \\ \hline\hline
$
\begin{array}{c}
\\
\mathcal{N}=3 \\
~
\end{array}
$ & $\frac{SU(3,n)}{SU(2,n)}~$ & $-$ & $\frac{SU(3,n)}{SU(3,n-1)}~$ \\ \hline
$
\begin{array}{c}
\\
\mathcal{N}=4 \\
~
\end{array}
$ & $\frac{SU(1,1)}{U(1)}\otimes \frac{SO(6,n)}{SO(4,n)}~$ & $\frac{SU(1,1)}{%
SO(1,1)}\otimes \frac{SO(6,n)}{SO(5,n-1)}~$ & $\frac{SU(1,1)}{U(1)}\otimes
\frac{SO(6,n)}{SO(6,n-2)}$ \\ \hline
$
\begin{array}{c}
\\
\mathcal{N}=5 \\
~
\end{array}
$ & $\frac{SU(1,5)}{SU(3)\otimes SU\left( 2,1\right) }$ & $-$ & $-$ \\ \hline
$
\begin{array}{c}
\\
\mathcal{N}=6 \\
~
\end{array}
$ & $\frac{SO^{\ast }(12)}{SU(4,2)}~$ & $\frac{SO^{\ast }(12)}{SU^{\ast }(6)}%
~$ & $\frac{SO^{\ast }(12)}{SU(6)}~$ \\ \hline
$
\begin{array}{c}
\\
\mathcal{N}=8 \\
~
\end{array}
$ & $\frac{E_{7\left( 7\right) }}{E_{6\left( 2\right) }}$ & $\frac{%
E_{7\left( 7\right) }}{E_{6\left( 6\right) }}$ & $-~$ \\ \hline
\end{tabular}
\end{center}
\caption{\textbf{Charge orbits of the real, symplectic }$R$ \textbf{%
representation of the }$U$\textbf{-duality group }$G$ \textbf{supporting BH
attractors with non-vanishing entropy in $3\leqslant \mathcal{N}\leqslant 8$%
, $d=4$ supergravities}}
\end{table}
\begin{table}[p]
\begin{center}
\begin{tabular}{|c||c|c|c|}
\hline
& $
\begin{array}{c}
\\
\frac{1}{\mathcal{N}}\text{-BPS} \\
\text{moduli space }\frac{\mathcal{H}}{\frak{h}}\text{ } \\
~
\end{array}
$ & $
\begin{array}{c}
\\
\text{non-BPS, }Z_{AB}\neq 0 \\
\text{moduli space }\frac{\widehat{\mathcal{H}}}{\widehat{\frak{h}}} \\
~
\end{array}
$ & $
\begin{array}{c}
\\
\text{non-BPS, }Z_{AB}=0 \\
\text{moduli space }\frac{\widetilde{\mathcal{H}}}{\widetilde{\frak{h}}} \\
~
\end{array}
$ \\ \hline\hline
$
\begin{array}{c}
\\
\mathcal{N}=3 \\
~
\end{array}
$ & $\frac{SU(2,n)}{SU(2)\otimes SU\left( n\right) \otimes U\left( 1\right) }%
~$ & $-$ & $\frac{SU(3,n-1)}{SU(3)\otimes SU\left( n-1\right) \otimes
U\left( 1\right) }~$ \\ \hline
$
\begin{array}{c}
\\
\mathcal{N}=4 \\
~
\end{array}
$ & $\frac{SO(4,n)}{SO(4)\otimes SO\left( n\right) }~$ & $SO(1,1)\otimes
\frac{SO(5,n-1)}{SO(5)\otimes SO\left( n-1\right) }~$ & $\frac{SO(6,n-2)}{%
SO(6)\otimes SO\left( n-2\right) }$ \\ \hline
$
\begin{array}{c}
\\
\mathcal{N}=5 \\
~
\end{array}
$ & $\frac{SU\left( 2,1\right) }{SU\left( 2\right) \otimes U\left( 1\right) }
$ & $-$ & $-$ \\ \hline
$
\begin{array}{c}
\\
\mathcal{N}=6 \\
~
\end{array}
$ & $\frac{SU(4,2)}{SU(4)\otimes SU\left( 2\right) \otimes U\left( 1\right) }%
~$ & $\frac{SU^{\ast }(6)}{USp\left( 6\right) }~$ & $-$ \\ \hline
$
\begin{array}{c}
\\
\mathcal{N}=8 \\
~
\end{array}
$ & $\frac{E_{6\left( 2\right) }}{SU\left( 6\right) \otimes SU\left(
2\right) }$ & $\frac{E_{6\left( 6\right) }}{USp\left( 8\right) }$ & $-~$ \\
\hline
\end{tabular}
\end{center}
\caption{\textbf{Moduli spaces of BH attractors with non-vanishing entropy
in $3\leqslant \mathcal{N}\leqslant 8$, $d=4$ supergravities (}$\frak{h}$%
\textbf{, }$\widehat{\frak{h}}$\textbf{\ and }$\widetilde{\frak{h}}$\textbf{%
\ are maximal compact subgroups of }$\mathcal{H}$\textbf{, }$\widehat{%
\mathcal{H}}$\textbf{\ and }$\widetilde{\mathcal{H}}$\textbf{, respectively)}
}
\end{table}
\newpage

\section*{\textbf{Acknowledgments}}

A. M. would like to thank the Department of Physics and Astronomy, UCLA and
the Department of Physics, Theory Unit Group at CERN, where part of this
work was done, for kind hospitality and stimulating environment.

The work of S.B. has been supported in part by the European Community Human
Potential Program under contract MRTN-CT-2004-005104 \textit{``Constituents,
fundamental forces and symmetries of the universe''}.

The work of S.F.~has been supported in part by the European Community Human
Potential Program under contract MRTN-CT-2004-005104 \textit{``Constituents,
fundamental forces and symmetries of the universe''}, in association with
INFN Frascati National Laboratories and by D.O.E.~grant DE-FG03-91ER40662,
Task C.

The work of R.K.~has been supported by NSF Grant PHY-0244728.\textbf{\ }

The work of A.M. has been supported by a Junior Grant of the \textit{%
``Enrico Fermi''} Center, Rome, in association with INFN Frascati National
Laboratories. \newpage \appendix \setcounter{equation}0
\def\theequation{I.\arabic{equation}}
\section{Appendix I}

Up to the third order of covariant differentiation included, the possible
independent 4-forms ($\left( 1,-1\right) $-K\"{a}hler weighted with respect
to $K$) on $CY_{4}$ beside $\hat{\Omega}_{4}$ are:
\begin{eqnarray}
D_{a}\hat{\Omega}_{4} &:&  \notag \\
&&  \notag \\
&&\left\{
\begin{array}{l}
\begin{array}{l}
a=0: \\
\\
D_{0}\hat{\Omega}_{4}=\left( D_{0}\hat{\Omega}_{1}\right) \wedge \hat{\Omega}%
_{3}=ie^{K_{1}}\overline{\hat{\Omega}}_{1}\wedge \hat{\Omega}_{3}=\left(
\overline{t}^{\overline{0}}-t^{0}\right) ^{-1}\overline{\hat{\Omega}}%
_{1}\wedge \hat{\Omega}_{3};
\end{array}
\\
\\
\\
\\
\begin{array}{l}
a=i: \\
\\
D_{i}\hat{\Omega}_{4}=\hat{\Omega}_{1}\wedge D_{i}\hat{\Omega}_{3}=e^{\frac{1%
}{2}K_{3}}\hat{\Omega}_{1}\wedge D_{i}\Omega _{3}=e^{\frac{1}{2}K_{3}}\hat{%
\Omega}_{1}\wedge \left[ \partial _{i}\Omega _{3}+\left( \partial
_{i}K_{3}\right) \Omega _{3}\right] = \\
\\
=\frac{1}{\sqrt{i\left( \overline{X}^{\Delta }F_{\Delta }-X^{\Delta }%
\overline{F}_{\Delta }\right) }}\hat{\Omega}_{1}\wedge \left\{
\begin{array}{l}
\left[ \partial _{i}X^{\Lambda }-\frac{\left( \overline{X}^{\Sigma }\partial
_{i}F_{\Sigma }-\left( \partial _{i}X^{\Sigma }\right) \overline{F}_{\Sigma
}\right) }{\overline{X}^{\Xi }F_{\Xi }-X^{\Xi }\overline{F}_{\Xi }}%
X^{\Lambda }\right] \alpha _{\Lambda }+ \\
\\
-\left[ \partial _{i}F_{\Lambda }-\frac{\left( \overline{X}^{\Sigma
}\partial _{i}F_{\Sigma }-\left( \partial _{i}X^{\Sigma }\right) \overline{F}%
_{\Sigma }\right) }{\overline{X}^{\Xi }F_{\Xi }-X^{\Xi }\overline{F}_{\Xi }}%
F_{\Lambda }\right] \beta ^{\Lambda }
\end{array}
\right\} .
\end{array}
\end{array}
\right.  \notag \\
&&  \label{DaOmega4}
\end{eqnarray}
\begin{align}
D_{a}D_{b}\hat{\Omega}_{4}& =D_{(a}D_{b)}\hat{\Omega}_{4}:  \notag \\
&  \notag \\
& \left\{
\begin{array}{l}
\left( a,b\right) =\left( 0,0\right) :D_{0}D_{0}\hat{\Omega}_{4}=0; \\
\\
\\
\begin{array}{l}
\left( a,b\right) =\left( 0,i\right) : \\
\\
D_{0}D_{i}\hat{\Omega}_{4}=D_{0}\hat{\Omega}_{1}\wedge D_{i}\hat{\Omega}_{3}=
\\
\\
=\frac{1}{\left( \overline{t}^{\overline{0}}-t^{0}\right) \sqrt{i\left(
\overline{X}^{\Delta }F_{\Delta }-X^{\Delta }\overline{F}_{\Delta }\right) }}%
\overline{\hat{\Omega}}_{1}\wedge \left\{
\begin{array}{l}
\left[ \partial _{i}X^{\Lambda }-\frac{\left( \overline{X}^{\Sigma }\partial
_{i}F_{\Sigma }-\left( \partial _{i}X^{\Sigma }\right) \overline{F}_{\Sigma
}\right) }{\overline{X}^{\Xi }F_{\Xi }-X^{\Xi }\overline{F}_{\Xi }}%
X^{\Lambda }\right] \alpha _{\Lambda }+ \\
\\
-\left[ \partial _{i}F_{\Lambda }-\frac{\left( \overline{X}^{\Sigma
}\partial _{i}F_{\Sigma }-\left( \partial _{i}X^{\Sigma }\right) \overline{F}%
_{\Sigma }\right) }{\overline{X}^{\Xi }F_{\Xi }-X^{\Xi }\overline{F}_{\Xi }}%
F_{\Lambda }\right] \beta ^{\Lambda }
\end{array}
\right\} ;
\end{array}
\\
\\
\\
\begin{array}{l}
\left( a,b\right) =\left( i,j\right) : \\
\\
D_{i}D_{j}\hat{\Omega}_{4}=\hat{\Omega}_{1}\wedge D_{i}D_{j}\hat{\Omega}%
_{3}=iC_{ijk}g^{k\overline{l}}\hat{\Omega}_{1}\wedge \overline{D}_{\overline{%
l}}\overline{\hat{\Omega}}_{3}= \\
\\
=-e^{-K_{1}}C_{ijk}g^{k\overline{l}}\overline{D}_{\overline{0}}\overline{%
\hat{\Omega}}_{1}\wedge \overline{D}_{\overline{l}}\overline{\hat{\Omega}}%
_{3}=-e^{-K_{1}}C_{ijk}g^{k\overline{l}}\overline{D}_{\overline{0}}\overline{%
D}_{\overline{l}}\overline{\hat{\Omega}}_{4}= \\
\\
=-e^{-K_{1}}C_{ijk}g^{k\overline{l}}\overline{D}_{\overline{l}}\overline{D}_{%
\overline{0}}\overline{\hat{\Omega}}_{4}=-i\left( \overline{t}^{\overline{0}%
}-t^{0}\right) C_{ijk}g^{k\overline{l}}\overline{D}_{\overline{l}}\overline{D%
}_{\overline{0}}\overline{\hat{\Omega}}_{4}.
\end{array}
\end{array}
\right.  \notag \\
&  \label{DaDbOmega4}
\end{align}
\begin{eqnarray}
&&
\begin{array}{l}
D_{a}D_{b}D_{c}\hat{\Omega}_{4}=D_{(a}D_{b}D_{c)}\hat{\Omega}_{4}: \\
\\
\\
\left\{
\begin{array}{l}
\left( a,b,c\right) =\left( 0,0,0\right) :D_{0}D_{0}D_{0}\hat{\Omega}_{4}=0;
\\
\\
\left( a,b,c\right) =\left( 0,0,i\right) :D_{0}D_{0}D_{i}\hat{\Omega}_{4}=0;
\\
\\
\left( a,b,c\right) =\left( 0,i,j\right) :\left\{
\begin{array}{l}
D_{0}D_{i}D_{j}\hat{\Omega}_{4}=D_{0}\hat{\Omega}_{1}\wedge D_{i}D_{j}\hat{%
\Omega}_{3}= \\
\\
=-e^{K_{1}}C_{ijk}g^{k\overline{l}}\overline{\hat{\Omega}}_{1}\wedge
\overline{D}_{\overline{l}}\overline{\hat{\Omega}}_{3}= \\
\\
=-e^{K_{1}}C_{ijk}g^{k\overline{l}}\overline{D}_{\overline{l}}\overline{\hat{%
\Omega}}_{4}=i\left( \overline{t}^{\overline{0}}-t^{0}\right)
^{-1}C_{ijk}g^{k\overline{l}}\overline{D}_{\overline{l}}\overline{\hat{\Omega%
}}_{4};
\end{array}
\right. \\
\\
\left( a,b,c\right) =\left( i,j,k\right) :\left\{
\begin{array}{l}
D_{i}D_{j}D_{k}\hat{\Omega}_{4}=-i\left( \overline{t}^{\overline{0}%
}-t^{0}\right) \left( D_{i}C_{jkl}\right) g^{l\overline{m}}\overline{D}_{%
\overline{m}}\overline{D}_{\overline{0}}\overline{\hat{\Omega}}_{4}+ \\
\\
-i\left( \overline{t}^{\overline{0}}-t^{0}\right) C_{jkl}g^{l\overline{m}%
}D_{i}\overline{D}_{\overline{m}}\overline{D}_{\overline{0}}\overline{\hat{%
\Omega}}_{4}= \\
\\
=-i\left( \overline{t}^{\overline{0}}-t^{0}\right) \left(
D_{i}C_{jkl}\right) g^{l\overline{m}}\overline{D}_{\overline{0}}\overline{D}%
_{\overline{m}}\overline{\hat{\Omega}}_{4}+ \\
\\
-i\left( \overline{t}^{\overline{0}}-t^{0}\right) C_{jkl}g^{l\overline{m}%
}\left( \overline{D}_{\overline{0}}\overline{\hat{\Omega}}_{1}\right) \wedge
D_{i}\overline{D}_{\overline{m}}\overline{\hat{\Omega}}_{3}= \\
\\
=-i\left( \overline{t}^{\overline{0}}-t^{0}\right) \left(
D_{i}C_{jkl}\right) g^{l\overline{m}}\overline{D}_{\overline{0}}\overline{D}%
_{\overline{m}}\overline{\hat{\Omega}}_{4}-i\left( \overline{t}^{\overline{0}%
}-t^{0}\right) C_{ijk}\overline{D}_{\overline{0}}\overline{\hat{\Omega}}_{4}.
\end{array}
\right.
\end{array}
\right.
\end{array}
\notag \\
&&  \label{DaDbDcOmega4}
\end{eqnarray}
\begin{eqnarray}
\overline{D}_{\overline{a}}D_{b}D_{c}\hat{\Omega}_{4} &=&\overline{D}_{%
\overline{a}}D_{(b}D_{c)}\hat{\Omega}_{4}:  \notag \\
&&  \notag \\
&&\left\{
\begin{array}{l}
\left( \overline{a},b,c\right) =\left( \overline{0},0,0\right) :\overline{D}%
_{\overline{0}}D_{0}D_{0}\hat{\Omega}_{4}=0; \\
\\
\\
\left( \overline{a},b,c\right) =\left( \overline{0},0,i\right) :\overline{D}%
_{\overline{0}}D_{0}D_{i}\hat{\Omega}_{4}=g_{0\overline{0}}D_{i}\hat{\Omega}%
_{4}=-\left( \overline{t}^{\overline{0}}-t^{0}\right) ^{-2}D_{i}\hat{\Omega}%
_{4}=e^{2K_{1}}D_{i}\hat{\Omega}_{4}; \\
\\
\\
\left( \overline{a},b,c\right) =\left( \overline{0},i,j\right) :\overline{D}%
_{\overline{0}}D_{i}D_{j}\hat{\Omega}_{4}=0; \\
\\
\\
\left( \overline{a},b,c\right) =\left( \overline{l},0,0\right) :\overline{D}%
_{\overline{l}}D_{0}D_{0}\hat{\Omega}_{4}=0; \\
\\
\\
\left( \overline{a},b,c\right) =\left( \overline{l},0,i\right) :\overline{D}%
_{\overline{l}}D_{0}D_{i}\hat{\Omega}_{4}=g_{i\overline{l}}D_{0}\hat{\Omega}%
_{4}; \\
\\
\\
\begin{array}{l}
\left( \overline{a},b,c\right) =\left( \overline{l},i,j\right) : \\
\\
\overline{D}_{\overline{l}}D_{i}D_{j}\hat{\Omega}_{4}=iC_{ijk}g^{k\overline{k%
}}\hat{\Omega}_{1}\wedge \overline{D}_{\overline{l}}\overline{D}_{\overline{k%
}}\overline{\hat{\Omega}}_{3}= \\
\\
=g^{k\overline{k}}C_{ijk}\overline{C}_{\overline{l}\overline{m}\overline{k}%
}g^{m\overline{m}}\hat{\Omega}_{1}\wedge D_{m}\hat{\Omega}_{3}= \\
\\
=g^{k\overline{k}}C_{ijk}\overline{C}_{\overline{l}\overline{m}\overline{k}%
}g^{m\overline{m}}D_{m}\hat{\Omega}_{4}= \\
\\
\overset{\text{\textit{SKG constraints}}}{=}\left( R_{i\overline{l}j%
\overline{m}}g^{m\overline{m}}+\delta _{j}^{m}g_{i\overline{l}}+\delta
_{i}^{m}g_{j\overline{l}}\right) D_{m}\hat{\Omega}_{4}.
\end{array}
\end{array}
\right.  \notag \\
&&  \label{DabDbDcOmega4}
\end{eqnarray}

Since the covariant derivatives of $\hat{\Omega}_{4}$ are often considered
in local ``flat'' coordinated in $M$, below we write the independent ones,
up to the third order included, by recalling Eqs. (\ref{M-bein-2-sol}) and (%
\ref{M-bein-inv-sol}) (implemented by Eqs. (\ref{funct-dep1})), Eqs. (\ref
{g-M}) and (\ref{g-M-inv}), and Eqs. (\ref{DaOmega4})-(\ref{DabDbDcOmega4}):
\begin{equation}
D_{A}\hat{\Omega}_{4}:\left\{
\begin{array}{l}
A=\underline{0}:D_{\underline{0}}\hat{\Omega}_{4}=e_{\underline{0}}^{a}D_{a}%
\hat{\Omega}_{4}=e_{\underline{0}}^{0}\left( D_{0}\hat{\Omega}_{1}\right)
\wedge \hat{\Omega}_{3}=\overline{\hat{\Omega}}_{1}\wedge \hat{\Omega}_{3};
\\
\\
A=I:D_{I}\hat{\Omega}_{4}=e_{I}^{a}D_{a}\hat{\Omega}_{4}=e_{I}^{i}D_{i}\hat{%
\Omega}_{4}=e_{I}^{i}\left( \hat{\Omega}_{1}\wedge D_{i}\hat{\Omega}%
_{3}\right) .
\end{array}
\right.  \label{DAOmega4}
\end{equation}
\begin{eqnarray}
&&
\begin{array}{l}
D_{A}D_{B}\hat{\Omega}_{4}=D_{(A}D_{B)}\hat{\Omega}_{4}: \\
\\
\\
\left\{
\begin{array}{l}
\left( A,B\right) =\left( \underline{0},\underline{0}\right) :D_{\underline{0%
}}D_{\underline{0}}\hat{\Omega}_{4}=e_{\underline{0}}^{a}e_{\underline{0}%
}^{b}D_{a}D_{b}\hat{\Omega}_{4}=\left( e_{\underline{0}}^{0}\right)
^{2}D_{0}D_{0}\hat{\Omega}_{4}=0; \\
\\
\\
\left( A,B\right) =\left( \underline{0},I\right) :D_{\underline{0}}D_{I}\hat{%
\Omega}_{4}=e_{\underline{0}}^{a}e_{I}^{b}D_{a}D_{b}\hat{\Omega}_{4}=e_{%
\underline{0}}^{0}e_{I}^{i}D_{0}D_{i}\hat{\Omega}_{4}=D_{\underline{0}}\hat{%
\Omega}_{1}\wedge D_{I}\hat{\Omega}_{3}=\overline{\hat{\Omega}}_{1}\wedge
D_{I}\hat{\Omega}_{3}; \\
\\
\\
\left( A,B\right) =\left( I,J\right) :\left\{
\begin{array}{l}
D_{I}D_{J}\hat{\Omega}_{4}=e_{I}^{a}e_{J}^{b}D_{a}D_{b}\hat{\Omega}_{4}= \\
\\
=e_{I}^{i}e_{J}^{j}D_{i}D_{j}\hat{\Omega}_{4}=e_{I}^{i}e_{J}^{j}\left( \hat{%
\Omega}_{1}\wedge D_{i}D_{j}\hat{\Omega}_{3}\right) = \\
\\
=-e_{I}^{i}e_{J}^{j}e^{-K_{1}}C_{ijk}g^{k\overline{l}}\overline{D}_{%
\overline{0}}\overline{D}_{\overline{l}}\overline{\hat{\Omega}}%
_{4}=iC_{IJK}\delta ^{K\overline{K}}\hat{\Omega}_{1}\wedge \overline{D}_{%
\overline{K}}\overline{\hat{\Omega}}_{3}= \\
\\
=iC_{IJK}\delta ^{K\overline{K}}\overline{D}_{\underline{\overline{0}}}%
\overline{\hat{\Omega}}_{1}\wedge \overline{D}_{\overline{K}}\overline{\hat{%
\Omega}}_{3}.
\end{array}
\right.
\end{array}
\right.
\end{array}
\notag \\
&&  \label{DADBOmega4}
\end{eqnarray}
\begin{eqnarray}
&&
\begin{array}{l}
D_{A}D_{B}D_{C}\hat{\Omega}_{4}=D_{(A}D_{B}D_{C)}\hat{\Omega}_{4}: \\
\\
\\
\left\{
\begin{array}{l}
\left( A,B,C\right) =\left( \underline{0},\underline{0},\underline{0}\right)
:D_{\underline{0}}D_{\underline{0}}D_{\underline{0}}\hat{\Omega}_{4}=e_{%
\underline{0}}^{a}e_{\underline{0}}^{b}e_{\underline{0}}^{c}D_{a}D_{b}D_{c}%
\hat{\Omega}_{4}=\left( e_{\underline{0}}^{0}\right) ^{3}D_{0}D_{0}D_{0}\hat{%
\Omega}_{4}=0; \\
\\
\\
\left( A,B,C\right) =\left( \underline{0},\underline{0},I\right) :D_{%
\underline{0}}D_{\underline{0}}D_{I}\hat{\Omega}_{4}=e_{\underline{0}}^{a}e_{%
\underline{0}}^{b}e_{I}^{c}D_{a}D_{b}D_{c}\hat{\Omega}_{4}=\left( e_{%
\underline{0}}^{0}\right) ^{2}e_{I}^{i}D_{0}D_{0}D_{i}\hat{\Omega}_{4}=0; \\
\\
\\
\left( A,B,C\right) =\left( \underline{0},I,J\right) :\left\{
\begin{array}{l}
D_{\underline{0}}D_{I}D_{J}\hat{\Omega}_{4}=e_{\underline{0}%
}^{a}e_{I}^{b}e_{J}^{c}D_{a}D_{b}D_{c}\hat{\Omega}_{4}=e_{\underline{0}%
}^{0}e_{I}^{i}e_{J}^{j}D_{0}D_{i}D_{j}\hat{\Omega}_{4}= \\
\\
=ie_{\underline{0}}^{0}e_{I}^{i}e_{J}^{j}\left( \overline{t}^{\overline{0}%
}-t^{0}\right) ^{-1}C_{ijk}g^{k\overline{l}}\overline{D}_{\overline{l}}%
\overline{\hat{\Omega}}_{4}=iC_{IJK}\delta ^{K\overline{K}}\overline{D}_{%
\overline{K}}\overline{\hat{\Omega}}_{4};
\end{array}
\right. \\
\\
\\
\left( A,B,C\right) =\left( I,J,K\right) :\left\{
\begin{array}{l}
D_{I}D_{J}D_{K}\hat{\Omega}_{4}=e_{I}^{a}e_{J}^{b}e_{K}^{c}D_{a}D_{b}D_{c}%
\hat{\Omega}_{4}=e_{I}^{i}e_{J}^{j}e_{K}^{k}D_{i}D_{j}D_{k}\hat{\Omega}_{4}=
\\
\\
=-ie_{I}^{i}e_{J}^{j}e_{K}^{k}\left( \overline{t}^{\overline{0}%
}-t^{0}\right) \left( D_{i}C_{jkl}\right) g^{l\overline{m}}\overline{D}_{%
\overline{0}}\overline{D}_{\overline{m}}\overline{\hat{\Omega}}_{4}+ \\
\\
-ie_{I}^{i}e_{J}^{j}e_{K}^{k}\left( \overline{t}^{\overline{0}}-t^{0}\right)
C_{ijk}\overline{D}_{\overline{0}}\overline{\hat{\Omega}}_{4}= \\
\\
=i\left( D_{I}C_{JKL}\right) \delta ^{L\overline{L}}\overline{D}_{\overline{%
\underline{0}}}\overline{D}_{\overline{L}}\overline{\hat{\Omega}}%
_{4}+iC_{IJK}\overline{D}_{\overline{\underline{0}}}\overline{\hat{\Omega}}%
_{4}.
\end{array}
\right.
\end{array}
\right.
\end{array}
\notag \\
&&  \label{DADBDCOmega4}
\end{eqnarray}
\begin{eqnarray}
&&
\begin{array}{l}
\overline{D}_{\overline{A}}D_{B}D_{C}\hat{\Omega}_{4}=\overline{D}_{%
\overline{A}}D_{(B}D_{C)}\hat{\Omega}_{4}: \\
\\
\\
\left\{
\begin{array}{l}
\left( \overline{A},B,C\right) =\left( \overline{\underline{0}},\underline{0}%
,\underline{0}\right) :\overline{D}_{\overline{\underline{0}}}D_{\underline{0%
}}D_{\underline{0}}\hat{\Omega}_{4}=\overline{e}_{\overline{\underline{0}}}^{%
\overline{a}}e_{\underline{0}}^{b}e_{\underline{0}}^{c}\overline{D}_{%
\overline{a}}D_{b}D_{c}\hat{\Omega}_{4}=\overline{e}_{\overline{\underline{0}%
}}^{\overline{0}}\left( e_{\underline{0}}^{0}\right) ^{2}\overline{D}_{%
\overline{0}}D_{0}D_{0}\hat{\Omega}_{4}=0; \\
\\
\\
\begin{array}{l}
\left( \overline{A},B,C\right) =\left( \overline{\underline{0}},\underline{0}%
,I\right) :\overline{D}_{\overline{\underline{0}}}D_{\underline{0}}D_{I}\hat{%
\Omega}_{4}=\overline{e}_{\overline{\underline{0}}}^{\overline{a}}e_{%
\underline{0}}^{b}e_{I}^{c}\overline{D}_{\overline{a}}D_{b}D_{c}\hat{\Omega}%
_{4}= \\
\\
=\overline{e}_{\overline{\underline{0}}}^{\overline{0}}e_{\underline{0}%
}^{0}e_{I}^{i}\overline{D}_{\overline{0}}D_{0}D_{i}\hat{\Omega}_{4}=%
\overline{e}_{\overline{\underline{0}}}^{\overline{0}}e_{\underline{0}%
}^{0}e_{I}^{i}g_{0\overline{0}}D_{i}\hat{\Omega}_{4}=D_{I}\hat{\Omega}_{4};
\end{array}
\\
\\
\\
\left( \overline{A},B,C\right) =\left( \overline{\underline{0}},I,J\right) :%
\overline{D}_{\overline{\underline{0}}}D_{I}D_{J}\hat{\Omega}_{4}=\overline{e%
}_{\overline{\underline{0}}}^{\overline{a}}e_{I}^{b}e_{J}^{c}\overline{D}_{%
\overline{a}}D_{b}D_{c}\hat{\Omega}_{4}=\overline{e}_{\overline{\underline{0}%
}}^{\overline{0}}e_{I}^{i}e_{J}^{j}\overline{D}_{\overline{0}}D_{i}D_{j}\hat{%
\Omega}_{4}=0; \\
\\
\\
\left( \overline{A},B,C\right) =\left( \overline{L},\underline{0},\underline{%
0}\right) :\overline{D}_{\overline{L}}D_{\underline{0}}D_{\underline{0}}\hat{%
\Omega}_{4}=\overline{e}_{\overline{L}}^{\overline{a}}e_{\underline{0}%
}^{b}e_{\underline{0}}^{c}\overline{D}_{\overline{a}}D_{b}D_{c}\hat{\Omega}%
_{4}=\overline{e}_{\overline{L}}^{\overline{l}}\left( e_{\underline{0}%
}^{0}\right) ^{2}\overline{D}_{\overline{l}}D_{0}D_{0}\hat{\Omega}_{4}=0; \\
\\
\\
\begin{array}{l}
\left( \overline{A},B,C\right) =\left( \overline{L},\underline{0},I\right) :%
\overline{D}_{\overline{L}}D_{\underline{0}}D_{I}\hat{\Omega}_{4}=\overline{e%
}_{\overline{L}}^{\overline{a}}e_{\underline{0}}^{b}e_{I}^{c}\overline{D}_{%
\overline{a}}D_{b}D_{c}\hat{\Omega}_{4}= \\
\\
=\overline{e}_{\overline{L}}^{\overline{l}}e_{\underline{0}}^{0}e_{I}^{i}%
\overline{D}_{\overline{l}}D_{0}D_{i}\hat{\Omega}_{4}=\overline{e}_{%
\overline{L}}^{\overline{l}}e_{\underline{0}}^{0}e_{I}^{i}g_{i\overline{l}%
}D_{0}\hat{\Omega}_{4}=\delta _{I\overline{L}}D_{\underline{0}}\hat{\Omega}%
_{4};
\end{array}
\\
\\
\\
\left( \overline{A},B,C\right) =\left( \overline{L},I,J\right) :\left\{
\begin{array}{l}
\overline{D}_{\overline{L}}D_{I}D_{J}\hat{\Omega}_{4}=\overline{e}_{%
\overline{L}}^{\overline{a}}e_{I}^{b}e_{J}^{c}\overline{D}_{\overline{a}%
}D_{b}D_{c}\hat{\Omega}_{4}=\overline{e}_{\overline{L}}^{\overline{l}%
}e_{I}^{i}e_{J}^{j}\overline{D}_{\overline{l}}D_{i}D_{j}\hat{\Omega}_{4}= \\
\\
=\overline{e}_{\overline{L}}^{\overline{l}}e_{I}^{i}e_{J}^{j}g^{k\overline{k}%
}C_{ijk}\overline{C}_{\overline{l}\overline{m}\overline{k}}g^{m\overline{m}%
}D_{m}\hat{\Omega}_{4}= \\
\\
\overset{\text{\textit{SKG constraints in local ``flat'' coords.}}}{=}\left(
\delta _{J}^{M}\delta _{I\overline{L}}+\delta _{I}^{M}\delta _{J\overline{L}%
}\right) D_{M}\hat{\Omega}_{4}.
\end{array}
\right.
\end{array}
\right.
\end{array}
\notag \\
&&  \label{DAbDBDCOmega4}
\end{eqnarray}
\setcounter{equation}0
\def\theequation{II.\arabic{equation}}
\section{Appendix II}

The \textit{``intersections''} among the elements of the set of 4-forms $%
\hat{\Omega}_{4}$,$~D_{0}\hat{\Omega}_{4}$,$~D_{i}\hat{\Omega}_{4}$,$%
~D_{0}D_{i}\hat{\Omega}_{4}$, $\overline{\hat{\Omega}}_{4}$,$~\overline{D}_{%
\overline{0}}\overline{\hat{\Omega}}_{4}$,$~\overline{D}_{\overline{i}}%
\overline{\hat{\Omega}}_{4}$ and$~\overline{D}_{\overline{0}}\overline{D}_{%
\overline{i}}\overline{\hat{\Omega}}_{4}$ in generic local ``curved'' and in
local ``flat'' coordinates of $M$ respectively read as follows:
\begin{eqnarray}
&&
\begin{array}{l}
\int_{CY_{4}}\hat{\Omega}_{4}\wedge \hat{\Omega}_{4}=0,~~\int_{CY_{4}}\hat{%
\Omega}_{4}\wedge D_{0}\hat{\Omega}_{4}=0,~~\int_{CY_{4}}\hat{\Omega}%
_{4}\wedge D_{i}\hat{\Omega}_{4}=0,~~\int_{CY_{4}}\hat{\Omega}_{4}\wedge
D_{0}D_{i}\hat{\Omega}_{4}=0;~ \\
\\
\int_{CY_{4}}\hat{\Omega}_{4}\wedge \overline{\hat{\Omega}}_{4}=1; \\
\\
\int_{CY_{4}}\hat{\Omega}_{4}\wedge \overline{D}_{\overline{0}}\overline{%
\hat{\Omega}}_{4}=0,~~\int_{CY_{4}}\hat{\Omega}_{4}\wedge \overline{D}_{%
\overline{i}}\overline{\hat{\Omega}}_{4}=0,~~\int_{CY_{4}}\hat{\Omega}%
_{4}\wedge \overline{D}_{\overline{0}}\overline{D}_{\overline{i}}\overline{%
\hat{\Omega}}_{4}=0;
\end{array}
\label{intersect1} \\
&&
\begin{array}{l}
\int_{CY_{4}}D_{i}\hat{\Omega}_{4}\wedge D_{j}\hat{\Omega}%
_{4}=0,~~\int_{CY_{4}}D_{i}\hat{\Omega}_{4}\wedge D_{0}\hat{\Omega}%
_{4}=0,~~\int_{CY_{4}}D_{i}\hat{\Omega}_{4}\wedge D_{0}D_{j}\hat{\Omega}%
_{4}=0;~ \\
\\
\int_{CY_{4}}D_{i}\hat{\Omega}_{4}\wedge \overline{D}_{\overline{j}}%
\overline{\hat{\Omega}}_{4}=-g_{i\overline{j}}; \\
\\
\int_{CY_{4}}D_{i}\hat{\Omega}_{4}\wedge \overline{D}_{\overline{0}}%
\overline{\hat{\Omega}}_{4}=0,~~\int_{CY_{4}}D_{i}\hat{\Omega}_{4}\wedge
\overline{D}_{\overline{0}}\overline{D}_{\overline{j}}\overline{\hat{\Omega}}%
_{4}=0;
\end{array}
\label{intersect2} \\
&&
\begin{array}{l}
\int_{CY_{4}}D_{0}\hat{\Omega}_{4}\wedge D_{0}\hat{\Omega}%
_{4}=0,~~\int_{CY_{4}}D_{0}\hat{\Omega}_{4}\wedge D_{0}D_{i}\hat{\Omega}%
_{4}=0; \\
\\
\int_{CY_{4}}D_{0}\hat{\Omega}_{4}\wedge \overline{D}_{\overline{0}}%
\overline{\hat{\Omega}}_{4}=-e^{2K_{1}}=\left( \overline{t}^{\overline{0}%
}-t^{0}\right) ^{-2}; \\
\\
\int_{CY_{4}}D_{0}\hat{\Omega}_{4}\wedge \overline{D}_{\overline{0}}%
\overline{D}_{\overline{i}}\overline{\hat{\Omega}}_{4}=0;
\end{array}
\label{intersect3} \\
&&
\begin{array}{l}
\int_{CY_{4}}D_{0}D_{i}\hat{\Omega}_{4}\wedge D_{0}D_{j}\hat{\Omega}_{4}=0;
\\
\\
\int_{CY_{4}}D_{0}D_{i}\hat{\Omega}_{4}\wedge \overline{D}_{\overline{0}}%
\overline{D}_{\overline{j}}\overline{\hat{\Omega}}_{4}=e^{2K_{1}}g_{i%
\overline{j}}=-\left( \overline{t}^{\overline{0}}-t^{0}\right) ^{-2}g_{i%
\overline{j}}.
\end{array}
\label{intersect4}
\end{eqnarray}
\begin{eqnarray}
&&
\begin{array}{l}
\int_{CY_{4}}\hat{\Omega}_{4}\wedge \hat{\Omega}_{4}=0,~~\int_{CY_{4}}\hat{%
\Omega}_{4}\wedge D_{\underline{0}}\hat{\Omega}_{4}=0,~~\int_{CY_{4}}\hat{%
\Omega}_{4}\wedge D_{I}\hat{\Omega}_{4}=0,~~\int_{CY_{4}}\hat{\Omega}%
_{4}\wedge D_{\underline{0}}D_{I}\hat{\Omega}_{4}=0;~ \\
\\
\int_{CY_{4}}\hat{\Omega}_{4}\wedge \overline{\hat{\Omega}}_{4}=1; \\
\\
\int_{CY_{4}}\hat{\Omega}_{4}\wedge \overline{D}_{\underline{\overline{0}}}%
\overline{\hat{\Omega}}_{4}=0,~~\int_{CY_{4}}\hat{\Omega}_{4}\wedge
\overline{D}_{\overline{I}}\overline{\hat{\Omega}}_{4}=0,~~\int_{CY_{4}}\hat{%
\Omega}_{4}\wedge \overline{D}_{\overline{\underline{0}}}\overline{D}_{%
\overline{I}}\overline{\hat{\Omega}}_{4}=0;
\end{array}
\label{flat-intersect1} \\
&&
\begin{array}{l}
\int_{CY_{4}}D_{I}\hat{\Omega}_{4}\wedge D_{J}\hat{\Omega}%
_{4}=0,~~\int_{CY_{4}}D_{I}\hat{\Omega}_{4}\wedge D_{\underline{0}}\hat{%
\Omega}_{4}=0,~~\int_{CY_{4}}D_{I}\hat{\Omega}_{4}\wedge D_{\underline{0}%
}D_{J}\hat{\Omega}_{4}=0;~ \\
\\
\int_{CY_{4}}D_{I}\hat{\Omega}_{4}\wedge \overline{D}_{\overline{J}}%
\overline{\hat{\Omega}}_{4}=-e_{I}^{i}\overline{e}_{\overline{J}}^{\overline{%
j}}g_{i\overline{j}}=-\delta _{I\overline{J}}; \\
\\
\int_{CY_{4}}D_{I}\hat{\Omega}_{4}\wedge \overline{D}_{\overline{\underline{0%
}}}\overline{\hat{\Omega}}_{4}=0,~~\int_{CY_{4}}D_{I}\hat{\Omega}_{4}\wedge
\overline{D}_{\overline{\underline{0}}}\overline{D}_{\overline{J}}\overline{%
\hat{\Omega}}_{4}=0;
\end{array}
\label{flat-intersect2} \\
&&
\begin{array}{l}
\int_{CY_{4}}D_{\underline{0}}\hat{\Omega}_{4}\wedge D_{\underline{0}}\hat{%
\Omega}_{4}=0,~~\int_{CY_{4}}D_{\underline{0}}\hat{\Omega}_{4}\wedge D_{%
\underline{0}}D_{I}\hat{\Omega}_{4}=0; \\
\\
\int_{CY_{4}}D_{\underline{0}}\hat{\Omega}_{4}\wedge \overline{D}_{\overline{%
\underline{0}}}\overline{\hat{\Omega}}_{4}=-\left| e_{\underline{0}%
}^{0}\right| ^{2}e^{2K_{1}}=-1; \\
\\
\int_{CY_{4}}D_{\underline{0}}\hat{\Omega}_{4}\wedge \overline{D}_{\overline{%
\underline{0}}}\overline{D}_{\overline{I}}\overline{\hat{\Omega}}_{4}=0;
\end{array}
\label{flat-intersect3} \\
&&
\begin{array}{l}
\int_{CY_{4}}D_{\underline{0}}D_{I}\hat{\Omega}_{4}\wedge D_{\underline{0}%
}D_{J}\hat{\Omega}_{4}=0; \\
\\
\int_{CY_{4}}D_{\underline{0}}D_{I}\hat{\Omega}_{4}\wedge \overline{D}_{%
\overline{\underline{0}}}\overline{D}_{\overline{J}}\overline{\hat{\Omega}}%
_{4}=\left| e_{\underline{0}}^{0}\right| ^{2}e_{I}^{i}\overline{e}_{%
\overline{J}}^{\overline{j}}e^{2K_{1}}g_{i\overline{j}}=\delta _{I\overline{J%
}}.
\end{array}
\label{flat-intersect4}
\end{eqnarray}
\setcounter{equation}0
\def\theequation{III.\arabic{equation}}
\section{Appendix III}

The complete Hodge-decomposition of the real, K\"{a}hler gauge-invariant
4-form $\frak{F}_{4}$ of Type IIB on $\frac{CY_{3}\times T^{2}}{\mathbb{Z}%
_{2}}$ in generic local ``curved'' coordinates in $M$ reads as follows:
\begin{eqnarray}
\frak{F}_{4} &=&\left[
\begin{array}{l}
Z\overline{\hat{\Omega}}_{4}-g^{a\overline{b}}\left( D_{a}Z\right) \overline{%
D}_{\overline{b}}\overline{\hat{\Omega}}_{4}+\left| e_{\underline{0}%
}^{0}\right| ^{2}g^{a\overline{b}}\left( D_{0}D_{a}Z\right) \overline{D}_{%
\overline{0}}\overline{D}_{\overline{b}}\overline{\hat{\Omega}}_{4}+ \\
\\
+\left| e_{\underline{0}}^{0}\right| ^{2}g^{b\overline{a}}\left( \overline{D}%
_{\overline{0}}\overline{D}_{\overline{a}}\overline{Z}\right) D_{0}D_{b}\hat{%
\Omega}_{4}-g^{b\overline{a}}\left( \overline{D}_{\overline{a}}\overline{Z}%
\right) D_{b}\hat{\Omega}_{4}+\overline{Z}\hat{\Omega}_{4}
\end{array}
\right] =  \notag \\
&&  \notag \\
&=&2Re\left[ \overline{Z}\hat{\Omega}_{4}-g^{a\overline{b}}\left( \overline{D%
}_{\overline{b}}\overline{Z}\right) D_{a}\hat{\Omega}_{4}+\left| e_{%
\underline{0}}^{0}\right| ^{2}g^{a\overline{b}}\left( \overline{D}_{%
\overline{0}}\overline{D}_{\overline{b}}\overline{Z}\right) D_{0}D_{a}\hat{%
\Omega}_{4}\right] =  \label{curved-FV-decomp2} \\
&&  \notag \\
&=&2Re\left[
\begin{array}{l}
\overline{Z}\hat{\Omega}_{1}\wedge \hat{\Omega}_{3}+ \\
\\
+\left( t^{0}-\overline{t}^{\overline{0}}\right) ^{2}\left( \overline{D}_{%
\overline{0}}\overline{Z}\right) \overline{\hat{\Omega}}_{1}\wedge \hat{%
\Omega}_{3}-g^{i\overline{j}}\left( \overline{D}_{\overline{j}}\overline{Z}%
\right) \hat{\Omega}_{1}\wedge D_{i}\hat{\Omega}_{3}+ \\
\\
+\left( t^{0}-\overline{t}^{\overline{0}}\right) g^{i\overline{j}}\left(
\overline{D}_{\overline{0}}\overline{D}_{\overline{j}}\overline{Z}\right)
\overline{\hat{\Omega}}_{1}\wedge D_{i}\hat{\Omega}_{3}
\end{array}
\right] =  \label{curved-FV-decomp3} \\
&&  \notag \\
&=&2Re\left[
\begin{array}{l}
\overline{Z}\hat{\Omega}_{1}\wedge \hat{\Omega}_{3}+ \\
\\
-\left| e_{\underline{0}}^{0}\right| ^{2}\left( \overline{D}_{\overline{0}}%
\overline{Z}\right) \overline{\hat{\Omega}}_{1}\wedge \hat{\Omega}%
_{3}-e_{I}^{i}\overline{e}_{\overline{J}}^{\overline{j}}\delta ^{I\overline{J%
}}\left( \overline{D}_{\overline{j}}\overline{Z}\right) \hat{\Omega}%
_{1}\wedge D_{i}\hat{\Omega}_{3}+ \\
\\
+\overline{e}_{\overline{\underline{0}}}^{\overline{0}}e_{I}^{i}\overline{e}%
_{\overline{J}}^{\overline{j}}\delta ^{I\overline{J}}\left( \overline{D}_{%
\overline{0}}\overline{D}_{\overline{j}}\overline{Z}\right) \overline{\hat{%
\Omega}}_{1}\wedge D_{i}\hat{\Omega}_{3}
\end{array}
\right] =  \label{curved-FV-decomp4} \\
&&  \notag \\
&=&2e^{K_{1}+K_{3}}Re\left[
\begin{array}{l}
\overline{W}\Omega _{1}\wedge \Omega _{3}+ \\
\\
-\left| e_{\underline{0}}^{0}\right| ^{2}\left( \overline{D}_{\overline{0}}%
\overline{W}\right) \overline{\Omega }_{1}\wedge \Omega _{3}-e_{I}^{i}%
\overline{e}_{\overline{J}}^{\overline{j}}\delta ^{I\overline{J}}\left(
\overline{D}_{\overline{j}}\overline{W}\right) \Omega _{1}\wedge D_{i}\Omega
_{3}+ \\
\\
+\overline{e}_{\overline{\underline{0}}}^{\overline{0}}e_{I}^{i}\overline{e}%
_{\overline{J}}^{\overline{j}}\delta ^{I\overline{J}}\left( \overline{D}_{%
\overline{0}}\overline{D}_{\overline{j}}\overline{W}\right) \overline{\Omega
}_{1}\wedge D_{i}\Omega _{3}
\end{array}
\right] .  \label{curved-FV-decomp5}
\end{eqnarray}

The evaluation of such identities along the constraints (\ref{FV-SUSY})
yields the supersymmetric FV AEs in $\mathcal{N}=1$, $d=4$ supergravity from
Type IIB on $\frac{CY_{3}\times T^{2}}{\mathbb{Z}_{2}}$ in local ``curved''
coordinates:
\begin{eqnarray}
&&
\begin{array}{l}
\frak{F}_{4}=\left[ Z\overline{\hat{\Omega}}_{4}+\left| e_{\underline{0}%
}^{0}\right| ^{2}g^{a\overline{b}}\left( D_{0}D_{a}Z\right) \overline{D}_{%
\overline{0}}\overline{D}_{\overline{b}}\overline{\hat{\Omega}}_{4}+\left|
e_{\underline{0}}^{0}\right| ^{2}g^{b\overline{a}}\left( \overline{D}_{%
\overline{0}}\overline{D}_{\overline{a}}\overline{Z}\right) D_{0}D_{b}\hat{%
\Omega}_{4}+\overline{Z}\hat{\Omega}_{4}\right] _{SUSY}= \\
\\
=2Re\left[ \overline{Z}\hat{\Omega}_{4}+\left| e_{\underline{0}}^{0}\right|
^{2}g^{a\overline{b}}\left( \overline{D}_{\overline{0}}\overline{D}_{%
\overline{b}}\overline{Z}\right) D_{0}D_{a}\hat{\Omega}_{4}\right] _{SUSY}=
\\
\\
=2Re\left[ \overline{Z}\hat{\Omega}_{1}\wedge \hat{\Omega}_{3}+\left( t^{0}-%
\overline{t}^{\overline{0}}\right) g^{i\overline{j}}\left( \overline{D}_{%
\overline{0}}\overline{D}_{\overline{j}}\overline{Z}\right) \overline{\hat{%
\Omega}}_{1}\wedge D_{i}\hat{\Omega}_{3}\right] _{SUSY}= \\
\\
=2Re\left[ \overline{Z}\hat{\Omega}_{1}\wedge \hat{\Omega}_{3}+\overline{e}_{%
\overline{\underline{0}}}^{\overline{0}}e_{I}^{i}\overline{e}_{\overline{J}%
}^{\overline{j}}\delta ^{I\overline{J}}\left( \overline{D}_{\overline{0}}%
\overline{D}_{\overline{j}}\overline{Z}\right) \overline{\hat{\Omega}}%
_{1}\wedge D_{i}\hat{\Omega}_{3}\right] _{SUSY}= \\
\\
=2e^{K_{1}+K_{3}}Re\left[ \overline{W}\Omega _{1}\wedge \Omega _{3}+%
\overline{e}_{\overline{\underline{0}}}^{\overline{0}}e_{I}^{i}\overline{e}_{%
\overline{J}}^{\overline{j}}\delta ^{I\overline{J}}\left( \overline{D}_{%
\overline{0}}\overline{D}_{\overline{j}}\overline{W}\right) \overline{\Omega
}_{1}\wedge D_{i}\Omega _{3}\right] _{SUSY}.
\end{array}
\notag \\
&&  \label{SUSY-FV-AEs2}
\end{eqnarray}

\end{document}